\documentclass[11pt,a4paper]{article}

\usepackage{jheppub}

\title{\boldmath Black holes with ${\mathfrak {su}}(N)$ gauge field hair and superconducting horizons}

\author{Ben~L.~Shepherd and Elizabeth Winstanley}

\affiliation{Consortium for Fundamental Physics, School of Mathematics and Statistics,\\
The University of Sheffield, Hicks Building, Hounsfield Road, Sheffield. S3 7RH United Kingdom.}

\emailAdd{E.Winstanley@sheffield.ac.uk}

\abstract{We present new planar dyonic black hole solutions of the ${\mathfrak {su}}(N)$ Einstein-Yang-Mills equations in asymptotically anti-de Sitter space-time, focussing on ${\mathfrak {su}}(2)$ and ${\mathfrak {su}}(3)$ gauge groups.
The magnetic part of the gauge field forms a condensate close to the planar event horizon.  We compare the free energy of a non-Abelian hairy black hole with that of an embedded Reissner-Nordstr\"om-anti-de Sitter (RN-AdS) black hole having the same Hawking temperature and electric charge.
We find that the hairy black holes have lower free energy. We present evidence that there is a phase transition at a critical temperature, above which the only solutions are embedded RN-AdS black holes. At the critical temperature, an RN-AdS black hole can decay into a hairy black hole, and it is thermodynamically favourable to do so.
Working in the probe limit, we compute the frequency-dependent conductivity, and find that enlarging the gauge group from ${\mathfrak {su}}(2)$ to ${\mathfrak {su}}(3)$ eliminates a divergence in the conductivity at nonzero frequency.}


\arxivnumber{1611.04162}

\begin{document}
\maketitle
\flushbottom

\section{Introduction}
\label{sec:intro}

Classical hairy black hole solutions of the Einstein-Yang-Mills (EYM) equations have many interesting properties (see, for example, \cite{Volkov:1998cc,Winstanley:2008ac,Winstanley:2015loa,Volkov:2016ehx} for some reviews).
In four-dimensional asymptotically flat space-time, with gauge group ${\mathfrak {su}}(2)$, nontrivial hairy black holes must have a purely magnetic gauge field configuration
\cite{Galtsov:1989ip,Bizon:1992pi} which is described by a single function $\omega $.
Numerical solutions of the asymptotically flat ${\mathfrak {su}}(2)$ EYM equations representing hairy black holes are parameterized by the event horizon radius and the number of zeros of $\omega $ \cite{Volkov:1989fi,Volkov:1990sva,Bizon:1990sr,Kuenzle:1990is}.
These black holes are ``hairy'' in the sense that the metric is not Schwarzschild near the horizon. At infinity, the metric approaches that of Schwarzchild space-time and the EYM black holes possess no charges to distinguish them from Schwarzchild black holes.
Since the discovery of the ${\mathfrak {su}}(2)$ EYM black holes \cite{Volkov:1989fi,Volkov:1990sva,Bizon:1990sr,Kuenzle:1990is}, an extensive literature on asymptotically flat black hole solutions of the EYM equations and various related matter models has developed \cite{Volkov:1998cc}.
For instance, the gauge group can be enlarged to ${\mathfrak {su}}(N)$ \cite{Galtsov:1991au,Kleihaus:1995tk,Kleihaus:1997rb,Kleihaus:1998qc,Mavromatos:1997zb,Ruan:2001dw}, in which case purely magnetic configurations are described by $N-1$ gauge field functions $\omega _{j}$.
However, one important property of all the purely magnetic, asymptotically flat, four-dimensional, EYM black holes is that they are dynamically unstable under small perturbations of the metric and gauge field \cite{Straumann:1990as,Galtsov:1991nk,Volkov:1994dq,Hod:2008ir,Lavrelashvili:1994rp,Volkov:1995np,Brodbeck:1994vu}.

Inspired by the AdS/CFT (anti-de Sitter space/conformal field theory) correspondence \cite{Maldacena:1997re,Witten:1998qj,Gubser:1998bc,Aharony:1999ti}, black holes with non-Abelian gauge field hair in four-dimensional asymptotically AdS space-times have also been extensively studied \cite{Winstanley:2008ac,Winstanley:2015loa}, both for gauge group ${\mathfrak {su}}(2)$ \cite{Winstanley:1998sn,Bjoraker:1999yd,Bjoraker:2000qd} and the larger ${\mathfrak {su}}(N)$ gauge group \cite{Baxter:2007at,Baxter:2007au,Baxter:2008pi,Baxter:2015gfa}.
In contrast with their asymptotically flat counterparts, there exist ${\mathfrak {su}}(2)$ purely magnetic hairy black holes for which the single gauge field function $\omega $ has no zeros \cite{Winstanley:1998sn,Bjoraker:1999yd,Bjoraker:2000qd}.
These nodeless black holes exist when the magnitude of the cosmological constant is sufficiently large (equivalently, the AdS radius of curvature is sufficiently small) and are particularly interesting because at least some of them are dynamically stable, that is, linearized mode perturbations of the metric and gauge field do not grow exponentially with time \cite{Winstanley:1998sn,Bjoraker:1999yd,Bjoraker:2000qd,Sarbach:2001mc,Winstanley:2001bs}.
The thermodynamics of these hairy black holes is studied in \cite{Mann:2006jc,Shepherd:2012sz,Fan:2014ixa,Kichakova:2015lza}. A detailed systematic study \cite{Kichakova:2015lza} revealed that while spherically symmetric embedded Abelian Reissner-Nordstr\"om-AdS (RN-AdS) black holes are thermodynamically unstable to the formation of non-Abelian gauge field hair, nonetheless the hairy black holes are globally thermodynamically unstable.
There are also purely magnetic ${\mathfrak {su}}(N)$ hairy black holes for which all of the $N-1$ magnetic gauge field functions $\omega _{j}$ have no zeros \cite{Baxter:2008pi}.
If the AdS radius of curvature is sufficiently small, it can be proven that at least some of these nodeless hairy black holes are dynamically stable \cite{Baxter:2015gfa}.

In four-dimensional asymptotically AdS space-time, unlike the situation in asymptotically flat space-time, dyonic EYM black holes exist \cite{Bjoraker:1999yd,Bjoraker:2000qd,Nolan:2012ax,Shepherd:2015dse} as well as the purely magnetic black holes discussed above.
In this case the gauge field has a nontrivial electric part as well as a magnetic part.
With gauge group ${\mathfrak {su}}(2)$, the existence of dynamically stable dyonic hairy black holes has recently been proven \cite{Nolan:2015vca}.
For the larger gauge group ${\mathfrak {su}}(N)$, numerical solutions of the EYM equations representing dyonic black holes have been found \cite{Shepherd:2015dse}, and it has been proven that, for sufficiently small AdS radius of curvature, there exist hairy black holes for which the $N-1$ magnetic gauge field functions $\omega _{j}$ are all nodeless \cite{Baxter:2015tda}.
It is likely that at least some of these nodeless dyonic ${\mathfrak {su}}(N)$ black holes are dynamically stable, but this remains to be proven.

The discussion thus far has been concerned only with static, spherically symmetric hairy black holes in four-dimensional EYM theory, either in asymptotically flat or asymptotically AdS space-time
(EYM black holes with gauge field hair also exist in higher-dimensional space-times - see the review \cite{Volkov:2006xt} for details and references - but we shall only consider four-dimensional black holes in this paper).
In four-dimensional asymptotically AdS space-time, it is well-known that static electrovacuum black holes do not need to have spherical event horizon topology
\cite{Birmingham:1998nr,Brill:1997mf,Lemos:1994fn,Lemos:1994xp,Lemos:1995cm,Vanzo:1997gw,Cai:1996eg,Mann:1996gj,Smith:1997wx,Mann:1997zn}.
Purely magnetic topological hairy black holes exist in both ${\mathfrak {su}}(2)$ \cite{VanderBij:2001ia} and, more generally, ${\mathfrak {su}}(N)$ \cite{Baxter:2014nka,Baxter:2015ffm} EYM theory in AdS.
With gauge group ${\mathfrak {su}}(2)$, all topological black holes are dynamically stable and the single gauge field function $\omega $ has no zeros \cite{VanderBij:2001ia}.  For the larger ${\mathfrak {su}}(N)$ gauge group (with $N>2$), the existence of nodeless topological black holes has been proven \cite{Baxter:2014nka}, at least some of which are dynamically stable \cite{Baxter:2015xfa}.
However, the phase space of purely magnetic topological black hole solutions is more complicated for the larger gauge group, and in particular it is possible for the magnetic gauge field functions to have zeros \cite{Baxter:2015ffm}.

The study of EYM black holes in AdS received fresh impetus in the context of holographic superconductors (see \cite{Hartnoll:2009sz,Herzog:2009xv,Horowitz:2010gk,Horowitz:2010nh,Kaminski:2010zu,Sachdev:2011wg,Benini:2012iq,Salvio:2013ja,Musso:2014efa,Cai:2015cya} for reviews of various aspects of holographic superconductors).
In the seminal paper \cite{Gubser:2008zu}, four-dimensional dyonic EYM black holes with planar event horizons and gauge group ${\mathfrak {su}}(2)$ are studied.
The electric part of the gauge field is described by a single function, and the magnetic part again by a single function $\omega $.
Unlike the purely magnetic case, for a dyonic gauge field configuration the magnetic gauge field function $\omega $ can have zeros.
This is crucial, since the solutions of particular interest for describing holographic superconductors are those for which $\omega $ vanishes on the AdS boundary.
When this happens, the magnetic part of the non-Abelian gauge field forms a condensate in a neighbourhood of the planar event horizon.
In \cite{Gubser:2008zu} it is shown that this happens when the black hole temperature is below a certain critical temperature $T_{C}$.
By computing the difference in free energy between an EYM black hole with a nontrivial condensate and an embedded planar RN-AdS black hole with the same temperature and electric charge, in \cite{Gubser:2008zu} it is shown that
the EYM black holes are thermodynamically favoured over the RN-AdS black holes.
If $(x,y)$ are the coordinates describing the planar event horizon, the ansatz for the gauge potential in \cite{Gubser:2008zu} has nonzero components in both the $x$ and $y$ directions, and corresponds to a model of a $p+ip$-wave superconductor \cite{Gubser:2008wv,Arias:2012py}.
Modelling $p$-wave superconductors with four-dimensional EYM black holes was initiated in \cite{Gubser:2008wv}, where the probe limit was considered (that is, the back-reaction of the gauge field on the space-time geometry was ignored).  The gauge potential ansatz in this case has no component in the $y$-direction of the planar event horizon. The gauge field is again described by two functions, one magnetic ($\omega $) and one electric.
As in the $p+ip$ case, the solutions of interest are those for which the magnetic gauge field function $\omega $ forms a condensate close to the planar event horizon, and vanishes on the AdS boundary.

In \cite{Gubser:2008wv} the frequency-dependent conductivities $\sigma _{xx}$ and $\sigma _{yy}$ (in the $x$ and $y$-directions respectively) were calculated in the probe limit from perturbations of the non-Abelian gauge field.
Both tend to constants for large frequencies, but exhibit different behaviour for lower frequencies.
The conductivity in the $x$-direction, $\sigma _{xx}$, has a pole in its imaginary part at a nonzero frequency (and the real part has a delta-function singularity at that value of the frequency).
The conductivity in the $y$-direction, $\sigma _{yy}$, shows gapped behaviour, its real part being very small for small frequency, rising rapidly to its large-frequency value.
In \cite{Gubser:2008wv}, some of the quasi-normal modes perturbing the black holes are also considered; this analysis suggests that the $p+ip$-wave configurations are dynamically unstable, while the $p$-wave configurations are dynamically stable.
The conductivity of the $p+ip$-wave solutions in the probe limit is studied in \cite{Roberts:2008ns}; the pole in the imaginary part at nonzero frequency persists.
There is also a pole in the imaginary part at zero frequency, corresponding to infinite DC conductivity.
A finite DC conductivity can be obtained in the probe limit using a non-Abelian gauge transformation \cite{Herzog:2014tpa}.

There is now a large literature on EYM black holes with superconducting horizons (see for example \cite{Cai:2015cya} for a recent review and a more complete list of references than we give here).
Considering four-dimensional planar EYM black holes with gauge group ${\mathfrak {su}}(2)$, the probe limit was further explored in papers including \cite{Roberts:2008ns,Gangopadhyay:2012gx,Arias:2014msa,Zeng:2009dr,Zeng:2010fs}, while the back-reaction of the gauge field on the black hole geometry is included in, for example, \cite{Basu:2009vv,Gubser:2010dm,Arias:2012py,Herzog:2014tpa}.
Aspects of holographic superconductors that have been studied in this model include fermion correlators \cite{Gubser:2010dm}, analytic approximations for the critical temperature \cite{Gangopadhyay:2012gx}, conductivity \cite{Roberts:2008ns,Basu:2009vv,Herzog:2014tpa}, thermodynamic phase diagrams \cite{Gubser:2010dm,Arias:2012py,Herzog:2014tpa}, superconducting coherence length \cite{Zeng:2009dr}, hydrodynamic modes \cite{Arias:2014msa} and fermionic spectral functions \cite{Giordano:2016tws}.
Higher-dimensional EYM black holes  with superconducting horizons have also been studied, see, for example,  \cite{Ammon:2009xh,Herzog:2009ci,Akhavan:2010bf,Cai:2010zm,Erdmenger:2012zu,Herzog:2014tpa,Manvelyan:2008sv}.

In the literature discussed above, a single degree of freedom (given by the function $\omega $) in the magnetic part of the non-Abelian gauge field breaks an Abelian symmetry when the condensate is present.
In this paper we explore the consequences of having more degrees of freedom breaking the Abelian symmetry.
Working in four space-time dimensions, we consider the EYM model with gauge group ${\mathfrak {su}}(N)$, and study planar AdS black holes with dyonic gauge field configurations.  Our solutions generalize those in \cite{Baxter:2015ffm} (where the gauge field is purely magnetic) and \cite{Shepherd:2015dse} (where spherically symmetric dyonic solutions are constructed).
As well as exploring the space of planar black hole solutions with a magnetic condensate, we focus on two aspects of holographic superconductors in this model, considering the effect of the larger gauge group on the free energy (and hence the thermodynamic stability of RN-AdS black holes with the same temperature and total electric charge) and the frequency-dependent conductivity.
Given the complexity of the fully coupled EYM field equations, we focus on the ${\mathfrak {su}}(2)$ and ${\mathfrak {su}}(3)$ gauge groups, comparing the properties of the planar dyonic EYM black holes in these two cases.
We generalize both the $p$-wave and $p+ip$-wave ansatze for the non-Abelian gauge field.
However, it turns out (see section \ref{sec:boundary}) that in the generalized $p$-wave case only embedded ${\mathfrak {su}}(2)$ solutions exist; there are no nontrivial ${\mathfrak {su}}(N)$ solutions.
Accordingly, we focus on the generalized $p+ip$-wave model, for which genuinely ${\mathfrak {su}}(N)$ planar EYM black holes can be found.

The outline of this paper is as follows. In section \ref{sec:eqs} we introduce the action for our ${\mathfrak {su}}(N)$ EYM model, gauge field and metric ansatze, and the field equations governing planar black holes.
Numerical solutions of the ${\mathfrak {su}}(2)$ and ${\mathfrak {su}}(3)$ field equations representing planar black holes with a non-Abelian gauge field condensate are presented in section \ref{sec:sols}.
Following \cite{Chrusciel:1987jr}, we define non-Abelian charges for these black holes and compute their free energy.
We study the difference in free energy between the non-Abelian EYM black holes and the embedded planar RN-AdS black holes having the same total charge.
We also study the critical temperature below which nontrivial EYM black holes exist.
To this point in the paper the system is fully back-reacting.  In section \ref{sec:pert} we ignore the back-reaction and work in the probe limit, studying oscillating non-Abelian gauge field perturbations of the embedded planar RN-AdS black holes.
From these perturbations we compute the frequency-dependent conductivities for ${\mathfrak {su}}(2)$ and ${\mathfrak {su}}(3)$ gauge groups.
Our conclusions are presented in section \ref{sec:conc}.

\section{General formalism}
\label{sec:eqs}

In this section we present the field equations for ${\mathfrak {su}}(N)$ EYM in AdS, together with our metric and gauge field ansatze, generalizing both the isotropic ($p+ip$-wave, ansatz I in \cite{Manvelyan:2008sv}) and anisotropic ($p$-wave, ansatz II in \cite{Manvelyan:2008sv}) ansatze for gauge group ${\mathfrak {su}}(2)$.
Since our form of the gauge field is different from that considered for planar ${\mathfrak {su}}(N)$ EYM black holes in \cite{Baxter:2015tda}, we explicitly check that our isotropic ansatz satisfies the required symmetry equations and hence is invariant under rotations in the plane.
From the field equations, we show that only the generalized isotropic ansatz has nontrivial ${\mathfrak {su}}(N)$ solutions.
We also discuss some trivial solutions and the scaling symmetries possessed by the field equations.

\subsection{Action and metric ansatz}
\label{sec:metric}

We consider four-dimensional ${\mathfrak {su}}(N)$ EYM theory with a negative cosmological constant $\Lambda $, described by the action
\begin{equation}
S = \int d^{4}x {\sqrt {-{\mathfrak {g}}}} \left[ \frac {1}{16\pi G} \left( R - 2\Lambda \right) - \frac {1}{4} {\text {Tr}} \,  {\mathfrak {F}}_{\mu \nu } {\mathfrak {F}}^{\mu \nu } \right] ,
\label{eq:action}
\end{equation}
where the metric has determinant $-{\mathfrak {g}}$, the quantity $R$ is the Ricci scalar, ${\mathfrak {F}}_{\mu \nu }$ is the non-Abelian field strength tensor and ${\text {Tr}}$ denotes a Lie algebra trace.
The field strength tensor ${\mathfrak {F}}_{\mu \nu }$ is given in terms of the non-Abelian gauge field potential $A_{\mu }$ by
\begin{equation}
{\mathfrak {F}}_{\mu \nu } = \partial _{\mu } A_{\nu } - \partial _{\nu } A_{\mu } + g\left[ A_{\mu }, A_{\nu } \right] ,
\label{eq:Fmunu}
\end{equation}
where $g$ is the gauge coupling constant.
The limit of large YM gauge coupling $g$ corresponds to the probe limit considered in, for example, \cite{Gubser:2008wv,Roberts:2008ns,Arias:2014msa}, where the back-reaction of the non-Abelian gauge field on the space-time geometry is ignored.
The AdS radius of curvature $\ell $ is given in terms of the cosmological constant $\Lambda $ by
\begin{equation}
\ell = {\sqrt { - \frac {\Lambda }{3}}}.
\label{eq:Ldef}
\end{equation}
Varying the action (\ref{eq:action}) gives the field equations
\begin{subequations}
\label{eq:fieldequations}
\begin{eqnarray}
R_{\mu \nu } - \frac {1}{2}Rg_{\mu \nu } + \Lambda g_{\mu \nu } & = & 8\pi G T_{\mu \nu } ,
\label{eq:EE}
\\
D_{\mu } {\mathfrak {F}}_{\nu }{}^{\mu } = \nabla _{\mu } {\mathfrak {F}}_{\nu }{}^{\mu } + g\left[ A_{\mu }, {\mathfrak {F}}_{\nu }{}^{\mu } \right] & = & 0 ,
\label{eq:YME}
\end{eqnarray}
\end{subequations}
where $R_{\mu \nu }$ is the Ricci tensor and the stress-energy tensor $T_{\mu \nu }$ is given by
\begin{equation}
T_{\mu \nu } = g^{\alpha \beta } {\text {Tr}} \, {\mathfrak {F}}_{\mu \alpha }{\mathfrak {F}}_{\nu \beta } - \frac {1}{4} g_{\mu \nu } {\text {Tr}} \, {\mathfrak {F}}_{\alpha \beta } {\mathfrak {F}}^{\alpha \beta }.
\end{equation}

Following \cite{Manvelyan:2008sv}, we start with a general metric ansatz describing a planar black hole
\begin{equation}
ds^{2} = -\sigma ^{2} \mu \, dt^{2} + r^{2}f^{2} dx^{2} + \frac {r^{2}}{f^{2}} dy^{2} + \mu ^{-1} dr^{2} ,
\label{eq:metric}
\end{equation}
where the metric functions $\sigma = \sigma (r)$, $\mu =\mu (r)$ and $f=f(r)$ depend on the coordinate $r$ only.
The coordinate $r$ is a radial coordinate, the planar event horizon will be located at $r=r_{h}$ and we have $r\rightarrow \infty $ as the AdS boundary is approached.
The planar coordinates $(x,y)$ describe surfaces parallel to the horizon.
It is convenient to further define a metric function $m(r)$ by
\begin{equation}
\mu (r) = -\frac {2m(r)}{r} - \frac {\Lambda r^{2}}{3}.
\label{eq:mdef}
\end{equation}
The functions $\sigma (r)$, $\mu (r)$ and $f(r)$ will be determined by the field equations.
If, however, we set $f(r)\equiv 1$, the line element (\ref{eq:metric}) possesses a ${\mathfrak {u}}(1)$ symmetry, corresponding to rotations in the $(x,y)$-plane.

\subsection{Gauge field ansatz}
\label{sec:gauge}

In \cite{Manvelyan:2008sv}, two different ansatze are used for the higher-dimensional gauge field potential.
For a four-dimensional space-time, the ansatze of \cite{Manvelyan:2008sv} give two ansatze for an ${\mathfrak {su}}(2)$ gauge field.
We generalize both of these to the larger gauge group ${\mathfrak {su}}(N)$ by taking the gauge field potential $A$ to have the form
\begin{equation}
gA  = gA_{\mu } dx^{\mu } = {\mathcal {A}} \, dt + \frac {i}{2} \left( C + C^{H} \right) dx + \frac {\zeta }{2} \left( C - C^{H} \right) dy ,
\label{eq:gauge}
\end{equation}
where $\zeta $ is a constant equal to either zero or unity, ${\mathcal {A}}$ and $C$ are $N\times N$ matrices depending only on the radial coordinate $r$, and $C^{H}$ is the Hermitian conjugate of $C$.
If we set the constant $\zeta =1$, then the ansatz (\ref{eq:gauge}) is a generalization of ansatz I of \cite{Manvelyan:2008sv},
and generalizes the isotropic $p+ip$-wave superconductor model considered in, for example,
\cite{Gubser:2008zu,Roberts:2008ns,Zeng:2010fs,Arias:2012py}.
On the other hand, the generalization of ansatz II of \cite{Manvelyan:2008sv} is achieved by setting $\zeta =0$ and generalizes the anisotropic $p$-wave superconductor model (see, for example, \cite{Gubser:2008wv,Zeng:2009dr,Basu:2009vv,Gubser:2010dm,Zeng:2010fs,Gangopadhyay:2012gx,Arias:2012py,Arias:2014msa,Herzog:2014tpa}).
If $\zeta =1$ then the metric function $f(r)\equiv 1$ (\ref{eq:metric}), but if $\zeta =0$ then $f(r)$ is determined by the field equations.

The electric part of the gauge potential (\ref{eq:gauge}) is given by
\begin{equation}
{\mathcal {A}} = -\sum _{p=1}^{N-1} h_{p}(r) H_{p} ,
\label{eq:Adef}
\end{equation}
where the $N-1$ scalar functions $h_{p}(r)$ depend on the radial coordinate $r$ only and the matrices $H_{p}$, $p =1,\ldots , N-1$ are generators of the Cartan subalgebra of ${\mathfrak {su}}(N)$, defined in a similar way to \cite{Brandt:1980em} but with different normalization \cite{Shepherd:2015dse}
\begin{equation}
\left[ H_{p} \right] _{j,k} = \frac {i}{{\sqrt {2p \left( p + 1 \right) }}}
\left(  p \delta _{j,p +1} \delta _{k,p +1} - \sum _{q=1}^{p}  \delta _{j,q}\delta _{k,q} \right) ,
\label{eq:Hdef}
\end{equation}
where $\delta _{j,k}$ is the Kronecker delta.
The magnetic part of the gauge potential (\ref{eq:gauge}) is determined by the upper-triangular matrix $C$, which has nonzero entries only immediately above the diagonal:
\begin{equation}
C_{j,j+1} = \omega _{j}(r), \qquad j =  1, \ldots , N-1,
\label{eq:Cdef}
\end{equation}
where the $N-1$ scalar functions $\omega _{j}(r)$ depend on the radial coordinate $r$ only.
We can expand $C+C^{H}$ and $C-C^{H}$ in terms of generators of the ${\mathfrak {su}}(N)$ Lie algebra as follows:
\begin{equation}
C+C^{H} = 2i\sum _{m=1}^{N-1} \omega _{m}(r) F_{m}, \qquad
C-C^{H} = -2\sum _{m=1}^{N-1} \omega _{m}(r) G_{m},
\label{eq:CFG}
\end{equation}
where the $N\times N$ matrices $F_{m}$ and $G_{m}$ are given by
\begin{equation}
\left[ F_{m} \right] _{j,k} = - \frac {i}{2} \left( \delta _{j,m+1}\delta _{k,m} + \delta _{j,m}\delta _{k,m+1} \right)  , \qquad
\left[ G_{m} \right] _{j,k} =  \frac {1}{2} \left( \delta _{j,m+1} \delta _{k,m} - \delta _{j,m} \delta _{k,m+1} \right) .
\label{eq:FGdef}
\end{equation}

If $\zeta =0$, the symmetries of the metric (\ref{eq:metric}) impose no additional constraints on the gauge potential ansatz (\ref{eq:gauge}), since we have already assumed that $\partial _{t}A=\partial _{x}A=\partial _{y}A=0$.
However, if $\zeta =1$, and $f(r)\equiv 1$, the line element (\ref{eq:metric}) has a ${\mathfrak {u}}(1)$ symmetry associated with rotations in the $(x,y)$-plane.
Since all physical quantities must be gauge-invariant, physical quantities calculated from the gauge potential (\ref{eq:gauge}) will be invariant under these ${\mathfrak {u}}(1)$ rotations if the effect of an infinitesimal space-time symmetry transformation on the gauge potential is equivalent to an infinitesimal gauge transformation \cite{Forgacs:1979zs} (see also \cite{Bergmann:1978fi,Harnad:1979in}).
For a general space-time symmetry, this requirement leads to a set of symmetry equations on the gauge potential ansatz which must be satisfied for the ansatz to be valid \cite{Forgacs:1979zs}.
We now derive the symmetry equations for the ${\mathfrak {u}}(1)$ rotations when $f(r)\equiv 1$, to verify that our ansatz (\ref{eq:gauge}) with $\zeta =1$ is valid.

Under an infinitesimal coordinate transformation $x^{\mu } \rightarrow x^{\mu } + \epsilon \xi ^{\mu }$, the gauge potential transforms as \cite{Forgacs:1979zs}
\begin{equation}
A_{\mu } \rightarrow A_{\mu } + \epsilon \left( \partial _{\mu } \xi ^{\nu } \right) A_{\nu } + \epsilon \xi ^{\nu } \left( \partial _{\nu } A_{\mu } \right) + {\mathcal {O}}(\epsilon ^{2}).
\label{eq:gencoord}
\end{equation}
For infinitesimal rotations in the $(x,y)$ plane, we have
\begin{equation}
\xi ^{\mu } = \left( 0, -y, x ,0\right) ,
\label{eq:rotation}
\end{equation}
giving
\begin{equation}
A_{\mu } \rightarrow A_{\mu } + \epsilon \left[  \left( \partial _{\mu } x \right) A_{y} - \left( \partial _{\mu } y \right) A_{x} + x\left( \partial _{y}A_{\mu } \right) - y \left( \partial _{x} A_{\mu } \right) \right] + {\mathcal {O}}(\epsilon ^{2}).
\label{eq:coordtrans}
\end{equation}
Applying an infinitesimal gauge transformation to the gauge potential gives
\begin{equation}
A_{\mu } \rightarrow A_{\mu } + \epsilon \left( \partial _{\mu } W - \left[ A_{\mu } , W \right] \right) ,
\label{eq:gaugetrans}
\end{equation}
where $W$ is an element of the ${\mathfrak {su}}(N)$ Lie algebra.
Comparing (\ref{eq:coordtrans}, \ref{eq:gaugetrans}) gives a set of four symmetry equations which must hold if an infinitesimal rotation in the $(x,y)$-plane
is to be equivalent to a gauge transformation:
\begin{subequations}
\label{eq:symmetry}
\begin{eqnarray}
\partial _{t} W - \left[ A_{t}, W \right] & = &
x\left( \partial _{y}A_{t} \right) - y \left( \partial _{x}A_{t} \right),
\label{eq:symmetry0}
\\
\partial _{x} W -\left[ A_{x}, W \right] & = &
A_{y} + x\left( \partial _{y} A_{x} \right) - y \left( \partial _{x}A_{x} \right) ,
\label{eq:symmetry1}
\\
\partial _{y}W - \left[ A_{y}, W \right] & =&
-A_{x} + x\left( \partial _{y}A_{y} \right) - y \left( \partial _{x}A_{y} \right) ,
\label{eq:symmetry2}
\\
\partial _{r}W - \left[ A_{r}, W \right] & = &
x\left( \partial _{y}A_{r} \right) - y \left( \partial _{x}A_{r} \right) .
\label{eq:symmetry3}
\end{eqnarray}
\end{subequations}
Our ansatz (\ref{eq:gauge}) is valid only if we can find some $W$ in the Lie algebra satisfying the equations (\ref{eq:symmetry}).

To this end, consider
\begin{equation}
W = \sum _{p=1}^{N-1} H_{p} {\sqrt {\frac {p\left( p+ 1 \right)}{2}}} ,
\label{eq:Wdef}
\end{equation}
where the matrices $H_{p}$ lie in the Cartan subalgebra and are given by (\ref{eq:Hdef}).
From our gauge potential ansatz (\ref{eq:gauge}), we automatically have $\partial _{t}A_{\mu } = \partial _{x}A_{\mu }=\partial _{y}A_{\mu }=0$ for all $\mu $ since $A_{\mu }$ depends only on the radial coordinate $r$.
We also have $\partial _{t}W=0=\partial _{r}W$ since $W$ (\ref{eq:Wdef}) does not depend on $t$ or $r$, and furthermore $\left[ A_{t},W \right] = 0 = \left[ A_{r}, W \right] $ since both $W$ and $A_{t}$ are in the Cartan subalgebra and $A_{r}=0$.
Therefore, with this choice of $W$, equations (\ref{eq:symmetry0}, \ref{eq:symmetry3}) are satisfied automatically.
The two remaining equations (\ref{eq:symmetry1}, \ref{eq:symmetry2}) become
\begin{equation}
\left[ A_{x} , W \right] = -A_{y}, \qquad \left[ A_{y}, W \right] = A_{x}.
\label{eq:symmetryreduced}
\end{equation}
To verify that these equations hold, we require the following commutation relations between the Lie algebra generators (\ref{eq:Hdef}, \ref{eq:FGdef}):
\begin{eqnarray}
\left[ F_{k}, H_{p} \right] & =  & \frac {1}{{\sqrt {2k}}} G_{k}\left( \delta _{p, k-1} {\sqrt {k-1}} - \delta _{p ,k}{\sqrt {k+1}} \right) ,
\nonumber \\
\left[ G_{k}, H_{p } \right] & = & \frac {1}{{\sqrt {2k}}} F_{k}\left( \delta _{p ,k} {\sqrt {k+1}} - \delta _{p ,k-1} {\sqrt {k-1}} \right) .
\label{eq:commutators}
\end{eqnarray}
Using these commutators, we find:
\begin{eqnarray}
\left[ A_{x}, W \right] & = &
-\frac {1}{2g} \left[ \sum _{m=1}^{N-1} \omega _{m}F_{m}, \sum _{p=1}^{N-1} H_{p} {\sqrt {2p\left( p + 1 \right) }} \right]
= \frac {1}{g} \sum _{k=1}^{N-1} \omega _{k}G_{k}
= -A_{y},
\nonumber \\
\left[ A_{y}, W \right] & = &
-\frac {1}{2g} \left[ \sum _{m=1}^{N-1} \omega _{m}G_{m}, \sum _{p=1}^{N-1} H_{p} {\sqrt {2p\left( p + 1 \right) }} \right]
= - \frac {1}{g} \sum _{k=1}^{N-1} \omega _{k}F_{k}
= A_{x},
\label{eq:A2Weq}
\end{eqnarray}
as required.
Therefore our ansatz (\ref{eq:gauge}) with $\zeta =1$ is compatible with rotations in the $(x,y)$-plane.

If we set $\omega _{k}\equiv 0$ for all $k$, then we have an embedded planar RN-AdS black hole with an Abelian ${\mathfrak {u}}(1)^{N-1}$ gauge field configuration given by the functions $h_{k}$, which in this case are independent (see section \ref{sec:trivial}).
Both ansatze (\ref{eq:gauge}) break this Abelian symmetry when at least one of the magnetic gauge field functions $\omega _{k}$ is nontrivial.
If $\zeta =1$, the additional ${\mathfrak {u}}(1)$ rotational symmetry in the $(x,y)$-plane is preserved even when the gauge field has a nontrivial magnetic part, but if $\zeta =0$ then the presence of nonzero $\omega _{k}$ also breaks this rotational symmetry.
We close this subsection by noting that our ansatz (\ref{eq:gauge}) is not the same as that considered in \cite{Baxter:2015tda} for planar ${\mathfrak {su}}(N)$ EYM black holes, due to our using a difference coordinate system, and also a different matrix basis for the electric part of the gauge field (\ref{eq:Adef}).

\subsection{Field equations and boundary conditions}
\label{sec:boundary}

Using the ansatz (\ref{eq:gauge}) for the gauge potential and the metric ansatz (\ref{eq:metric}), the Einstein equations (\ref{eq:EE}) take the form
(see \cite{Shepherd} for a detailed derivation):
\begin{subequations}
\label{eq:staticfieldequations}
\begin{eqnarray}
m'	& = &
\frac{\mu r^2 f'^2}{2f^2} + \alpha^2 \sum_{k=1}^{N-1} \left\{ \frac{\omega_k^2}{2\sigma^2 \mu}\left(\sqrt{\frac{k+1}{2k}}h_k - \sqrt{\frac{k-1}{2k}}h_{k-1}\right)^2 \left( \frac{1}{f^2} + \zeta^2 f^2 \right) \right\}
\nonumber \\ & &
+ \alpha^2 \sum_{k=1}^{N-1} \left\{\frac{r^2 h_k'^2}{2\sigma^2} + \frac{\mu \omega_k'^2}{2}\left( \frac{1}{f^2} + \zeta^2 f^2 \right) + \frac{k(k+1)\zeta^2}{4r^2} \left( \frac{\omega_k^2}{k} - \frac{\omega_{k+1}^2}{k+1} \right)^2 \right\},
\label{eq:mprime}
\\
\sigma' & = & \frac{r\sigma f'^2}{f^2} + \alpha^2 \sum_{k=1}^{N-1} \left\{ \frac{\omega_k^2}{2\sigma \mu^2 r}\left(\sqrt{\frac{k+1}{2k}}h_k - \sqrt{\frac{k-1}{2k}}h_{k-1}\right)^2 \left( \frac{1}{f^2} + \zeta^2 f^2 \right) \right\}
\nonumber \\ & &
 + \alpha^2 \sum_{k=1}^{N-1} \left\{   \frac{\sigma \omega_k'^2}{r}  \left( \frac{1}{f^2} + \zeta^2 f^2 \right)  \right\},
\label{eq:sigmaprime}
\\
f'' & = &   \alpha^2 \left( \frac{1}{f^2} - \zeta^2 f^2 \right) \sum_{k=1}^{N-1} \left\{ \frac{2\omega_k^2 h_k^2}{k(k+1)\sigma^2 \mu^2 r^2} - \frac{\omega_k'^2}{r^2}\right\}
 - f'\left(\frac{\sigma'}{\sigma} + \frac{\mu'}{\mu} + \frac{2}{r} - \frac{f'}{f} \right),
 \nonumber \\
\label{eq:fprimeprime}
\end{eqnarray}
and the Yang-Mills equations (\ref{eq:YME}) are (again, derived in \cite{Shepherd})
\begin{eqnarray}
h_k'' & = & h_k'\left( \frac{\sigma'}{\sigma} - \frac{2}{r}\right)
+ \frac{\sqrt{k(k+1)}}{2\mu r^2}\frac{\omega_k^2}{k} \left(\sqrt{\frac{k+1}{k}}h_k - \sqrt{\frac{k-1}{k}}h_{k-1}\right) \left(\frac{1}{f^2} + \zeta^2 f^2 \right)
\nonumber \\ & &
 + \frac{\sqrt{k(k+1)}}{2\mu r^2}\frac{\omega_{k+1}^2}{k+1}\left(\sqrt{\frac{k}{k+1}}h_k - \sqrt{\frac{k+2}{k+1}}h_{k+1}\right) \left(\frac{1}{f^2} +\zeta^2 f^2 \right),
 \label{eq:hprimeprime}
 \\
0 & = & \omega_k''  +  \omega_k'\left( \frac{\sigma'}{\sigma} + \frac{\mu'}{\mu} - \frac{2f'}{f}\right)   +  \frac{\omega_k }{\sigma^2 \mu^2}\left(\sqrt{\frac{k+1}{2k}}h_k - \sqrt{\frac{k-1}{2k}}h_{k-1}\right)^2
\nonumber \\ & &
+ \frac{\zeta^2f^2\omega_k}{2\mu r^2} \left( \omega_{k-1}^2 - 2\omega_k^2 + \omega_{k+1}^2 \right)  ,
\label{eq:wprimeprime}
\end{eqnarray}
along with a constraint equation
\begin{equation}
0  = \left( \omega_k \omega_{k+1}' - \omega_{k+1}\omega_k' \right)\left( \frac{1}{f^2} - \zeta^2 f^2 \right) ,
\label{eq:constraint}
\end{equation}
\end{subequations}
where we have defined the constant
\begin{equation}
\alpha ^{2} = \frac {4\pi G}{g^{2}}.
\label{eq:alphadef}
\end{equation}

In the $N=2$ case, the field equations (\ref{eq:staticfieldequations}) reduce to the $d=4$ equations in \cite{Manvelyan:2008sv}, with $\zeta =1$ corresponding to ansatz I and $\zeta =0$ corresponding to ansatz II.
The constraint equation (\ref{eq:constraint}) is satisfied trivially in the $\zeta =1$, $f\equiv 1$ case.
If $\zeta =0$, the constraint equation (\ref{eq:constraint}) implies that, if the functions $\omega _{k}$ are nonzero, then they are all scalar multiples of each other.
Then, to obtain a consistent set of equations for the $\omega _{k}$, it must be the case that the electric gauge field functions $h_{k}$ are also scalar multiples of each other.
This gives an embedded ${\mathfrak {su}}(2)$ solution, see section \ref{sec:trivial}.

The field equations (\ref{eq:staticfieldequations}) are singular at the black hole event horizon $r=r_{h}$, where $\mu (r_{h})=0$ and as $r\rightarrow \infty $.
We therefore need to derive suitable boundary conditions on the field variables in neighbourhoods of these singular points.
In this paper we consider only nonextremal black holes with nonzero surface gravity and Hawking temperature, for which we require that
\begin{equation}
\mu '(r_{h}) = -\Lambda r_{h} - \frac {2m'(r_{h})}{r_{h}} >0,
\label{eq:murhprime}
\end{equation}
and hence $m'(r_{h})<-\Lambda r_{h}^{2}$.
In order for physical quantities to be regular at the event horizon, it must be the case that $h_{k}(r_{h})=0$.
We assume that all field variables have regular Taylor series expansions in a neighbourhood of the horizon.
We then find
\begin{eqnarray}
m(r) & = & \frac{r_h^3}{2\ell ^2} + m'(r_h)(r-r_h) + \mathcal{O}(r-r_h)^2,
\nonumber\\
f(r) & = & f(r_h) + \mathcal{O}(r-r_h)^2, \nonumber\\
\sigma(r) & = & \sigma(r_h) + \sigma'(r_h)(r-r_h) + \mathcal{O}(r-r_h)^2,
\nonumber\\
h_k(r) & = & h_k'(r_h)(r-r_h) + \mathcal{O}(r-r_h)^2,
\nonumber \\
\omega_k(r) & = & \omega_k(r_h) + \omega_k'(r_h)(r-r_h) + \mathcal{O}(r-r_h)^2,
\label{eq:horizon}
\end{eqnarray}
where
\begin{eqnarray}
\omega_k'(r_h) & = &
\frac{\zeta^2 \ell ^2 f(r_h)^2 \omega_k(r_h)\left[2\omega_k(r_h)^2 - \omega_{k-1}(r_h)^2-\omega_{k+1}(r_h)^2\right]}{2r_h^2\left[3r_h - 2m'(r_h)\ell ^2\right]},
\nonumber\\
m'(r_h)  & = &
\alpha^2 \sum_{k=1}^{N-1} \left\{ \frac{r_h^2 h_k'^2}{2\sigma(r_h)^2} + \frac{k(k+1)\zeta^2}{4r_h^2}\left( \frac{\omega_k(r_h)^2}{k} - \frac{\omega_{k+1}(r_h)^2}{k+1}\right)^2\right\},
\nonumber\\
\sigma'(r_h) & = &
\alpha^2\left[\frac{1}{f(r_h)} + \zeta^2 f(r_h)^2\right]\sum_{k=1}^{N-1}\left[\frac{2\omega_k(r_h)^2 h_k'(r_h)^2 r_h \ell ^4}{k(k+1)\sigma(r_h)\left(3r_h^2 - 2m'(r_h)\ell ^2\right)^2}
+\frac{\sigma(r_h)\omega_k'(r_h)^2}{r_h}\right] .
\nonumber \\
\label{eq:horizon1}
\end{eqnarray}
Near the horizon, the hairy black holes are parameterized by the $2\left( N-1 \right)$ constants $\omega _{k}(r_{h})$ and $h_{k}'(r_{h})$, together with the event horizon radius $r_{h}$ and AdS radius of curvature $\ell $.
Although $\sigma (r_{h})$ is a free parameter in the expansions (\ref{eq:horizon}), in practice it is determined by the boundary conditions at infinity (see section \ref{sec:num}).

As $r\rightarrow \infty $, the metric (\ref{eq:metric}) approaches that of pure AdS, so we require $f\rightarrow 1$, $\sigma \rightarrow 1$ as $r\rightarrow \infty $.
Assuming that the field variables have regular Taylor series expansions for large $r$, we find
\begin{eqnarray}
m(r)       	& = & m_0 - \frac {\alpha^2}{r}\sum_{k=1}^{N-1}\left[ \frac{\omega_{k,\infty}^2\ell^2}{2}\left(\sqrt{\frac{k+1}{2k}}h_{k, \infty} - \sqrt{\frac{k-1}{2k}}h_{k-1,\infty}\right)^2\left(1+\zeta^2\right) \right]
\nonumber \\ &  &
-  \frac{\alpha^2}{r}\sum_{k=1}^{N-1}\left[\frac{k(k+1)\zeta^2}{4}\left(\frac{\omega_{k,\infty }^2}{k} - \frac{\omega_{k+1,\infty }^2}{k+1}\right)^2 + \frac{h_{k,1}^2}{2} + \frac{c_{k,1}^{2}}{2\ell^2}\left(1+\zeta^2\right) \right] + \mathcal{O}\left(\frac{1}{r^2}\right) ,
\nonumber\\
\sigma(r)  	& = & 1 - \frac{\left(1+\zeta^2\right)}{4r^{2}}\alpha^2\sum_{k=1}^{N-1}\left[\ell^4\omega_{k,\infty}^2\left(\sqrt{\frac{k+1}{2k}}h_{k, \infty} - \sqrt{\frac{k-1}{2k}}h_{k-1,\infty}\right)^2 + c_{k,1}^2 \right]
\nonumber \\ & &
 + \mathcal{O}\left(\frac{1}{r^5}\right) ,
\nonumber \\
f(r) & = & 1 + \frac{f_3}{r^3} + \mathcal{O}\left(\frac{1}{r^4}\right),
\nonumber\\
\omega_k(r) & = & \omega _{k,\infty } +  \frac{c_{k,1}}{r} + \mathcal{O}\left(\frac{1}{r^2}\right),
\nonumber\\
h_k(r)  & = & h_{k,\infty} + \frac{h_{k,1}}{r} + \mathcal{O}\left(\frac{1}{r^2}\right) ,
\label{eq:infinity}
\end{eqnarray}
where the constant $f_{3}$ is unconstrained.
As well as $f_{3}$, $m_{0}$ and $\ell $, the above expansions are determined by the $4\left( N -1 \right)$ arbitrary parameters $\omega _{k,\infty }$, $c_{k,1}$, $h_{k,\infty }$ and $h_{k,1}$.
Our primary interest in this paper is solutions for which $\omega _{k,\infty }=0$, so that the magnetic part of the non-Abelian gauge field forms a condensate in a region near the planar event horizon.
In the ${\mathfrak {su}}(2)$ case, the constant
$h_{k,\infty }$  is then interpreted as the chemical potential of the thermal state in the dual CFT \cite{Gubser:2008zu}, while
the constant $c_{k,1}$ is an order parameter interpreted as a component of the boundary current \cite{Gubser:2008zu} (see section \ref{sec:thermo} for more details).

\subsection{Trivial solutions}
\label{sec:trivial}

Closed form solutions of the EYM equations (\ref{eq:staticfieldequations}) cannot easily be found in general, and numerical analysis is required. However, there are a number of trivial solutions which we now outline.

First, consider the planar Schwarzschild-AdS black hole with metric
\begin{equation}
ds^2 = -\mu _{S} dt^2 + r^2 \left[ dx^2 + dy^2 \right] +  \mu _{S}^{-1}dr^2,
\label{eq:SAdS}
\end{equation}
where the metric function $\mu _{S}$ is given by
\begin{equation}
\mu _{S} = -\frac {2M_{S}}{r} - \frac {\Lambda r^{2}}{3} ,
\label{eq:muS}
\end{equation}
and $M_{S}$ is a constant.
This is a solution of the field equations (\ref{eq:staticfieldequations}) on setting $\sigma \equiv 1$, $f\equiv 1$, and requiring that $m'\equiv 0$.
We set the electric part of the gauge field to vanish, $h_{k}\equiv 0$, and then it must be the case that $\omega _{k}'\equiv 0$ and
\begin{equation}
\sum_{k=1}^{N-1}  k(k+1) \left( \frac{\omega_k^2}{k} - \frac{\omega_{k+1}^2}{k+1} \right)^2 = 0 .
\label{eq:SAdSomega}
\end{equation}
This is solved by taking
\begin{equation}
\omega _{k} \equiv \pm {\mathcal {W}} {\sqrt {k}}
\label{eq:SAdSomegasol}
\end{equation}
where ${\mathcal {W}}$ is a constant independent of $k$.
In the spherically symmetric case \cite{Shepherd:2015dse}, the constant ${\mathcal {W}}$ is fixed by the field equations, but in the planar case the constant ${\mathcal {W}}$ is arbitrary due to the scaling symmetries discussed in section \ref{sec:scaling}.

The second trivial solution is RN-AdS with metric
\begin{equation}
ds^2 = -\mu_{RN}dt^2 + r^2 \left[ dx^2 + dy^2 \right]  + \mu_{RN}^{-1}dr^2,
\label{eq:RNAdS}
\end{equation}
where
\begin{equation}
\mu_{RN} = -\frac{2M_{RN}}{r}+ \frac{\alpha ^{2}Q_{RN}^2}{r^2} - \frac{\Lambda r^2}{3},
\label{eq:muRN}
\end{equation}
and the mass $M_{RN}$ and charge $Q_{RN}$ are constants.
Again we set $\sigma \equiv 1$, $f\equiv 1$, but now $\omega _{k}\equiv 0$ for all $k$.
Equation (\ref{eq:hprimeprime}) then reduces to
\begin{equation}
h_{k}'' = -\frac {2h_{k}'}{r}
\label{eq:RNheqn}
\end{equation}
which has solution
\begin{equation}
h_{k,RN}(r) = b_{k}-\frac {a_{k}}{r} ,
\label{eq:RNgauge}
\end{equation}
for constants $a_{k}$ and $b_{k}$.
For comparison with the ${\mathfrak {su}}(N)$ solutions for which the electric gauge field functions $h_{k}$ must vanish on the horizon $r=r_{h}$, it is convenient to choose the constants $a_{k}$ and $b_{k}$ in (\ref{eq:RNgauge}) such that $h_{k,RN}(r_{h})=0$, corresponding to a choice of gauge.
Then we have $a_{k}=b_{k}r_{h} = r_{h}^{2}h_{k,RN}'(r_{h})$ and
\begin{equation}
h_{k,RN}(r) = \left( 1-\frac {r_{h}}{r} \right) r_{h}h_{k,RN}'(r_{h}).
\label{eq:RNgauge1}
\end{equation}
Substituting in (\ref{eq:mprime}) and comparing with the derivative $m'(r)$ obtained from (\ref{eq:muRN}), we find
 that the electric charge $Q_{RN}$ is given by
\begin{equation}
Q_{RN}^{2} = \sum _{k=1}^{N-1} a_{k}^{2} = \sum _{k=1}^{N-1}r_{h}^{4} \left[h_{k,RN}'(r_{h}) \right] ^{2}.
\label{eq:QRN}
\end{equation}

We can also embed any solution of the ${\mathfrak {su}}(2)$ equations into the ${\mathfrak {su}}(N)$ field equations (\ref{eq:staticfieldequations}).
To see this, we start by setting
\begin{equation}
\omega _{k}(r) = {\mathfrak {A}}_{k}\omega (r), \qquad
h_{k}(r) = {\mathfrak {B}}_{k}h(r),
\end{equation}
where ${\mathfrak {A}}_{k}$ and ${\mathfrak {B}}_{k}$ are constants.
Substituting into the Einstein equations (\ref{eq:mprime}, \ref{eq:sigmaprime}, \ref{eq:fprimeprime}), we obtain the following constraints on the constants ${\mathfrak {A}}_{k}$ and ${\mathfrak {B}}_{k}$:
\begin{subequations}
\label{eq:su2conds}
\begin{multline}
\sum_{k=1}^{N-1} {\mathfrak {A}}_k^2\left(\sqrt{\frac{k+1}{2k}}{\mathfrak {B}}_k - \sqrt{\frac{k-1}{2k}}{\mathfrak {B}}_{k-1}\right)^2 =
\sum_{k=1}^{N-1}{\mathfrak {A}}_k^2 \\
= \sum_{k=1}^{N-1}{\mathfrak {B}}_k^2
= \sum_{k=1}^{N-1}\frac{k(k+1)}{2}\left(
\frac{{\mathfrak {A}}_k^2}{k} - \frac{{\mathfrak {A}}_{k+1}^2}{k+1}\right)^2 ,
\label{eq:su21}
\end{multline}
while in order to obtain consistent equations for $h_{k}$ and $\omega _{k}$ from the Yang-Mills equations (\ref{eq:hprimeprime}, \ref{eq:wprimeprime}) we require
\begin{eqnarray}
1 & = & \left(\sqrt{\frac{k+1}{2k}}{\mathfrak {B}}_k - \sqrt{\frac{k-1}{2k}}{\mathfrak {B}}_{k-1}\right)^2
= \frac{2{\mathfrak {A}}_k^2 - {\mathfrak {A}}_{k+1}^2 - {\mathfrak {A}}_{k-1}^2}{2} \nonumber\\
  & = & {\sqrt{\frac{k}{2\left(k+1\right)}}}{\mathfrak {A}}_{k+1}^2\left(\sqrt{\frac{k}{2(k+1)}}
  - \sqrt{\frac{k+2}{2(k+1)}}\frac{{\mathfrak {B}}_{k+1}}{{\mathfrak {B}}_k}\right) \nonumber\\
  & & + {\sqrt {\frac{(k+1)}{2k}}}{\mathfrak {A}}_k^2\left(\sqrt{\frac{k+1}{2k}} - \sqrt{\frac{k-1}{2k}}\frac{{\mathfrak {B}}_{k-1}}{{\mathfrak {B}}_k}\right) .
  \label{eq:su22}
\end{eqnarray}
\end{subequations}
We can solve (\ref{eq:su2conds}) by taking \cite{Shepherd:2015dse}
\begin{equation}
{\mathfrak {A}}_{k} = {\sqrt {k\left( N - k\right)}}, \qquad {\mathfrak {B}}_{k} = {\sqrt {\frac {1}{2} k\left( k + 1 \right)}} .
\label{eq:su2constants}
\end{equation}
We now define rescaled variables as follows \cite{Shepherd:2015dse}
\begin{equation}
R = \lambda_N^{-1}r, \quad \tilde{m} = \lambda_N^{-1}m, \quad \tilde{h} = \lambda_N h, \quad \tilde{\Lambda} = \lambda_N^2 \Lambda,
\label{eq:scaledsu2}
\end{equation}
where $\omega $, $f$, $\mu $, $\sigma $ and $\alpha $ are unchanged, and
\begin{equation}
\lambda_N^2 = \sum_{k=1}^{N-1}{\mathfrak {A}}_k^2 = \sum _{k=1}^{N-1} {\mathfrak {B}}_{k}^{2}= \frac{1}{6}N(N^2 - 1) .
\label{eq:lambdaN}
\end{equation}
The static field equations (\ref{eq:staticfieldequations}) then reduce to
\begin{eqnarray}
\frac{d\tilde{m}}{dR}     & = & \frac{\mu R^2 }{2f^2}\left(\frac{df}{dR}\right)^2 +  \alpha^2 \left\{ \left( \frac{1}{f^2} + \zeta^2 f^2 \right)
\left[ \frac{\omega^2 \tilde{h}^2}{2\sigma^2 \mu} + \frac{\mu}{2}\left(\frac{dw}{dR}\right)^2 \right] +
\frac{R^2 }{2\sigma^2}\left(\frac{d\tilde{h}}{dR}\right)^2
\frac{\zeta^2\omega^4}{2R^2}  \right\},
\nonumber  \\
\frac{d\sigma}{dR}        & = & \frac{R\sigma}{f^2}\left(\frac{df}{dR}\right)^2 + \alpha^2 \left( \frac{1}{f^2} + \zeta^2 f^2 \right) \left[ \frac{\omega^2 \tilde{h}^2}{2R\sigma\mu^2}
+ \frac{\sigma}{R}\left(\frac{d\omega}{dR}\right)^2 \right],
\nonumber \\
\frac{d^2f}{dR^2}         & = & \alpha^2 \left( \frac{1}{f^2} - \zeta^2 f^2 \right)  \left\{ \frac{\omega^2 h^2}{\sigma^2 \mu^2 R^2} -
\frac{1}{R^2}\left(\frac{d\omega}{dR}\right)^2\right\}
- \frac{df}{dR}\left(\frac{1}{\sigma}\frac{d\sigma}{dR} + \frac{1}{\mu}\frac{d\mu}{dR} + \frac{2}{R} - \frac{1}{f}\frac{df}{dR} \right),
                          \nonumber \\
\frac{d^2\tilde{h}}{dR^2} & = & \frac{d\tilde{h}}{dR}\left( \frac{1}{\sigma}\frac{d\sigma}{dR} - \frac{2}{R}\right) + \frac{\tilde{h}\omega^2}{\mu R^2}
\left( \frac{1}{f^2} + \zeta^2 f^2 \right), \nonumber \\
0 	& = &\frac{d^2\omega}{dR}  +  \frac{d\omega}{dR}\left( \frac{1}{\sigma}\frac{d\sigma}{dR} + \frac{1}{\mu}\frac{d\mu}{dR} - \frac{2}{f}\frac{df}{dR}\right)
+  \frac{\omega}{\mu}\left( \frac{\tilde{h}^2}{\sigma^2\mu} - \frac{\zeta^2\omega^2 f^2}{R^2}\right) ,
\label{eq:su2staticequations}
\end{eqnarray}
which are precisely the $\mathfrak{su}(2)$ field equations in terms of the new variables (\ref{eq:scaledsu2}).
We note that the constants (\ref{eq:su2constants}) are not the same as those used in \cite{Baxter:2015tda} to embed planar ${\mathfrak {su}}(2)$ EYM black holes into ${\mathfrak {su}}(N)$ EYM; this is because our gauge field ansatz (\ref{eq:gauge}) differs from that in \cite{Baxter:2015tda}.

\subsection{Scaling symmetries}
\label{sec:scaling}

The static EYM field equations (\ref{eq:staticfieldequations}) possess several scaling symmetries which can be used to reduce the number of parameters.
First, the field equations (\ref{eq:staticfieldequations}) are invariant under the transformation
\begin{equation}
r \to \lambda r, \qquad m \to \lambda m, \qquad \ell \to \lambda \ell, \qquad h_k \to \lambda^{-1} h_k, \qquad \alpha \to \lambda \alpha ,
\label{eq:scaling1}
\end{equation}
with $\sigma $,  $f$ and $\omega _{k}$ unchanged.
Hence by transforming the variables using $\lambda =\alpha ^{-1}$ we can effectively set $\alpha =1$ in (\ref{eq:alphadef}).

The second transformation under which the field equations (\ref{eq:staticfieldequations}) remain invariant is
\begin{equation}
r \to \lambda r, \qquad \omega_k \to \lambda \omega_k, \qquad h_k \to \lambda h_k, \qquad m \to \lambda^3 m,
\label{eq:scaling2}
\end{equation}
with $\ell$,  $\sigma $ and $f$ unchanged.  Under the transformation (\ref{eq:scaling2}), we have $\mu \to \lambda ^{2}\mu $.
By setting $\lambda = r_{h}^{-1}$, we can use the transformation (\ref{eq:scaling2}) to set the event horizon radius $r_{h}=1$ without loss of generality.

We then have two remaining symmetries, the first of which is
\begin{equation}
h_{k} \to \lambda h_{k}, \qquad \sigma \to \lambda \sigma ,
\label{eq:scaling3}
\end{equation}
(with all other variables unchanged) which can be used to set $\sigma (\infty )=1$ by taking $\lambda =\sigma (\infty )^{-1}$, and the second of which is $\omega _{k}\to -\omega _{k}$ (for each $k$ independently), which means that we can restrict attention to $\omega _{k}(r_{h})>0$ without loss of generality.

\section{Planar ${\mathfrak {su}}(N)$ EYM black holes}
\label{sec:sols}

In this section we present numerical black hole solutions of the static field equations (\ref{eq:staticfieldequations}).  For the remainder of this paper, we set $\zeta =1$ in (\ref{eq:gauge}) (and hence the metric function $f\equiv 1$ in (\ref{eq:metric})), since there are no genuinely ${\mathfrak {su}}(N)$ solutions in the $\zeta =0$ case, only embedded ${\mathfrak {su}}(2)$ solutions.
We use the scaling symmetry (\ref{eq:scaling1}) to set $\alpha =1$ and (\ref{eq:scaling2}) to set the event horizon radius $r_{h}=1$ without loss of generality.
Given the complexity of the field equations (\ref{eq:staticfieldequations}), we focus on the $N=2$ and $N=3$ cases.
In this section we consider the fully coupled EYM system, including the back-reaction of the non-Abelian gauge field on the space-time metric.
Planar EYM black holes with gauge group ${\mathfrak {su}}(2)$ and $\zeta =1$ have been previously studied in, for example, \cite{Gubser:2008zu,Arias:2012py}.
We compare the new ${\mathfrak {su}}(3)$ solutions presented here with the properties of those ${\mathfrak {su}}(2)$ solutions.

\subsection{Numerical solutions}
\label{sec:num}

To solve the field equations (\ref{eq:staticfieldequations}) numerically, we used a Bulirsch-Stoer algorithm in C++ \cite{Press:1992zz}.
Since the field equations are singular at the event horizon, we start integrating at $r-r_{h}\sim 10^{-7}$, using the expansions (\ref{eq:horizon}) as initial conditions.
We integrate outwards until the field variables have converged to within a suitable tolerance.
We require $\sigma (\infty )=1$ for the space-time to be asymptotically AdS, but numerically it is easier to take $\sigma (r_{h})=1$, which in general means that $\sigma (\infty )\neq 1$.
We then use the rescaling (\ref{eq:scaling3}) with $\lambda = \sigma (\infty )^{-1}$ after we have performed the integration, so that the rescaled variables satisfy the correct boundary conditions as $r\rightarrow \infty $.
The static solutions are then given in terms of the parameters $h_{k}'(r_{h})$, $\omega _{k}(r_{h})$ and the cosmological constant $\Lambda $ (or, equivalently, the AdS radius of curvature $\ell $ (\ref{eq:Ldef})).

In the ${\mathfrak {su}}(2)$ case, it is straightforward to show that the electric gauge field function $h(r)$ is monotonic and is nonzero outside the event horizon; the proof is identical to that in \cite{Shepherd:2015dse} for the spherically symmetric case. For larger gauge group, defining new quantities ${\mathcal {E}}_{k}(r)$ as follows \cite{Shepherd:2015dse}:
\begin{equation}
{\mathcal {E}}_{k}(r) = {\sqrt {\frac {k+1}{2k}}}h_{k}(r) - {\sqrt {\frac {k-1}{2k}}}h_{k-1}(r),
\end{equation}
then it can be proven \cite{Baxter:2015tda} that the ${\mathcal {E}}_{k}$ are monotonic functions which are zero only on the event horizon.
In our numerical work we found that the electric gauge field functions $h_{k}(r)$ are also monotonic for all the solutions investigated, but we were unable to prove that this is a general result.
However, since the ${\mathcal {E}}_{k}$ electric gauge field functions have no zeros, we therefore label our solutions by $n_{k}$, the number of zeros of the magnetic gauge field function $\omega _{k}$.

\begin{figure}
\begin{center}
\includegraphics[width=8cm,angle=270]{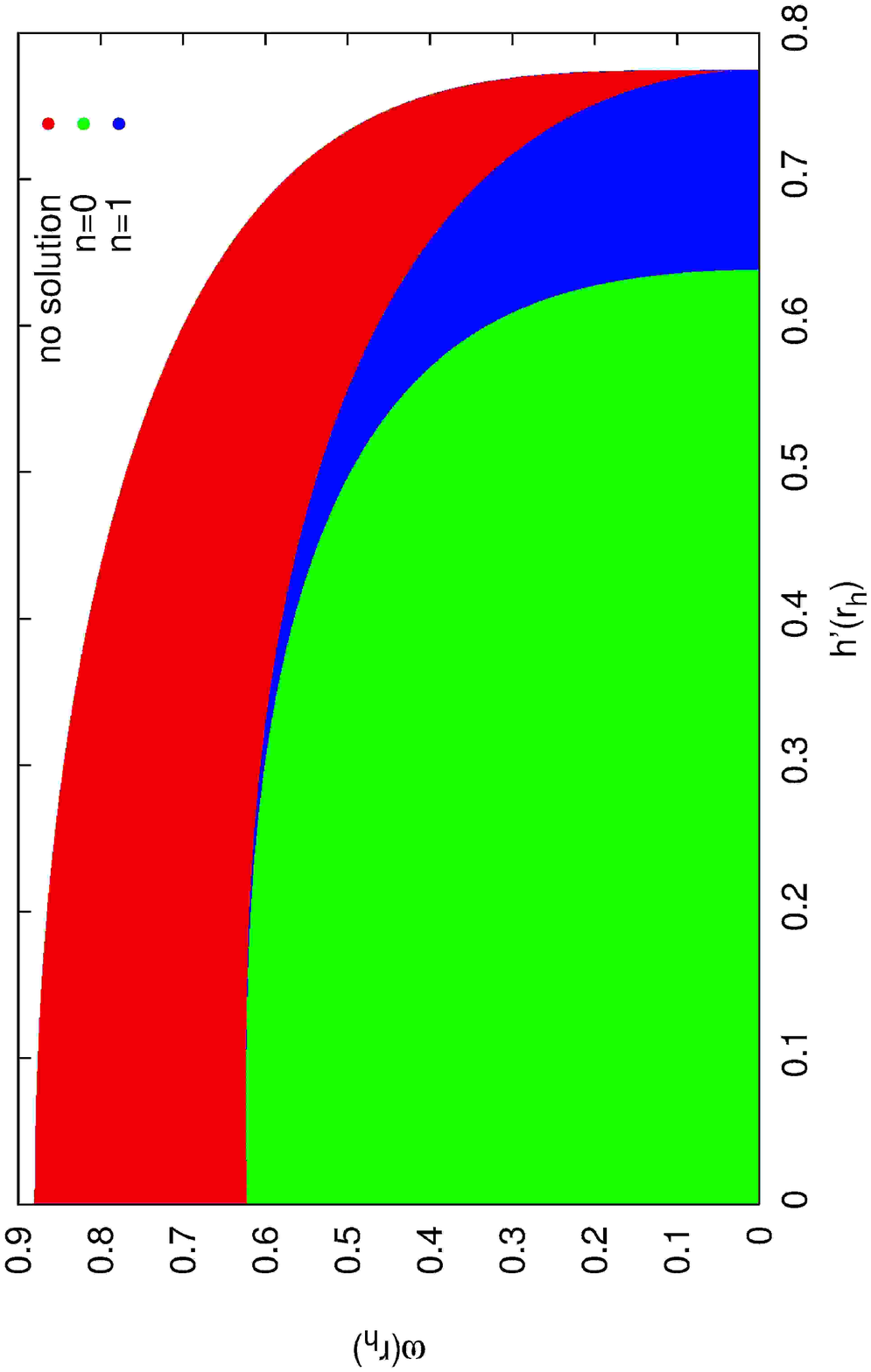}
\includegraphics[width=8cm,angle=270]{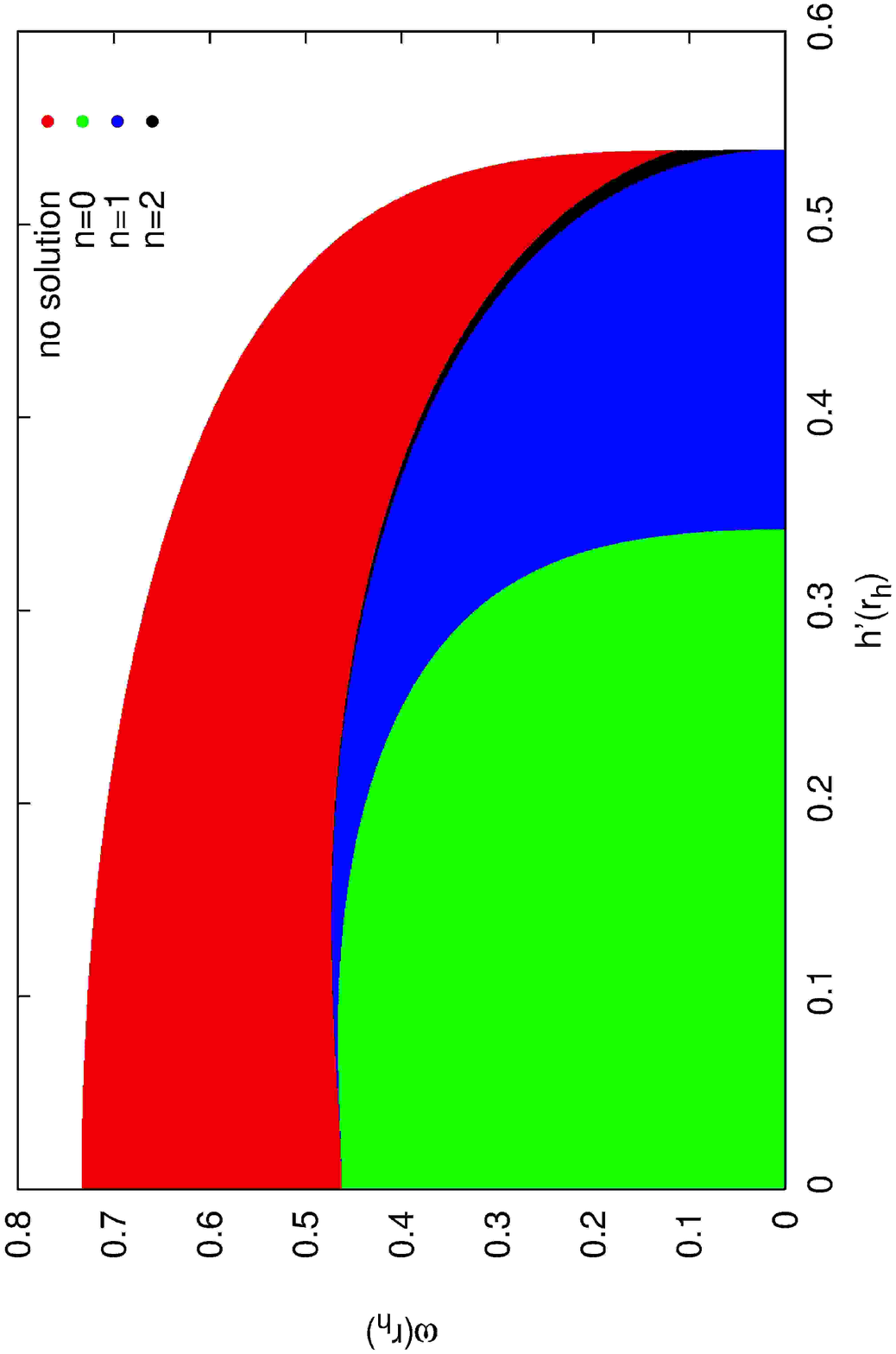}
\end{center}
\caption{Solution spaces for ${\mathfrak {su}}(2)$ planar black holes with $\Lambda = -0.6$ (top) and $\Lambda =-0.3$ (bottom),
colour-coded by $n$, the number of zeros of the gauge field function $\omega (r)$.
The red ``no solution'' region is where the condition (\ref{eq:murhprime}) for a nonextremal event horizon is satisfied, but we do not find black hole solutions.
We are interested in solutions for which $\omega (r)$ has no zeros, and $\omega (r)\rightarrow 0$ as $r\rightarrow \infty $.
These solutions lie on the boundary between the green $n=0$ and blue $n=1$ regions.}
\label{fig:su2phase}
\end{figure}

In figure \ref{fig:su2phase} we show the solution spaces for ${\mathfrak {su}}(2)$ black holes with cosmological constant $\Lambda =-0.6$ (top plot) and $\Lambda  =-0.3$ (lower plot).
With $\Lambda $ fixed, the black hole solutions are parameterized by $\omega (r_{h})$ and $h'(r_{h})$, which are the axes in the plots in figure \ref{fig:su2phase}.
The solution spaces in figure \ref{fig:su2phase} are colour-coded by the number of zeros $n$ of the single magnetic gauge field function $\omega (r)$.
In each plot there is a red region denoted ``no solution'', where the condition (\ref{eq:murhprime}) for a nonextremal event horizon is satisfied, but we do not find black hole solutions.
Unlike the situation for purely magnetic planar ${\mathfrak {su}}(2)$ black holes, where $\omega (r)$ must have no zeros \cite{VanderBij:2001ia}, in the dyonic case we find solutions for which $\omega (r)$ has one or more zeros.
We focus on those solutions where the magnetic gauge field function $\omega (r)\rightarrow 0$ as $r\rightarrow \infty $, which lie on the boundary of the blue $n=1$ and green $n=0$ regions in each plot in figure \ref{fig:su2phase}.
When $\Lambda = -0.3$, there are also solutions where $\omega (r)\rightarrow 0$ as $r\rightarrow \infty $ on the boundary of the black $n=2$ and blue $n=1$ regions.  However, for these solutions $\omega (r)$ has a zero between the event horizon and infinity, and therefore these solutions are excited states relative to the nodeless solutions.  For the rest of this paper, we focus our attention on solutions for which $\omega (r)$ has no zeros but vanishes as $r\rightarrow \infty $, so that it forms a condensate in a region close to the planar event horizon.

\begin{figure}
\begin{center}
\includegraphics[width=8cm,angle=270]{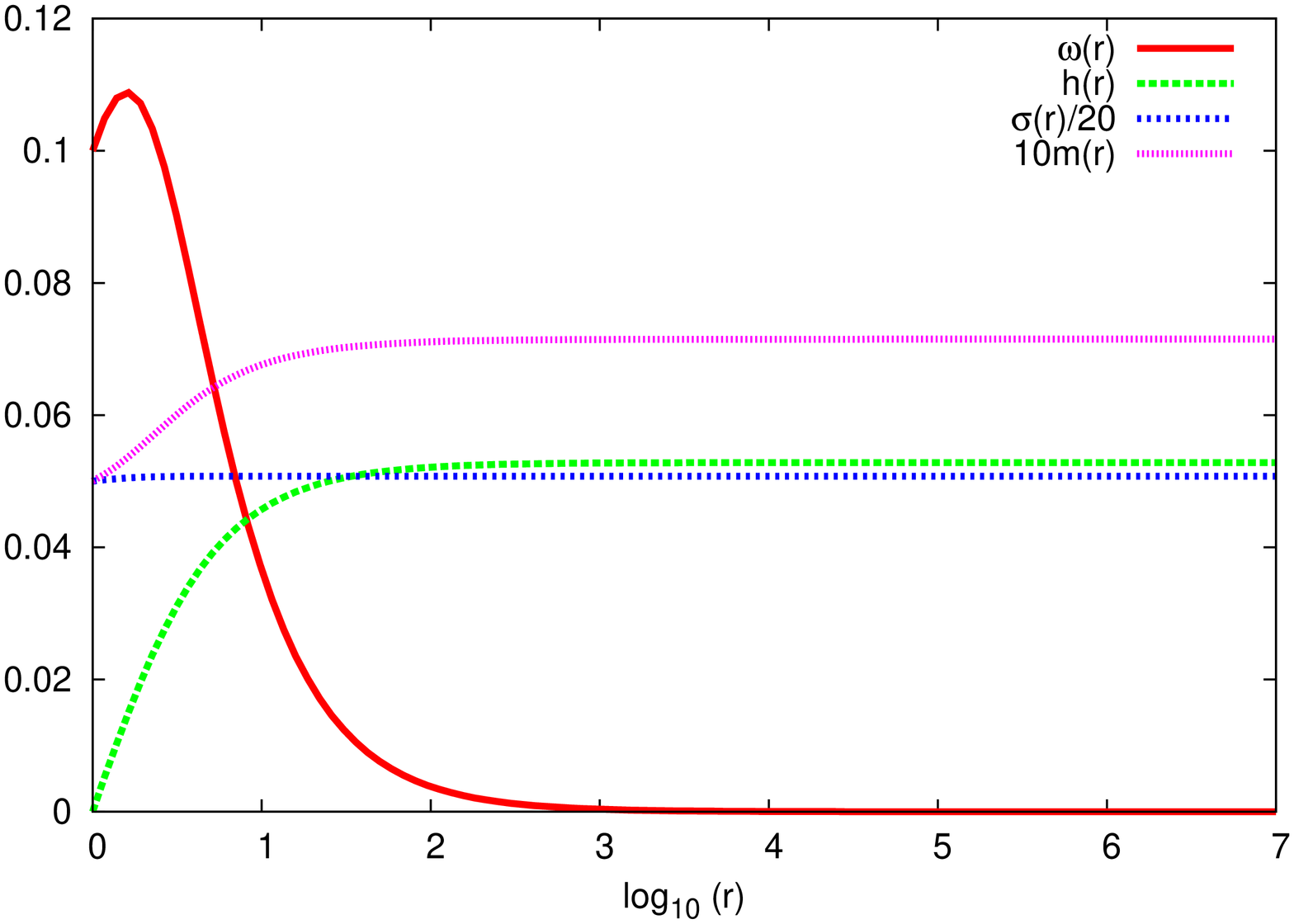}
\end{center}
\caption{Typical ${\mathfrak {su}}(2)$ planar black hole with $\Lambda = -0.03$, $\omega (r_{h})=0.1$. The value of $h'(r_{h})$ has been fixed by the requirement that $\omega (r)\rightarrow 0$ as $r\rightarrow \infty $.}
\label{fig:su2example}
\end{figure}

We find that solutions where $\omega \rightarrow 0 $ as $r\rightarrow \infty $ only exist if $\left| \Lambda \right| $ is not too large.  Indeed, it has been proven \cite{Baxter:2015tda} that for $\left| \Lambda \right| $ sufficiently large, it must be the case that $\omega (r)$ has no zeros (and is nonzero at infinity).
To find the solutions for which $\omega \rightarrow 0 $ as $r\rightarrow \infty $, we use the GSL root-finding algorithm \cite{Alken}.
For fixed $\Lambda $ (with $\left| \Lambda \right| $ sufficiently small), there is a continuous range of values of $\omega (r_{h})$ which give solutions for which $\omega (r)\rightarrow 0$ as $r\rightarrow \infty $, and for each such $\omega (r_{h})$ we find a unique value of $h'(r_{h})$ such that $\omega (\infty )=0$.
A typical ${\mathfrak {su}}(2)$ solution is shown in figure \ref{fig:su2example}, for $\Lambda = -0.03$ and $\omega (r_{h})=0.1$.
The magnetic gauge field function $\omega (r)$ is increasing close to the horizon, then has a maximum before decreasing to zero at infinity.
The electric gauge field function $h(r)$ is monotonically increasing from zero at the horizon to its asymptotic value.
To show the metric functions $m(r)$ and $\sigma (r)$ clearly on the same figure, we have scaled them, plotting $10m(r)$ and $\sigma (r)/20$.
From figure \ref{fig:su2example} we see that $m(r)$ takes small values and is monotonically increasing.  On the other hand $\sigma (r)$ takes larger values, but varies very little between the event horizon and infinity.

\begin{figure}
\begin{center}
\includegraphics[width=8cm,angle=270]{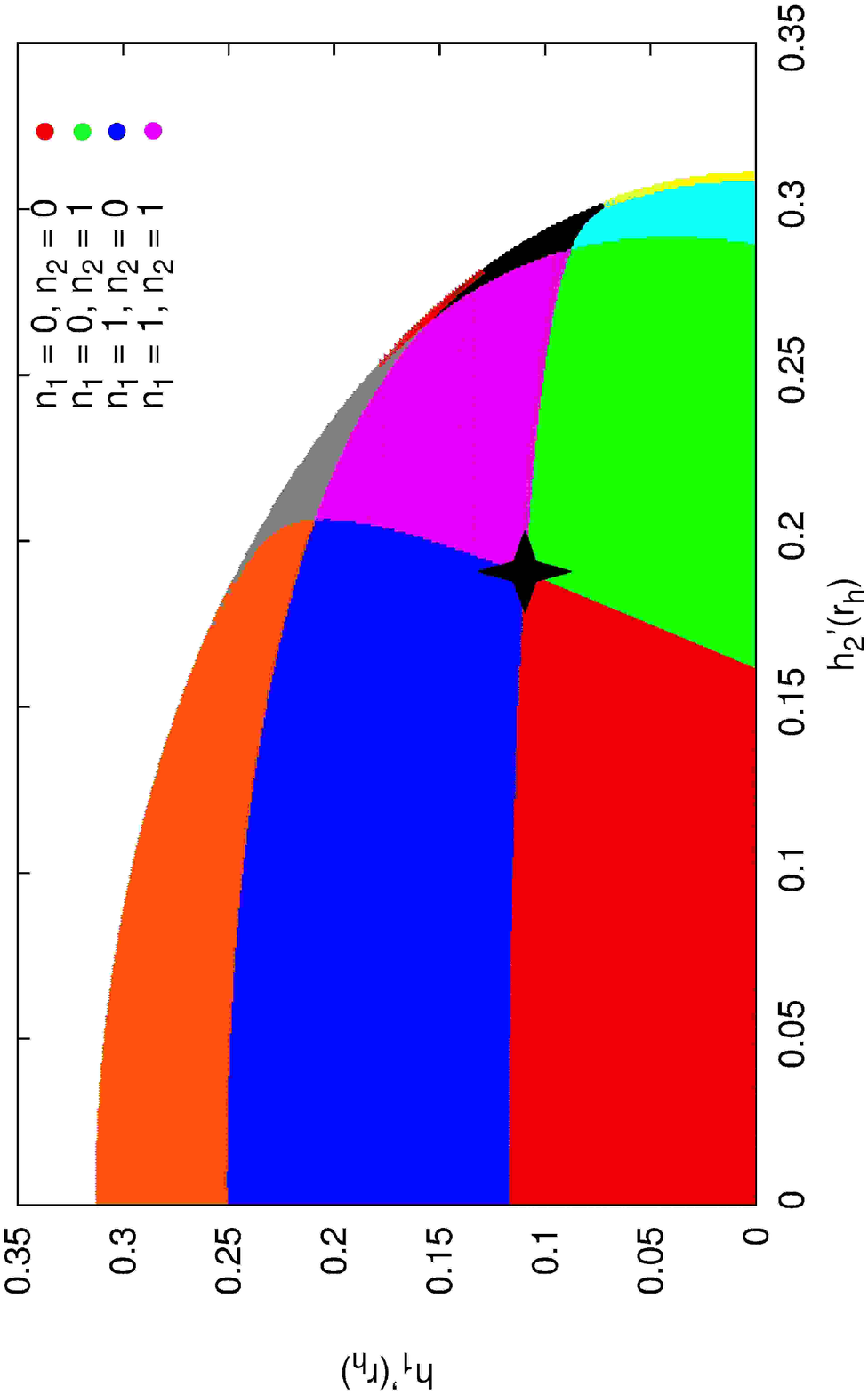}
\includegraphics[width=8cm,angle=270]{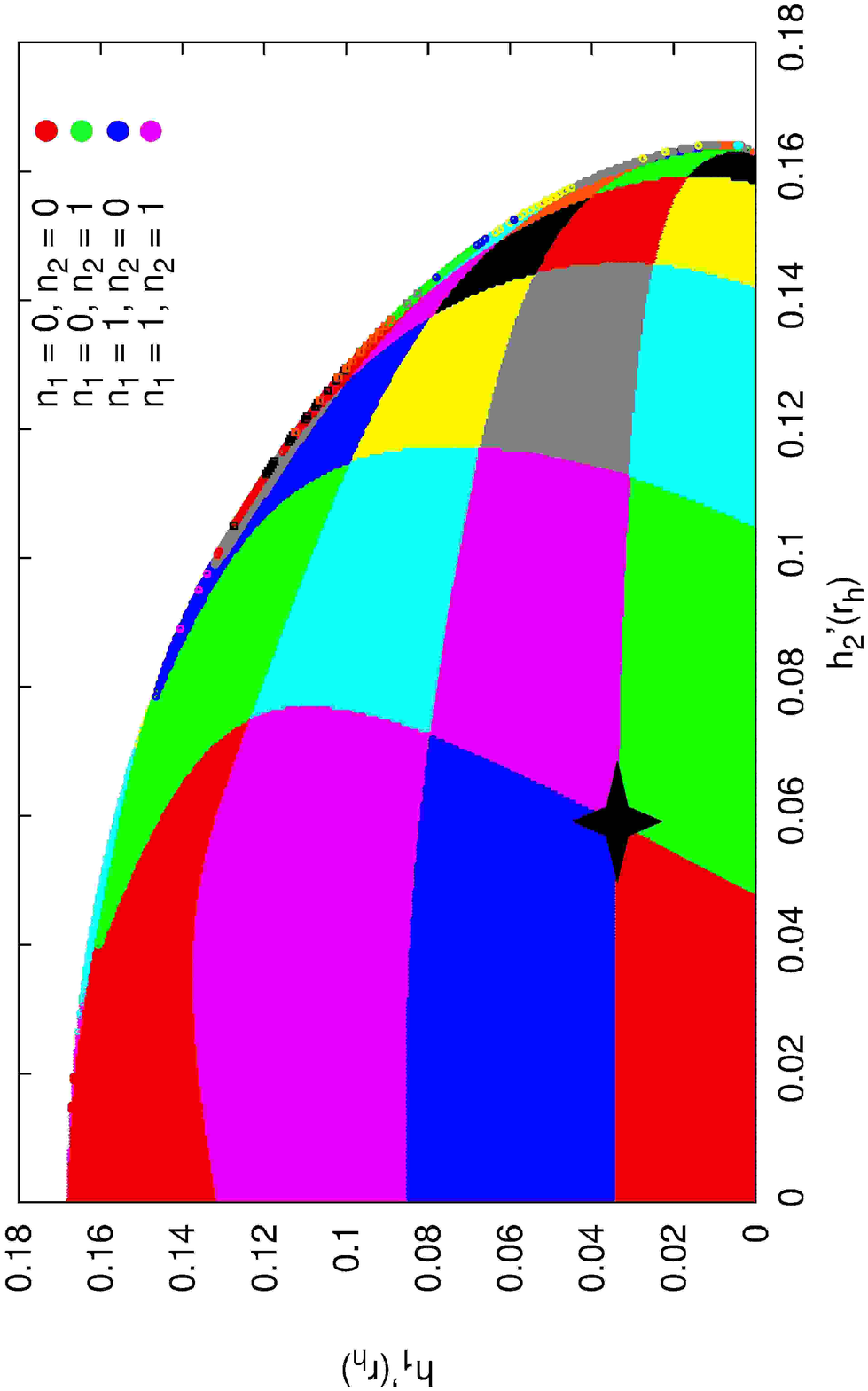}
\end{center}
\caption{Solution spaces for ${\mathfrak {su}}(3)$ planar black holes with $\omega _{1}(r_{h})=0.1=\omega _{2}(r_{h})$ and $\Lambda = -0.1$ (top), $\Lambda = -0.03$ (bottom).
All coloured regions correspond to nontrivial black hole solutions.
The solution space is colour-coded by $(n_{1},n_{2})$, the numbers of zeros of the gauge field functions $\omega _{1}(r)$ and $\omega _{2}(r)$ respectively.
We have explicitly indicated those regions where $(n_{1},n_{2})$ take the values $(0,0)$, $(1,0)$, $(0,1)$ and $(1,1)$, but we find solutions with other combinations of $(n_{1}, n_{2})$.
The nodeless solution where $\omega _{1}(r)\rightarrow 0$ and $\omega _{2}(r) \rightarrow 0$ as $r\rightarrow \infty $ is marked in each case by a black star.}
\label{fig:su3phase}
\end{figure}

In figure \ref{fig:su3phase} we show the solution spaces for ${\mathfrak {su}}(3)$ planar black hole solutions with $\Lambda = -0.1$ (top plot) and $\Lambda = -0.03$ (lower plot).
With $\Lambda $ fixed, there are four parameters describing the black holes: $\omega _{1}(r_{h})$, $\omega _{2}(r_{h})$, $h_{1}'(r_{h})$ and $h_{2}'(r_{h})$.
To produce two-dimensional plots, in figure \ref{fig:su3phase} we have fixed $\omega _{1}(r_{h})=0.1=\omega _{2}(r_{h})$.
Similar diagrams are found for other values of $\omega _{1}(r_{h})$ and $\omega _{2}(r_{h})$.
The solution space is colour-coded according to the values of $n_{1}$ and $n_{2}$, the numbers of zeros of the magnetic gauge field functions $\omega _{1}(r)$ and $\omega _{2}(r)$ respectively.
As was found for spherically symmetric dyonic black holes \cite{Shepherd:2015dse}, the solution space of planar ${\mathfrak {su}}(3)$ black holes shown in figure \ref{fig:su3phase} is complicated. We find regions with many different combinations of $(n_{1},n_{2})$.
For the lower value of $\left| \Lambda \right| $, there are more regions (and more combinations of $(n_{1},n_{2})$).
As in the ${\mathfrak {su}}(2)$ case, our interest is in solutions where $\omega _{1}(r)$ and $\omega _{2}(r)$ both tend to zero as $r\rightarrow \infty $, but have no zeros between the event horizon and infinity.
These solutions arise at the boundaries of the regions where $(n_{1},n_{2})=(0,0)$, $(0,1)$, $(1,0)$ and $(1,1)$.
We have therefore explicitly indicated in figure \ref{fig:su3phase} only those regions with these combinations of $(n_{1},n_{2})$.
In figure \ref{fig:su3phase}, a black star shows the location of the solutions where both $\omega _{1}(r)$ and $\omega _{2}(r)$ tend to zero as $r\rightarrow \infty $.

\begin{figure}
\begin{center}
\includegraphics[width=8cm,angle=270]{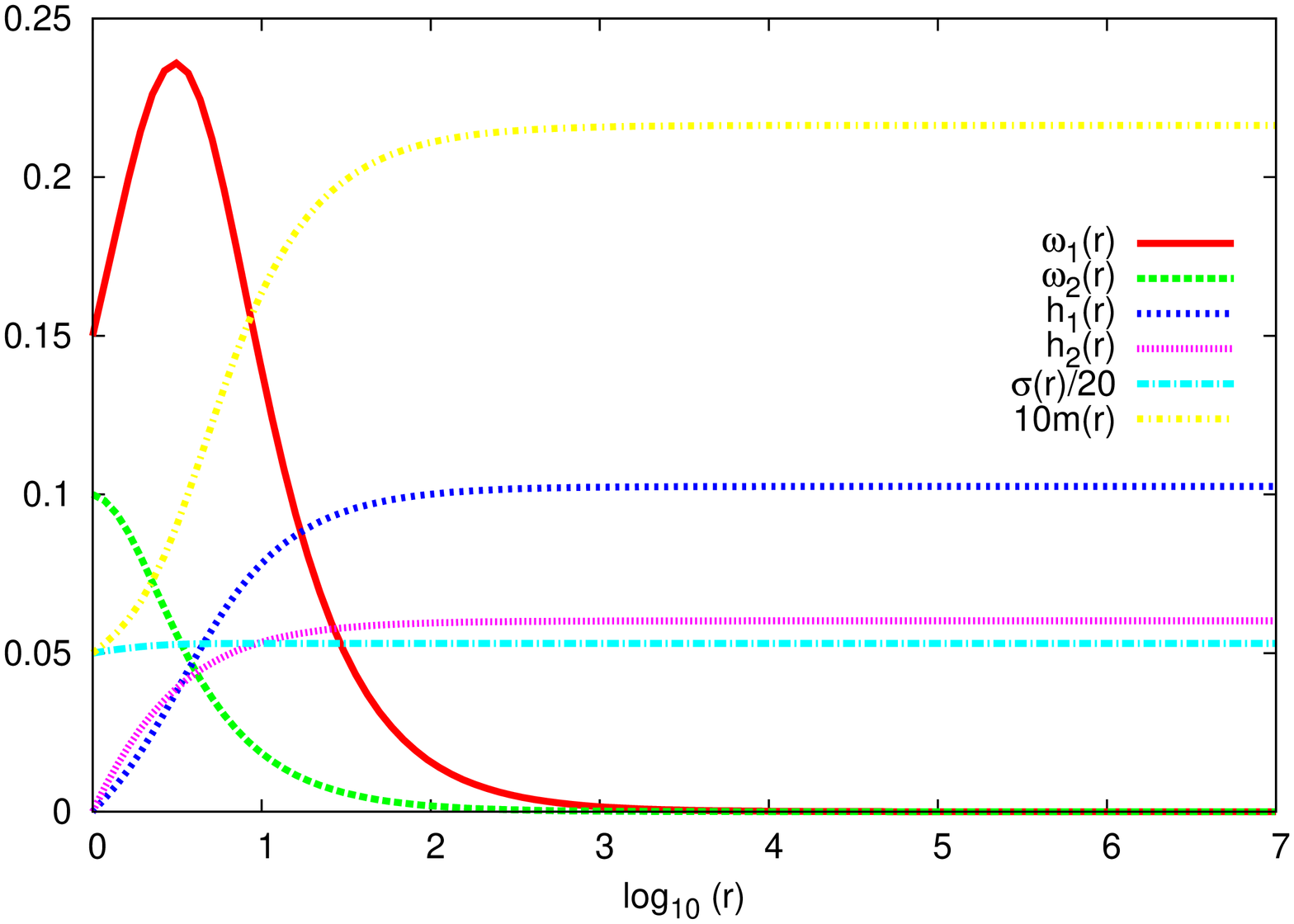}
\end{center}
\caption{Typical ${\mathfrak {su}}(3)$ planar black hole with $\Lambda = -0.03$, $\omega _{1}(r_{h})=0.15$, $\omega _{2}(r_{h})=0.1$. The values of $h_{1}'(r_{h})$ and $h_{2}'(r_{h})$ have been fixed by the requirement that $\omega _{1}(r)\rightarrow 0$ and $\omega _{2}(r) \rightarrow 0 $ as $r\rightarrow \infty $.}
\label{fig:su3example}
\end{figure}

As in the ${\mathfrak {su}}(2)$ case, the solutions we seek (where $\omega _{1},\omega _{2} \rightarrow 0$ as $r\rightarrow \infty $ and have no zeros between the event horizon and infinity) exist only for sufficiently small $\left| \Lambda \right| $, since both $\omega _{1}$ and $\omega _{2}$ are nodeless when $\left| \Lambda \right| $ is sufficiently large \cite{Baxter:2015tda}.
As seen in figure \ref{fig:su3phase}, for fixed $\Lambda $, $\omega _{1}(r_{h})$ and $\omega _{2}(r_{h})$, there is a single combination $(h_{1}'(r_{h}),h_{2}'(r_{h}))$ which gives a solution for which $\omega _{1}(\infty )=0$, $\omega _{2}(\infty )=0$.
We find the relevant values of $h_{1}'(r_{h})$ and $h_{2}'(r_{h})$ using the GSL root-finding algorithm \cite{Alken}.
A typical ${\mathfrak {su}}(3)$ planar black hole is shown in figure \ref{fig:su3example}, for $\Lambda = -0.03$, $\omega _{1}(r_{h})=0.15$ and $\omega _{2}(r_{h})=0.1$.
For these particular values of the parameters, we find that $\omega _{1}(r)$ is increasing close to the horizon, has a maximum and then decreases as $r$ increases, while $\omega _{2}(r)$ is monotonically decreasing. The two electric gauge field functions, $h_{1}(r)$ and $h_{2}(r)$, are both monotonically increasing.
As in figure \ref{fig:su2example}, we have scaled the metric functions $m(r)$ and $\sigma (r)$ in figure \ref{fig:su3example}.  Their properties are similar to the ${\mathfrak {su}}(2)$ example solution in figure \ref{fig:su2example}; in particular, $m(r)$ takes small values and is monotonically increasing; while $\sigma (r)$ is close to unity everywhere outside the event horizon.

\subsection{Physical quantities}
\label{sec:thermo}

We now calculate various physical quantities associated with the static solutions presented in the previous subsection.
In particular, we examine the $N-1$ conserved electric charges possessed by the non-Abelian gauge field, the Hawking temperature, and the free energy.

\begin{figure}
\begin{center}
\includegraphics[width=8cm,angle=270]{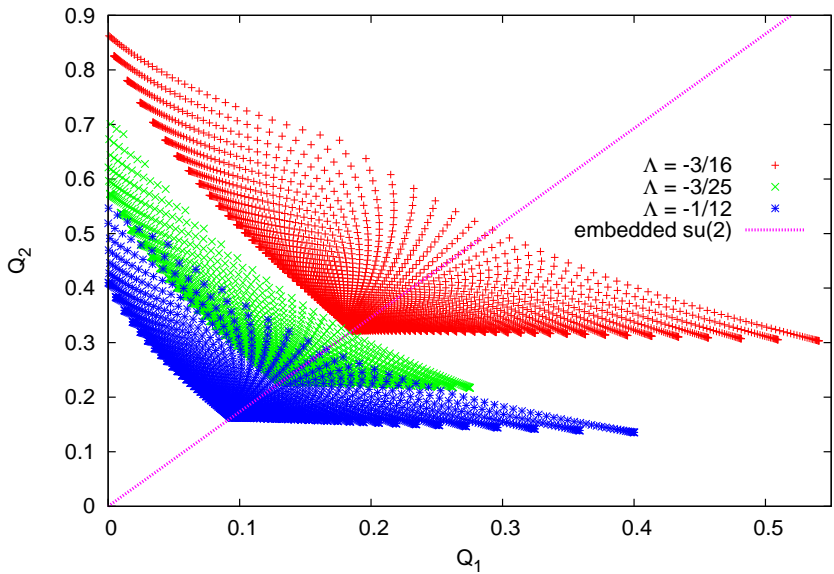}
\end{center}
\caption{Electric charges $Q_{1}$, $Q_{2}$ (\ref{eq:electriccharge}) of planar ${\mathfrak {su}}(3)$ black holes for $\ell = {\sqrt {-3/\Lambda }}=4,5,6$,
with charges for embedded planar ${\mathfrak {su}}(2)$ black holes (with $Q_{2}={\sqrt {3}}Q_{1}$) overlaid. Each point corresponds to a black hole solution found numerically.}
\label{fig:su3charges}
\end{figure}

Since the Lie algebra ${\mathfrak {su}}(N)$ has rank $N-1$, we can define $N-1$ gauge invariant electric charges $Q_{k}$ associated with the gauge potential (\ref{eq:gauge}), as follows \cite{Chrusciel:1987jr} (see also \cite{Brandt:1980em,Chan:1980xv,Creighton:1995au,Goddard:1976qe,Kleihaus:2001ti,Sudarsky:1992ty,Tafel:1982zs,Oh:1987kb,Mann:2006jc,Corichi:1999nw,Kleihaus:2002ee} for definitions of charges for non-Abelian gauge groups):
\begin{equation}
Q_j = \frac{1}{4\pi}\sup_{{\mathfrak {h}}} k\left( X, \int_{\Sigma_\infty} {\mathfrak {h}}^{-1} \ast {\mathfrak {F}} {\mathfrak {h}} \right),
\label{eq:electriccharge}
\end{equation}
where $X$ is an element of the Cartan subalgebra of ${\mathfrak {su}}(N)$, the supremum is taken over all possible
gauge transformations ${\mathfrak {h}}$, the integral is taken over a surface $\Sigma _{\infty }$ at spatial infinity, and $k(*,*)$ denotes the ${\mathfrak {su}}(N)$ Killing form.
On $\Sigma _{\infty}$ the dual field strength is given by
\begin{equation}
\ast {\mathfrak {F}} = -r^2\sigma\sum_{k=1}^{N-1}h_k'H_k\,dx\,dy,
\end{equation}
since we are considering static configurations and $dr=0$ on $\Sigma _{\infty }$.
The integrand in (\ref{eq:electriccharge}) takes its maximal value when ${\mathfrak {h}}^{-1}\ast {\mathfrak {F}} {\mathfrak {h}}$ is a member of the Cartan subalgebra \cite{Chrusciel:1987jr}, but since in our case $\ast {\mathfrak {F}}$ is already in the Cartan subalgebra there is no need to perform a gauge transformation to find the supremum.
A natural choice for a basis of the Cartan subalgebra is the $N-1$ diagonal generators $H_{k}$ (\ref{eq:Hdef}), in which case we find
\begin{equation}
Q_j \propto A_0\lim_{r \to \infty}\sigma(r)r^2h_j'(r),
\end{equation}
where $A_{0}$ is the unit area of $\Sigma _{\infty }$ and we are free to choose the normalization.
We follow the conventions of \cite{Shepherd:2012sz} and define
\begin{equation}
Q_j  =\frac{1}{g}\lim_{r \to \infty} r^2 h_j'(r).
\label{eq:electricQ}
\end{equation}
We will also define a total electric charge $Q$ by
\begin{equation}
Q^2 = \sum_{j=1}^{N-1}Q_j^2.
\label{eq:Qeff}
\end{equation}
In figure \ref{fig:su3charges}, we show scatter plots of the electric charges $(Q_{1},Q_{2})$ for ${\mathfrak {su}}(3)$ black holes with $\ell = {\sqrt {-3/\Lambda }}=4,5,6$. Each coloured point in figure \ref{fig:su3charges} corresponds to a numerical black hole solution, and the points for different values of $\ell $ have different colours.
We have also shown the straight line $Q_{2}={\sqrt {3}}Q_{1}$, along which embedded ${\mathfrak {su}}(2)$ solutions lie.
For fixed $\ell $ (equivalently, fixed $\Lambda $), the electric charges $(Q_{1}, Q_{2})$ lie in a region of the plane which is bounded below by two curves, one of which is almost parallel to the $Q_{1}$ axis, while the other has $Q_{2}$ increasing as $Q_{1}$ decreases. These two bounding curves meet at an apex, which is on the line of embedded ${\mathfrak {su}}(2)$ solutions.

In general the magnetic part of the gauge field also carries $N-1$ conserved charges ${\tilde {Q}}_{k}$, defined analogously to (\ref{eq:electriccharge}) \cite{Chrusciel:1987jr}. Following \cite{Shepherd:2012sz}, for planar black holes these charges take the form
\begin{equation}
{\tilde {Q}}_k = \frac{\sqrt{2k(k+1)}}{2}\left(\frac{\omega_{k+1}(\infty)^2}{k+1} - \frac{\omega_k(\infty)^2}{k}\right) .
\end{equation}
However, since we are considering solutions in which all the magnetic gauge field functions $\omega _{k}\rightarrow 0$ as $r\rightarrow \infty $, we find that all the magnetic charges vanish.

To define the mass of our planar black holes, we use the counterterm formalism of \cite{Balasubramanian:1999re}, which is unaffected by the ${\mathfrak {su}}(N)$ gauge field.  Since we are considering solutions to the equations of motion, the quasi-local stress-energy tensor on the boundary is given by \cite{Balasubramanian:1999re}
\begin{equation}
T_B^{\mu\nu} = \frac{1}{2} \left( \Theta^{\mu\nu} - \Theta \gamma^{\mu\nu} -\frac{2}{\ell }\gamma^{\mu\nu} - \ell {\tilde {G}}^{\mu\nu} \right),
\label{eq:boundaryTmunu}
\end{equation}
where $\gamma^{\mu\nu}$ is the boundary metric, ${\tilde {G}}^{\mu\nu}$ is the Einstein tensor on the boundary, and the extrinsic curvature $\Theta^{\mu\nu}$ is given by
\begin{equation}
\Theta^{\mu\nu} = -\frac{1}{2}\left( \nabla^\mu \hat{n}^\nu + \nabla^\nu \hat{n}^\mu \right),
\end{equation}
where $\hat{n}^\mu$ is the outward pointing normal to surfaces of constant radial coordinate $r$.
We define the mass $M$ to be
\begin{equation}
M  =  \int_{\Sigma_\infty} \ell r T_{tt} \, dx \, dy = \frac{A_0 m_0}{4\pi G}
\end{equation}
where $A_0$ is the unit area of the surface $\Sigma_{\infty}$ and the constant $m_{0}=\lim _{r\rightarrow \infty }m(r)$ (\ref{eq:infinity}).

The Bekenstein-Hawking entropy of our planar black holes is given by
\begin{equation}
S=\frac {A_{0}}{4G},
\label{eq:entropy}
\end{equation}
and the Hawking temperature $T$ is
\begin{equation}
T = \frac {\mu ' (r_{h})\sigma (r_{h})}{4\pi }.
\label{eq:THawking}
\end{equation}
We then define the free energy of each planar black hole to be
\begin{equation}
F = M - TS = \frac {A_{0}}{4\pi G} \left[ m_{0} - \frac {1}{4} \mu '(r_{h})\sigma (r_{h}) \right] .
\label{eq:freeenergy}
\end{equation}

We now investigate whether a non-Abelian ${\mathfrak {su}}(N)$ black hole is thermodynamically favoured over an embedded planar RN-AdS black hole, working in the canonical ensemble with fixed electric charge in order to compare our results with those in \cite{Gubser:2008zu,Manvelyan:2008sv}
(see for example \cite{Roberts:2008ns,Gubser:2010dm,Arias:2012py,Herzog:2014tpa} for work using the grand canonical ensemble with fixed chemical potential rather than fixed charge).  If we consider an RN-AdS black hole with the same temperature and effective charge as an ${\mathfrak {su}}(N)$ black hole, and denote its free energy by $F_{RN}$, then the ${\mathfrak {su}}(N)$ black hole will be thermodynamically favoured when
\begin{equation}
\Delta F = F- F_{RN}<0.
\end{equation}
We can determine the event horizon radius $r_{h}^{RN}$ of the relevant RN-AdS black hole using the requirement that the effective charge and Hawking temperature are the same as the non-Abelian solutions.
Using the RN-AdS metric function (\ref{eq:muRN}), and the Hawking temperature (\ref{eq:THawking}) with $\sigma (r)\equiv 1$, we find that the Hawking temperature of an embedded RN-AdS black hole with electric charge $Q_{RN}$ is given by
\begin{equation}
T_{RN} = -\frac {1}{4\pi } \left( \frac {\alpha ^{2}Q_{RN}^{2}}{\left(r_{h}^{RN}\right) ^{3}} - \frac {3r_{h}^{RN}}{\ell ^{2}} \right) .
\label{eq:THawkingRN}
\end{equation}
Given $T_{RN}$, we solve (\ref{eq:THawkingRN}) for the event horizon radius $r_{h}^{RN}$, where the electric charge is given by (\ref{eq:Qeff}).
Since the metric function $\mu _{RN}(r)$ (\ref{eq:muRN}) vanishes at the event horizon, we have
\begin{equation}
M_{RN} = \frac{Q_{RN}^2}{2r_h^{RN}} + \frac{\left(r_h^{RN}\right)^3}{2\ell ^2},
\end{equation}
so we can write the free energy of the embedded Reissner-Nordstr\"{o}m black hole as
\begin{equation}
F_{RN} = \frac{A_0}{4\pi G} \left( \frac{3\alpha^2Q_{RN}^2}{4r_h^{RN}} - \frac{\left(r_h^{RN}\right) ^3}{4\ell^2} \right),
\end{equation}
and hence
\begin{equation}
\Delta F = \frac{A_0}{4\pi G} \left(m_0 - \frac{\mu '(r_{h}) \sigma(r_{h})}{4} - \frac{3\alpha^2Q_{RN}^2}{4r_h^{RN}} + \frac{ \left( r_h^{RN}\right) ^3}{4\ell ^2} \right),
\label{eq:deltaF}
\end{equation}
where $r_{h}=1$ is the event horizon radius of the non-Abelian black hole.

Since our non-Abelian black hole solutions are known only numerically, we calculate $\Delta F$ (\ref{eq:deltaF}) numerically. After finding a non-Abelian black hole solution of the field equations (\ref{eq:staticfieldequations}) (see section \ref{sec:sols}), we use the GSL root-finding algorithm \cite{Alken}
to solve (\ref{eq:THawkingRN}) for $r_{h}^{RN}$ and hence calculate the difference in free energy between the non-Abelian solutions and the embedded RN-AdS black hole with the same temperature and electric charge using (\ref{eq:deltaF}).

\begin{figure}
\begin{center}
\includegraphics[width=8cm,angle=270]{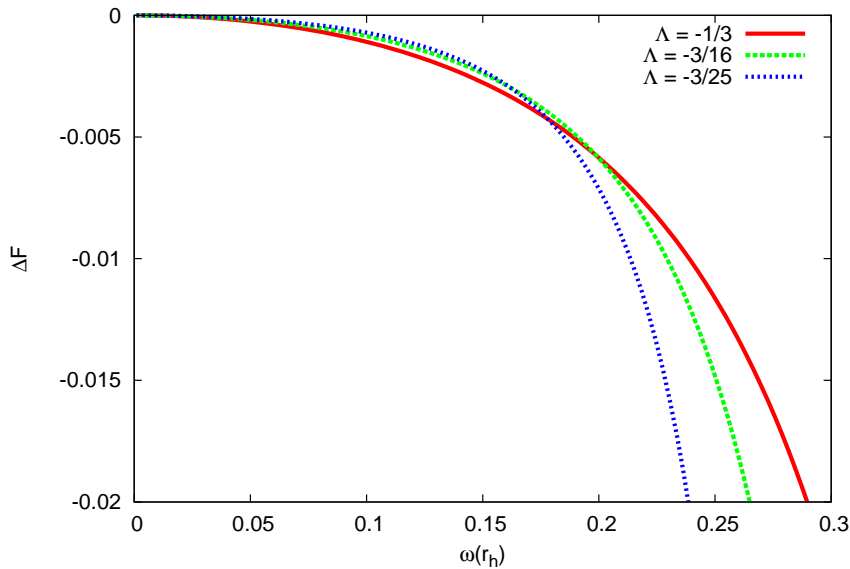}
\end{center}
\caption{Difference in free energy $\Delta F$ between planar ${\mathfrak {su}}(2)$ solutions and RN-AdS black holes with the same Hawking temperature and electric charge, plotted against the value of the gauge field function on the horizon $\omega (r_{h})$, for $\ell = {\sqrt {-3/\Lambda }}=3,4,5$.
We note that $\Delta F<0$ for all $\omega (r_{h})>0$, so that an ${\mathfrak {su}}(2)$ black hole is thermodynamically favoured over the corresponding RN-AdS black hole. This is in agreement with the results in \cite{Gubser:2008zu}.}
\label{fig:su2deltaF}
\end{figure}

First we consider ${\mathfrak {su}}(2)$ black holes. In figure \ref{fig:su2deltaF} we plot $\Delta F$ (\ref{eq:deltaF}) against the value of the magnetic gauge field function at the horizon $\omega (r_{h})$ for various values of $\ell = {\sqrt {-3/\Lambda }}$.
As $\omega (r_{h})\rightarrow 0$, the non-Abelian black holes approach the embedded Abelian RN-AdS black hole (\ref{eq:RNAdS}) since the field equations (\ref{eq:staticfieldequations}) ensure that $\omega (r) \equiv 0$ if $\omega (r_{h})=0$.
As expected from \cite{Gubser:2008zu}, the difference in free energy $\Delta F$ is negative for all solutions with nonzero $\omega (r_{h})$.
Therefore all nontrivial ${\mathfrak {su}}(2)$ non-Abelian black holes are thermodynamically favoured over the embedded RN-AdS black holes with the same temperature and charge.

\begin{figure}
\begin{center}
\includegraphics[width=8cm,angle=270]{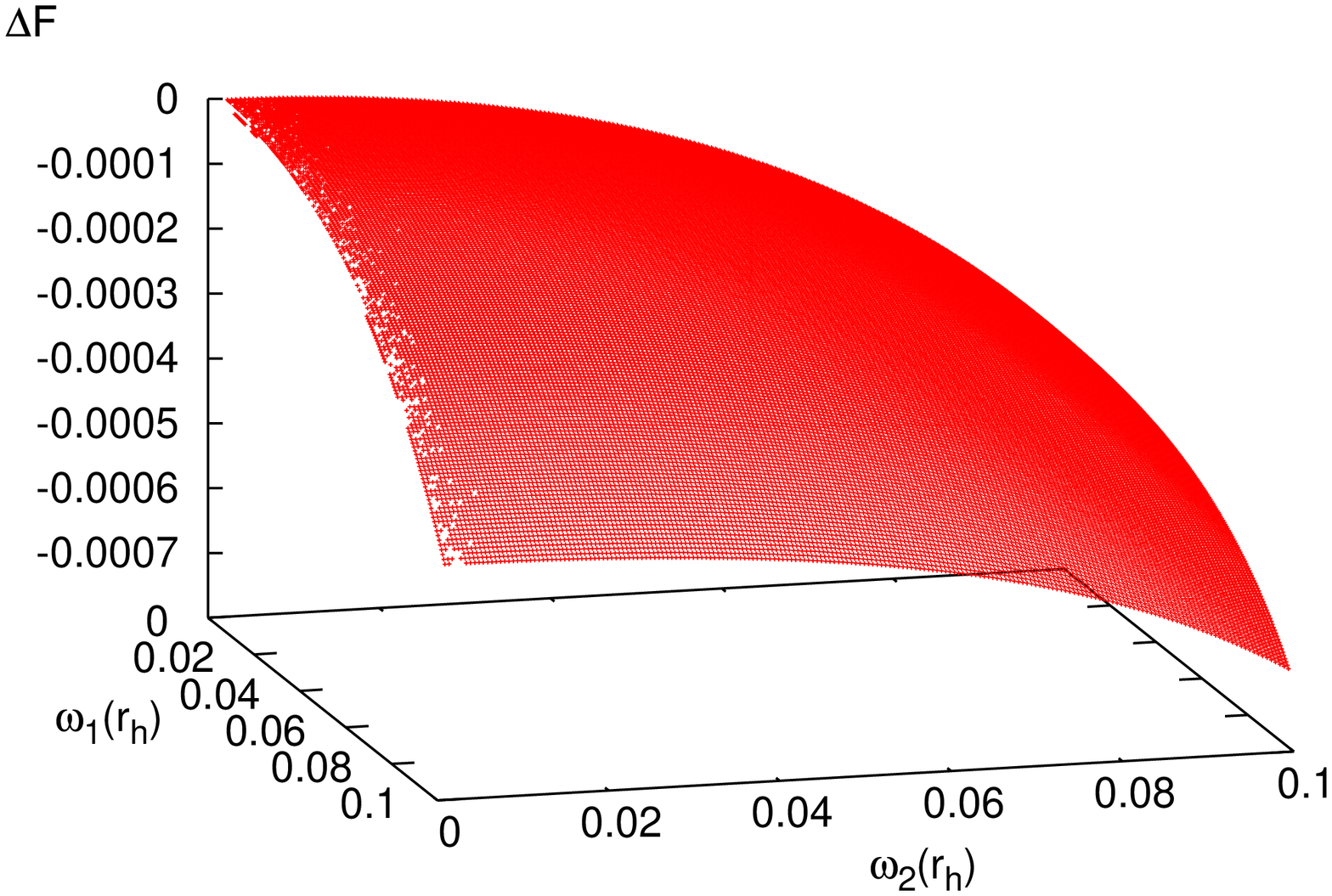}
\end{center}
\caption{Difference in free energy $\Delta F$ between planar ${\mathfrak {su}}(3)$ solutions and RN-AdS black holes with the same Hawking temperature and electric charge, plotted against the values of the gauge field functions on the horizon $\omega_{1} (r_{h})$, $\omega _{2}(r_{h})$, for $\ell = 5$.
As in the ${\mathfrak {su}}(2)$ case, we have $\Delta F<0$ for all $\omega _{1}(r_{h}),\omega _{2}(r_{h})>0$, so that the hairy black holes are thermodynamically favoured over the corresponding RN-AdS black hole.}
\label{fig:su3deltaFw}
\end{figure}

Now we turn to the ${\mathfrak {su}}(3)$ case.
In figure \ref{fig:su3deltaFw} we plot $\Delta F$ (\ref{eq:deltaF}) against $\omega _{1}(r_{h})$ and $\omega _{2}(r_{h})$ for $\ell = {\sqrt {-3/\Lambda }}=5$. Plots for other values of $\ell $ are qualitatively similar.
As in the ${\mathfrak {su}}(2)$ case, as both $\omega _{1}(r_{h})$ and $\omega _{2}(r_{h})$ approach zero, the black holes approach the embedded Abelian RN-AdS solution (\ref{eq:RNAdS}).
We find that $\Delta F<0$ for all black holes with nonzero $\omega _{1}(r_{h})$ and $\omega _{2}(r_{h})$.
Therefore, as in the ${\mathfrak {su}}(2)$ case, the non-Abelian ${\mathfrak {su}}(3)$ black holes are thermodynamically favoured over the Abelian RN-AdS black hole with the same temperature and charge (\ref{eq:Qeff}).

\begin{figure}
\begin{center}
\includegraphics[width=8cm,angle=270]{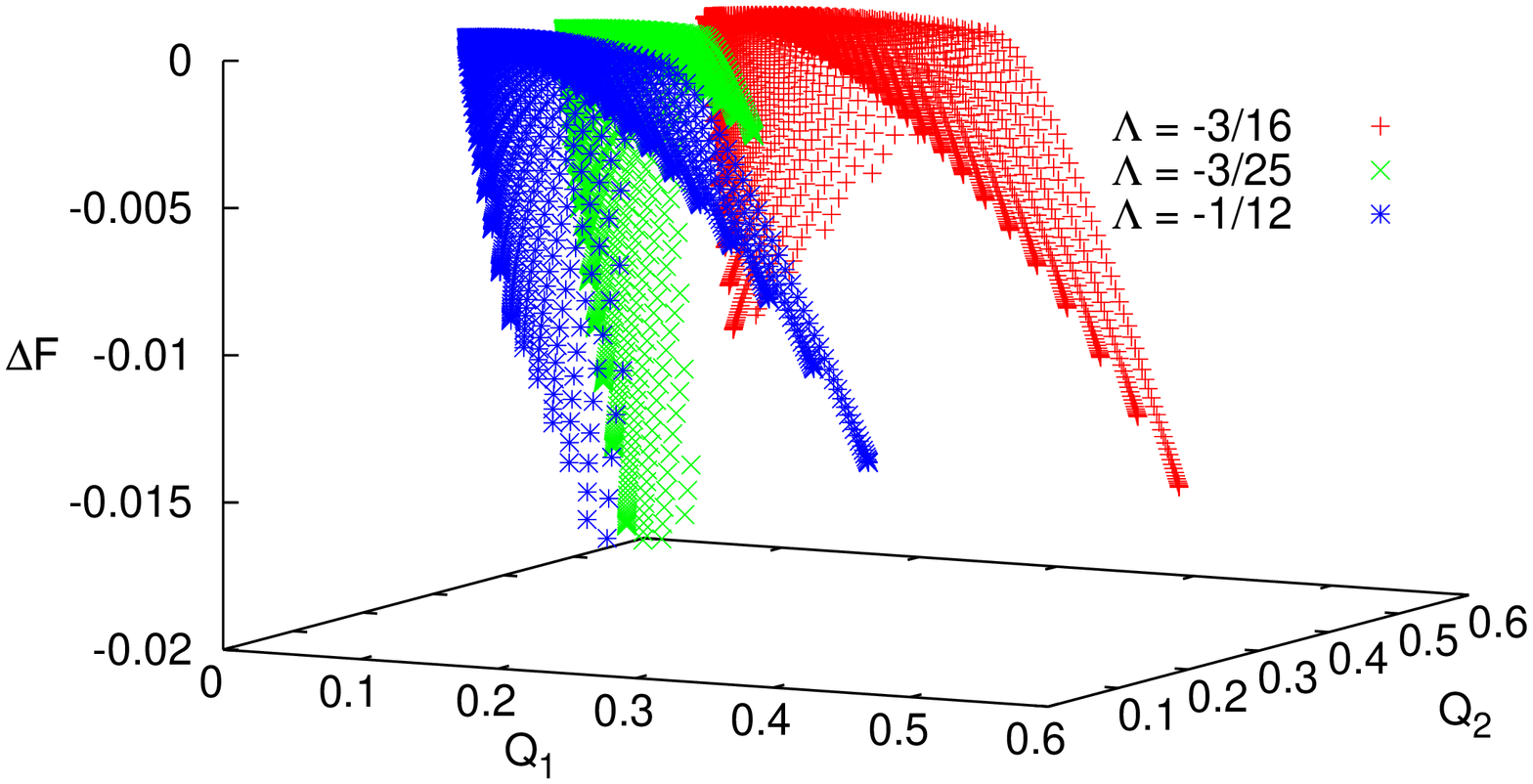}
\end{center}
\caption{Difference in free energy $\Delta F$ between planar ${\mathfrak {su}}(3)$ solutions and RN-AdS black holes with the same Hawking temperature and electric charge, plotted against the electric charges $Q_{1}$, $Q_{2}$, for $\ell = {\sqrt {-3/\Lambda }} =4, 5, 6$. Each point corresponds to a black hole solution found numerically.}
\label{fig:su3deltaFQ}
\end{figure}

For ${\mathfrak {su}}(3)$ non-Abelian gauge field configurations, as well as the effective charge $Q$ (\ref{eq:Qeff}), there are two electric charges $Q_{1}$, $Q_{2}$, given by (\ref{eq:electricQ}).  We now explore how $\Delta F$ (\ref{eq:deltaF}) depends on these electric charges.
In figure \ref{fig:su3deltaFQ} we plot the same charge data as in figure \ref{fig:su3charges} together with the difference in free energy $\Delta F$ (\ref{eq:deltaF}).
For each value of $\ell $ shown, there is a surface in $(Q_{1},Q_{2},\Delta F)$ space which has a fold along the line corresponding to embedded ${\mathfrak {su}}(2)$ black holes.  This means that $\left| \Delta F\right| $ is smaller for embedded ${\mathfrak {su}}(2)$ black holes than it is for genuinely ${\mathfrak {su}}(3)$ solutions.  We deduce that genuinely ${\mathfrak {su}}(3)$ solutions are thermodynamically favoured over embedded ${\mathfrak {su}}(2)$ solutions.  This is to be expected: in the ${\mathfrak {su}}(3)$ case there are more possible field configurations giving the same effective charge $Q$ (\ref{eq:Qeff}), and therefore more chance of finding a configuration with a lower free energy.

In the ${\mathfrak {su}}(2)$ case, there is a phase transition at a critical temperature $T_{C}$, above which only the embedded Abelian RN-AdS solutions exist, and below which the non-Abelian EYM black holes exist and are thermodynamically preferred.
In addition, there is a current on the boundary \cite{Gubser:2008zu,Manvelyan:2008sv}, given by
\begin{equation}
J = - \lim _{r\rightarrow \infty } r^{2}\omega '(r).
\label{eq:su2current}
\end{equation}
The holographic interpretation for the current (\ref{eq:su2current}) is as an order parameter \cite{Gubser:2008zu}, which is zero at temperatures at and above the phase transition, $T\ge T_{C}$.
Since $\omega (r) \equiv 0$ for the Abelian RN-AdS black hole embedded in ${\mathfrak {su}}(2)$ EYM theory, it is clear that $J$ (\ref{eq:su2current}) vanishes for RN-AdS solutions.
Plots of the current $J$ as a function of black hole temperature $T$ for ${\mathfrak {su}}(2)$ EYM black holes with $\zeta =1$ can be found in \cite{Roberts:2008ns} for the probe limit and \cite{Gubser:2008zu,Arias:2012py} with back-reaction included (see also \cite{Gubser:2010dm,Arias:2012py,Arias:2014msa,Herzog:2014tpa} for the anisotropic case with $\zeta = 0$).

\begin{figure}
\begin{center}
\includegraphics[width=8cm,angle=270]{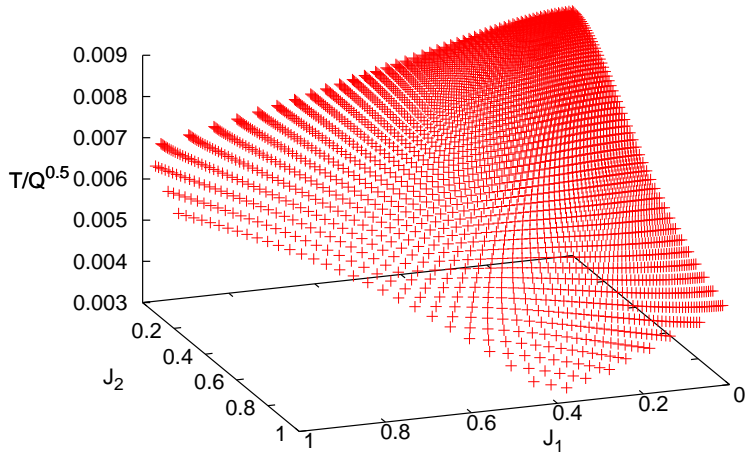}
\end{center}
\caption{Hawking temperature divided by the square root of the effective charge for planar ${\mathfrak {su}}(3)$ black holes with $\ell = {\sqrt {-3/\Lambda }}=4$, plotted against the components of the vector order parameter $J_{1}$, $J_{2}$. Each point corresponds to a black hole solution found numerically.}
\label{fig:JT}
\end{figure}

For gauge group ${\mathfrak {su}}(N)$, there are $N-1$ magnetic gauge field functions $\omega _{k}(r)$, and therefore $N-1$ currents $J_{k}$, $k=1,\ldots ,N-1$, given by
\begin{equation}
J_{k} = -\lim _{r\rightarrow \infty }r^{2} \omega _{k}'(r), \qquad k =1,\ldots , N-1.
\label{eq:suNcurrent}
\end{equation}
As in the ${\mathfrak {su}}(2)$ case, we expect to find a phase transition at some critical temperature $T_{C}$, above which only the embedded Abelian RN-AdS solutions exist, and below which the ${\mathfrak {su}}(N)$ non-Abelian black holes exist and are thermodynamically preferred.
In the ${\mathfrak {su}}(N)$ case, the embedded Abelian RN-AdS black hole has $\omega _{k}(r) \equiv 0$ for all $k$ (see section \ref{sec:trivial}), and hence
we expect that the currents $J_{k}$ (\ref{eq:suNcurrent}) will vanish for all $k$ at the phase transition.
We therefore consider the $J_{k}$ to be components of a vector order parameter, the length of which is zero at the phase transition, that is, we anticipate that
\begin{equation}
J^{2} = \sum _{k=1}^{N-1} J_{k}^{2}
\begin{cases}
=0 & {\mbox {for $T\ge T_{C}$,}} \\
\neq 0 & {\mbox {for $T<T_{C}$.}}
\end{cases}
\label{eq:Jlength}
\end{equation}
To test this hypothesis, in figure \ref{fig:JT} we plot the quantity $T/Q^{0.5}$, where $Q$ is the effective charge (\ref{eq:Qeff}) against the components of our vector order parameter $(J_{1}, J_{2})$ (\ref{eq:suNcurrent}) for ${\mathfrak {su}}(3)$ black holes with $\ell = {\sqrt {-3/\Lambda }}= 4$.
The quantity $T/Q^{0.5}$ is plotted because it is invariant under the rescaling (\ref{eq:scaling3}).
In figure \ref{fig:JT} it can be seen that the maximum temperature is approached as the length of the vector order parameter $J={\sqrt {J_{1}^{2}+J_{2}^{2}}}$ goes to zero and the RN-AdS solution is approached.
For temperatures below this maximum temperature, we find nonzero values for the order parameters $J_{1}$ and $J_{2}$.
Similar results were found in the ${\mathfrak {su}}(2)$ case \cite{Gubser:2008zu,Roberts:2008ns,Arias:2012py}.
We investigate whether this maximum temperature is indeed the critical temperature $T_{C}$ in the next subsection.

\subsection{Perturbations of the Reissner-Nordstr\"om solution}
\label{sec:RNpert}

We expect to find a phase transition between the embedded planar RN-AdS black hole and a nontrivial ${\mathfrak {su}}(N)$ planar hairy black hole when the temperature decreases below the critical temperature $T_{C}$.
For this to happen, as well as the planar hairy black hole having lower free energy, it must be the case that the planar RN-AdS black hole admits a static ${\mathfrak {su}}(N)$ perturbation when $T=T_{C}$.
If such a static perturbation exists, then the planar RN-AdS can decay into the planar hairy ${\mathfrak {su}}(N)$ black hole when it becomes thermodynamically favourable to do so.

We now investigate whether the planar RN-AdS black hole does indeed have a static ${\mathfrak {su}}(N)$ perturbation.
To this end, consider the planar embedded RN-AdS black hole with metric (\ref{eq:RNAdS}) and static ${\mathfrak {su}}(N)$ gauge field perturbations $\delta h_{p}$, $\delta \omega _{k}$, so that the ${\mathfrak {su}}(N)$ gauge potential takes the form
\begin{equation}
-gA = \sum _{p=1}^{N-1} \left[ h_{p,RN}(r) + \delta h_{p}(r)\right] H_{p} dt + \sum _{k=1}^{N-1}  F_{k} \delta \omega _{k}(r) dx +
\sum _{k=1}^{N-1} G_{k}\delta \omega _{k}(r) dy ,
\end{equation}
where $h_{p,RN}$ are the equilibrium forms of the electric gauge field functions $h_{p}$, given by (\ref{eq:RNgauge1}) and the matrices $H_{p}$, $F_{k}$ and $G_{k}$ are defined by (\ref{eq:Hdef}, \ref{eq:FGdef}).

We consider the back-reaction of the perturbations $\delta h_{p}$, $\delta \omega _{k}$ on the metric, which takes the form
\begin{eqnarray}
ds^2  & =  & -\left[1 + \delta \sigma(r) \right]^2 \left[\mu_{RN}(r) + \delta \mu(r) \right] dt^2
 + r^2 \left[  dx^2 + dy^2 \right] + \left[\mu_{RN}(r) + \delta \mu(r) \right]^{-1} dr^2
 \nonumber \\
      & = & -\left[\mu_{RN}(r) + \delta \mu(r) + 2\mu_{RN}(r)\delta\sigma(r)\right]  dt^2
      + r^2 \left[ dx^2 + dy^2 \right] + \frac{\mu_{RN}(r) - \delta \mu(r)}{\mu_{RN}(r)^2} dr^2,
\nonumber \\
\label{eq:RNpertsmetric}
\end{eqnarray}
to first order in the perturbations, where $\mu _{RN}$ is given by (\ref{eq:muRN}).
Defining a new metric perturbation $\delta m$ by
\begin{equation}
\delta \mu(r) = -\frac{2 \delta m(r)}{r},
\label{eq:RNpertmu}
\end{equation}
the linearized EYM equations for the perturbations are
\begin{subequations}
\begin{eqnarray}
\delta m' & = &\alpha^2 r^2\sum_{k=1}^{N-1}\left[ 2h_{k,RN}'\delta h_k' - 2 \left( h_{k,RN}'\right)^2\delta\sigma \right] ,
\label{eq:RNpertdeltamprime}\\
\delta\sigma' & = &0,
\label{eq:RNpertdeltasigmaprime}\\
0 		& = & \delta\omega_k'' + \frac{\mu_{RN}'\delta\omega_k'}{\mu_{RN}} + \frac{\delta\omega_k}{\mu _{RN}}\left(\sqrt{\frac{k+1}{2k}}h_{k,RN} - \sqrt{\frac{k-1}{2k}}h_{k-1,RN}\right),
\label{eq:RNpertdeltaomegak''}\\
\delta h_k'' 	& = & h_{k, RN}' \delta\sigma'  -\frac{2}{r}\delta h_k' = - \frac {2}{r}\delta h_{k}' .
\label{eq:RNpertdeltahk''}
\end{eqnarray}
\end{subequations}
Therefore, to first order in the perturbations, there is no coupling between the electric gauge field perturbations $\delta h_{k}$ and the magnetic gauge field perturbations $\delta \omega _{k}$.
The equation (\ref{eq:RNpertdeltahk''}) is identical to that (\ref{eq:RNheqn}) satisfied by the equilibrium electric gauge field (\ref{eq:RNgauge1}), and hence the electric gauge field perturbation corresponds simply to a perturbation of $h_{k,RN}'(r_{h})$.
The equation for $\delta m'$ (\ref{eq:RNpertdeltamprime}) can readily be integrated, and its solution corresponds to a perturbation of the RN-AdS mass $M_{RN}$ (\ref{eq:muRN}) since $\delta \sigma $ is a constant from (\ref{eq:RNpertdeltasigmaprime}).

This leaves the equation governing magnetic gauge field perturbations (\ref{eq:RNpertdeltaomegak''}).  We use the GSL root-finding algorithm \cite{Alken} to solve this equation numerically, seeking solutions where the perturbations $\delta \omega _{k}$ vanish at infinity.
This boundary condition gives an eigenvalue problem for the constants $h_{k,RN}'(r_{h})$.
Once these constants are determined, the charge of the RN-AdS solution is computed from (\ref{eq:QRN}), and its temperature from (\ref{eq:THawkingRN}).
We expect that the temperature for which the RN-AdS black hole admits this static perturbation is the critical temperature $T_{C}$, and that non-Abelian ${\mathfrak {su}}(N)$ planar black holes exist only at temperatures below $T_{C}$.

\begin{figure}
\begin{center}
\includegraphics[width=8cm,angle=270]{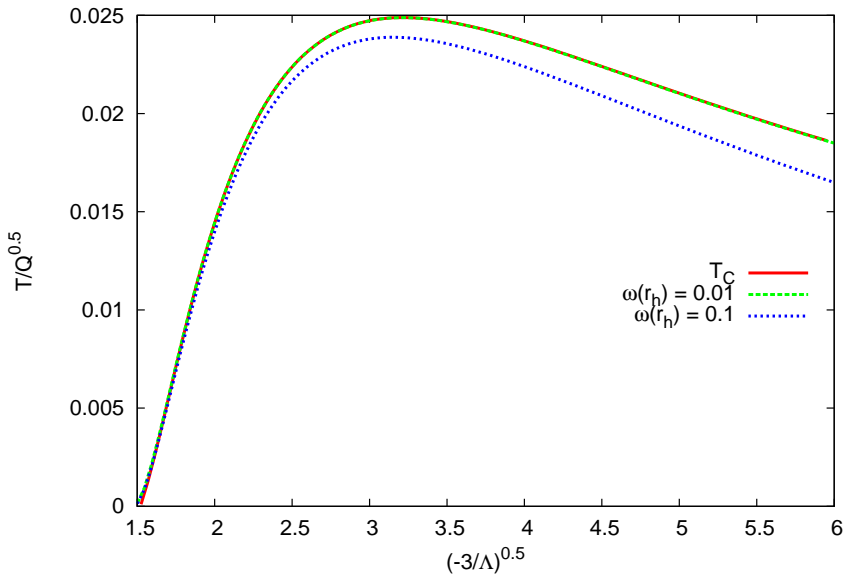}
\end{center}
\caption{Temperature divided by the square root of electric charge for planar ${\mathfrak {su}}(2)$ black holes with $\omega (r_{h})=0.1, 0.01$, against the AdS radius $\ell = {\sqrt {-3/\Lambda }}$.  The critical temperature $T_{C}$ is also shown.
The curve for $\omega (r_{h})=0.01$ lies very slightly below the critical temperature curve.
Our results are in agreement with those in \cite{Gubser:2008zu}.}
\label{fig:TCsu2}
\end{figure}

\begin{figure}
\begin{center}
\includegraphics[width=8cm,angle=270]{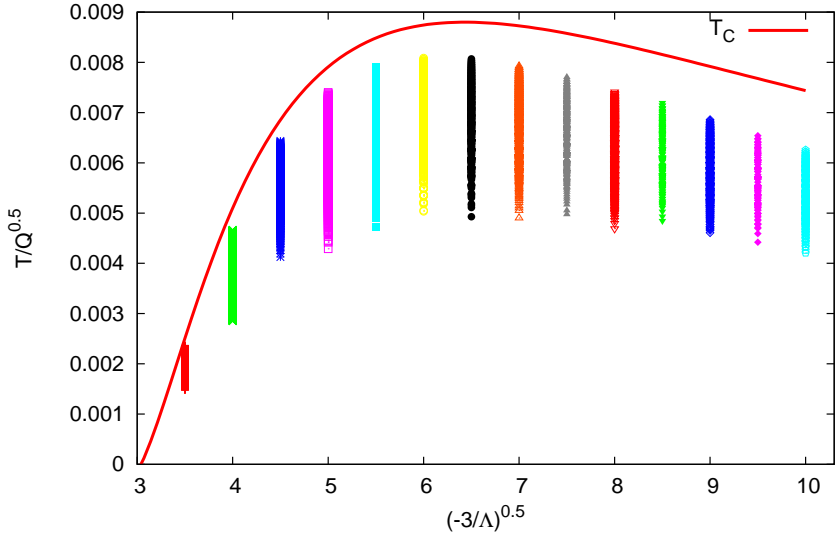}
\end{center}
\caption{Temperature divided by the square root of the effective charge for planar ${\mathfrak {su}}(3)$ black holes, together with the critical temperature $T_{C}$.
Each dot on the vertical lines corresponds to an ${\mathfrak {su}}(3)$ planar black hole.}
\label{fig:TCsu3}
\end{figure}

In figure \ref{fig:TCsu2} we consider ${\mathfrak {su}}(2)$ EYM black holes. We show the scale-invariant quantity $T/Q^{0.5}$ against the length scale $\ell = {\sqrt {-3/\Lambda }}$ for $\omega (r_{h})=0.1$ and $\omega (r_{h})=0.01$, together with the critical temperature $T_{C}$ for ${\mathfrak {su}}(2)$ perturbations of RN-AdS. The curve for $\omega (r_{h})=0.01$ lies very slightly below the $T_{C}$ curve.
As expected from \cite{Gubser:2008zu}, we find that ${\mathfrak {su}}(2)$ solutions exist only for temperatures below the critical temperature $T_{C}$, and that the critical temperature is approached as $\omega (r_{h})\rightarrow 0$.

We consider ${\mathfrak {su}}(3)$ EYM black holes in figure \ref{fig:TCsu3}.  Here we have chosen discrete values of $\ell = {\sqrt {-3/\Lambda }}$ and scanned over a range of values of $\omega _{1}(r_{h})$ and $\omega _{2}(r_{h})$ for each $\ell $. Each point on the vertical lines in figure \ref{fig:TCsu3} corresponds to a numerical ${\mathfrak {su}}(3)$ black hole.  The curve in figure \ref{fig:TCsu3} is the critical temperature $T_{C}$ for ${\mathfrak {su}}(3)$ perturbations of RN-AdS.
As in the ${\mathfrak {su}}(2)$ case, we find that nontrivial ${\mathfrak {su}}(3)$ black holes exist only for temperatures below the critical temperature, $T\le T_{C}$, and that the temperature approaches $T_{C}$ as $\omega _{1}(r_{h})$ and $\omega _{2}(r_{h})$ approach zero.
However, it becomes increasingly difficult to distinguish numerically between nodeless solutions and those with nodes when $\omega _{1}(r_{h})$ and $\omega _{2}(r_{h})$ are close to zero, and for this reason we were unable to find solutions very close to the phase transition.

In this subsection we have shown that there is a phase transition at a critical temperature $T_{C}$, at which the embedded Abelian RN-AdS solution can decay into ${\mathfrak {su}}(2)$ or ${\mathfrak {su}}(3)$ black holes, which exist at temperatures below $T_{C}$.
The critical temperature is approached as the non-Abelian solutions approach the RN-AdS black hole, so the length of the vector order parameter (\ref{eq:Jlength}) tends to zero as the phase transition is approached from below, and vanishes above the critical temperature.
In the previous subsection we showed that ${\mathfrak {su}}(2)$ and ${\mathfrak {su}}(3)$ non-Abelian black holes are thermodynamically favoured over embedded Abelian RN-AdS solutions with the same effective charge (\ref{eq:Qeff}).
Furthermore, the ${\mathfrak {su}}(3)$ black holes are thermodynamically favoured over the embedded ${\mathfrak {su}}(2)$ black holes, which implies that the RN-AdS solution will decay into the most complicated possible configuration.

\section{Gauge field perturbations}
\label{sec:pert}

In this section we follow the procedure of \cite{Gubser:2008wv} to compute the frequency-dependent conductivity of the ${\mathfrak {su}}(N)$ black holes by applying a time-dependent perturbation to the gauge field.
We follow \cite{Gubser:2008wv,Roberts:2008ns,Arias:2014msa,Herzog:2014tpa} by working in the probe limit, ignoring the back-reaction of the gauge field on the space-time metric.
Even in this limit, the perturbation equations for an ${\mathfrak {su}}(N)$ gauge field are rather complicated, and we anticipate that including the back-reaction would lead to a formidable set of equations to be solved (see, for example, \cite{Erdmenger:2012zu} for the full ${\mathfrak {su}}(2)$ perturbation equations including back-reaction, in the five-dimensional case).
For gauge group ${\mathfrak {su}}(2)$, gauge field perturbations in the probe limit have been used to compute the frequency-dependent conductivity in both the isotropic $\zeta =1$ case \cite{Roberts:2008ns} and the anisotropic $\zeta =0$ case \cite{Gubser:2008wv,Arias:2014msa,Herzog:2014tpa}.
The work in \cite{Roberts:2008ns} uses a different coordinate system to ours, so in this section we compute the conductivity in both the ${\mathfrak {su}}(2)$ and ${\mathfrak {su}}(3)$ cases and compare the results.

\subsection{Ansatz and field equations}
\label{sec:EMansatz}

We apply an oscillating perturbation with frequency $\xi $ to the non-Abelian gauge field.
We generalize the ${\mathfrak {su}}(2)$ ansatz of \cite{Gubser:2008wv} to ${\mathfrak {su}}(N)$ by taking
\begin{eqnarray}
-gA & = & \sum_{p=1}^{N-1}\left( h_p H_p + e^{-i \xi t}\delta u_p F_p + e^{-i \xi t}\delta v_p G_p\right)dt
    + \sum_{k=1}^{N-1}\left(\omega_k F_k + e^{-i\xi t}\delta h_{1,k}H_k\right)dx
    \nonumber \\ & &
   + \sum_{k=1}^{N-1}\left(\omega_k G_k + e^{-i\xi t}\delta h_{2,k} H_k\right)dy,
\label{eq:SccondA}
\end{eqnarray}
where $h_{p}$ and $\omega _{k}$ are, respectively, the background electric and magnetic gauge field functions and $\delta u_{p}$, $\delta v_{p}$, $\delta h_{1,k}$ and $\delta h_{2,k}$ are the perturbations.
The matrices $H_{p}$, $F_{k}$ and $G_{k}$ are defined by (\ref{eq:Hdef}, \ref{eq:FGdef}).
In the probe limit, the space-time is fixed to be the planar Schwarzschild-AdS black hole with metric (\ref{eq:SAdS}) and metric function $\mu _{S}$ (\ref{eq:muS}).
The background electric and magnetic gauge field functions satisfy the equilibrium YM equations (\ref{eq:hprimeprime}, \ref{eq:wprimeprime}) with $\sigma (r) \equiv 1$, $\mu (r)=\mu _{S}(r)$, $f(r)\equiv 1$ and $\zeta =1$.

In terms of new complex variables
\begin{equation}
U_k = \delta u_k + i\delta v_k, \qquad V_k = \delta u_k - i\delta v_k, \qquad
C_k = \delta h_{1, k} + i\delta h_{2,k}, \qquad D_k = \delta h_{1,k} - i \delta h_{2,k},
\label{eq:ABCDdef}
\end{equation}
the $4(N-1)$ Yang-Mills equations for the $4(N-1)$ perturbations can be written in the form
(see \cite{Shepherd} for a detailed derivation)
\begin{subequations}
\label{eq:condYM}
\begin{eqnarray}
U_k''  & = & -\frac{2}{r} U_k' + \frac{1}{\mu _{S} r^2}\left[\frac{\omega_{k+1}}{2}(U_k \omega_{k+1} - U_{k+1} \omega_k) + \frac{\omega_{k-1}}{2}(U_k \omega_{k-1} - U_{k-1} \omega_k) \right]
\nonumber\\
& & + \frac{\omega_k}{\mu _{S} r^2}\left(\sqrt{\frac{k-1}{2k}}h_{k-1} - \sqrt{\frac{k+1}{2k}}h_k\right)\left(\sqrt{\frac{k+1}{2k}}C_k - \sqrt{\frac{k-1}{2k}}C_{k-1}\right)
\nonumber\\
& & + \frac{(k+1)\omega_k}{2\mu _{S} r^2}\left(\frac{U_k \omega_k}{k} - \frac{U_{k+1} \omega_{k+1}}{k+1}\right) + \frac{(k-1)\omega_{k-1}}{2 \mu _{S} r^2}\left(\frac{U_k \omega_k}{k} - \frac{U_{k-1} \omega_{k-1}}{k-1}\right)
\nonumber\\
& & - \frac{\xi \omega_k}{\mu _{S} r^2} \left( \sqrt{\frac{k-1}{2k}}C_{k-1} - \sqrt{\frac{k+1}{2k}}C_k\right),
\label{eq:condAk}
\end{eqnarray}
\begin{eqnarray}
V_k''  & = & -\frac{2}{r} V_k' + \frac{1}{\mu _{S} r^2}\left[\frac{\omega_{k+1}}{2}(V_k \omega_{k+1} - V_{k+1} \omega_k) + \frac{\omega_{k-1}}{2}(V_k \omega_{k-1} - V_{k-1} \omega_k) \right]
\nonumber\\
& & + \frac{\omega_k}{\mu _{S} r^2}\left(\sqrt{\frac{k-1}{2k}}h_{k-1} - \sqrt{\frac{k+1}{2k}}h_k\right)\left(\sqrt{\frac{k+1}{2k}}D_k - \sqrt{\frac{k-1}{2k}}D_{k-1}\right)
\nonumber\\
& &
+ \frac{(k+1)\omega_k}{2\mu _{S} r^2}\left(\frac{V_k \omega_k}{k} - \frac{V_{k+1} \omega_{k+1}}{k+1}\right) + \frac{(k-1)\omega_{k-1}}{2 \mu _{S} r^2}\left(\frac{V_k \omega_k}{k} - \frac{V_{k-1} \omega_{k-1}}{k-1}\right)
\nonumber\\
& &
+ \frac{\xi \omega_k}{\mu _{S} r^2} \left( \sqrt{\frac{k-1}{2k}}D_{k-1} - \sqrt{\frac{k+1}{2k}}D_k\right),
\label{eq:condBk}
\end{eqnarray}
\begin{eqnarray}
0 & = & C_k'' + \frac{\mu _{S}'}{\mu _{S}}C_k' + \sqrt{\frac{k+1}{2k}}\frac{U_k \omega_k}{\mu _{S}^2}\left(\sqrt{\frac{k-1}{2k}}h_{k-1} - \sqrt{\frac{k+1}{2k}}h_k\right) \nonumber\\
& &
+ \sqrt{\frac{k}{2(k+1)}}\frac{U_{k+1} \omega_{k+1}}{\mu _{S}^2}\left(\sqrt{\frac{k+2}{2(k+1)}}h_{k+1} - \sqrt{\frac{k}{2(k+1)}}h_k\right)
\nonumber\\
&  &
+ \sqrt{\frac{k+1}{2k}}\frac{\omega_k^2}{\mu _{S} r^2}\left(\sqrt{\frac{k-1}{2k}}C_{k-1} - \sqrt{\frac{k+1}{2k}}C_k \right)
\nonumber\\ & &
+ \sqrt{\frac{k}{2(k+1)}}\frac{\omega_{k+1}^2}{\mu _{S} r^2}\left(\sqrt{\frac{k+2}{2(k+1)}}C_{k+1} - \sqrt{\frac{k}{2(k+1)}} C_k \right)
\nonumber\\
& &
+ \frac{\xi}{\mu _{S}^2}\left(\sqrt{\frac{k+1}{2k}}U_k \omega_k - \sqrt{\frac{k}{2(k+1)}}U_{k+1}\omega_{k+1} + \xi C_k\right),
\label{eq:condCk}
\end{eqnarray}
\begin{eqnarray}
0 & = &
D_k'' + \frac{\mu _{S}'}{\mu _{S}}D_k' + \sqrt{\frac{k+1}{2k}}\frac{V_k \omega_k}{\mu _{S}^2}\left(\sqrt{\frac{k-1}{2k}}h_{k-1} - \sqrt{\frac{k+1}{2k}}h_k\right)
\nonumber\\
  & & + \sqrt{\frac{k}{2(k+1)}}\frac{V_{k+1} \omega_{k+1}}{\mu _{S}^2}\left(\sqrt{\frac{k+2}{2(k+1)}}h_{k+1} - \sqrt{\frac{k}{2(k+1)}}h_k\right)
  \nonumber\\
  & & + \sqrt{\frac{k+1}{2k}}\frac{\omega_k^2}{\mu _{S} r^2}\left(\sqrt{\frac{k-1}{2k}}D_{k-1} - \sqrt{\frac{k+1}{2k}}D_k \right)
  \nonumber\\
  &  & + \sqrt{\frac{k}{2(k+1)}}\frac{\omega_{k+1}^2}{\mu _{S} r^2}\left(\sqrt{\frac{k+2}{2(k+1)}}D_{k+1} - \sqrt{\frac{k}{2(k+1)}} D_k \right)
  \nonumber\\
  & & - \frac{\xi}{\mu _{S}^2}\left(\sqrt{\frac{k+1}{2k}}V_k \omega_k - \sqrt{\frac{k}{2(k+1)}}V_{k+1}\omega_{k+1} - \xi D_k\right),
  \label{eq:condDk}
\end{eqnarray}
\end{subequations}
where $k=1,2,...,N-1$. We also have $2(N-2)$ zeroth order constraint equations, which are given by
\begin{subequations}
\label{eq:condconst0}
\begin{eqnarray}
0 & = & \frac{h_k}{\sqrt{2k(k+1)}}\left[ \left(1-k \right) U_k \omega_{k+1} -\left( k + 2\right) U_{k+1} \omega_k\right]
 + \sqrt{\frac{k+2}{2(k+1)}}h_{k+1}\left(2U_k \omega_{k+1} - U_{k+1} \omega_k\right)
\nonumber\\
  & &
  + \sqrt{\frac{k-1}{2k}}h_{k-1}\left( 2U_{k+1} \omega_k - U_k \omega_{k+1}\right) + \xi\left( U_k \omega_{k+1} - U_{k+1} \omega_k\right)
 \nonumber\\
  & &
  + \frac{\mu _{S} \omega_k \omega_{k+1}}{r^2}\left(\sqrt{\frac{1}{2k(k+1)}}\left(2k+1\right) C_k - \sqrt{\frac{k+2}{2(k+1)}} C_{k+1} - \sqrt{\frac{k-1}{2k}} C_{k-1}\right) ,
    \label{eq:condconst0a}
    \\
0 & = & \frac{h_k}{\sqrt{2k(k+1)}}\left[\left( 1-k \right) V_k \omega_{k+1} -\left( k + 2\right) V_{k+1} \omega_k\right]  + \sqrt{\frac{k+2}{2(k+1)}}h_{k+1}\left(2V_k \omega_{k+1} - V_{k+1} \omega_k\right)
\nonumber\\
  & &
  + \sqrt{\frac{k-1}{2k}}h_{k-1}\left( 2V_{k+1} \omega_k - V_k \omega_{k+1}\right) - \xi\left( V_k \omega_{k+1} - V_{k+1} \omega_k\right)
    \nonumber\\
  & &
  + \frac{\mu _{S} \omega_k \omega_{k+1}}{r^2}\left(\sqrt{\frac{1}{2k(k+1)}} \left( 2k+1\right) D_k - \sqrt{\frac{k+2}{2(k+1)}} D_{k+1}
 - \sqrt{\frac{k-1}{2k}} D_{k-1}\right) ,
  \label{eq:condconst0b}
\end{eqnarray}
\end{subequations}
where $k=1,2,...,N-2$  (the $k=N-1$ equations vanish since $\omega_N = U_N = 0$), and $2(N-1)$ first order constraint equations,
\begin{subequations}
\label{eq:condconst1}
\begin{eqnarray}
0 & = &
\xi U_{k}'+
\sqrt{\frac{k+1}{2k}}\left(h_k U_k' - U_k h_k'\right) + \sqrt{\frac{k-1}{2k}}\left(U_k h_{k-1}' - h_{k-1}U_k'\right)
\nonumber\\
  &\, & + \frac{\mu _{S}}{r^2}\left[\sqrt{\frac{k+1}{2k}}\left(\omega_k C_k' - C_k \omega_k'\right) + \sqrt{\frac{k-1}{2k}}\left( C_{k-1}\omega_k' - \omega_k C_{k-1}'\right)\right] ,
\label{eq:condconst1a}
\\
0 & = &
-\xi V_{k}'+
\sqrt{\frac{k+1}{2k}}\left(h_k V_k' - V_k h_k'\right) + \sqrt{\frac{k-1}{2k}}\left(V_k h_{k-1}' - h_{k-1}V_k'\right)
\nonumber\\
  &\, & + \frac{\mu _{S}}{r^2}\left[\sqrt{\frac{k+1}{2k}}\left(\omega_k D_k' - D_k \omega_k'\right) + \sqrt{\frac{k-1}{2k}}\left( D_{k-1}\omega_k' - \omega_k D_{k-1}'\right)\right] ,
\label{eq:condconst1b}
\end{eqnarray}
\end{subequations}
where $k=1,2,...,N-1$. If we differentiate the first order constraints (\ref{eq:condconst1}), we find that they propagate, in other words
if the equations (\ref{eq:condconst1}) are satisfied at one point in space, they will be satisfied everywhere as long as (\ref{eq:condYM}--\ref{eq:condconst0}) are satisfied everywhere. However, this is not the case for the zeroth order constraints, so equations (\ref{eq:condconst0}) must be implemented directly. This is achieved by using the zeroth order constraints to write $2(N-2)$ variables in terms of the other $2N$ variables, leaving $2N$ independent variables.

\subsection{Conductivity for ${\mathfrak {su}}(2)$ perturbations}
\label{sec:condsu2}

In the ${\mathfrak {su}}(2)$ case, the zeroth order constraints (\ref{eq:condconst0}) vanish since $U_{N}=V_{N}=\omega _{N}=0$, leaving just the four equations of motion (\ref{eq:condYM}) and two first order constraints (\ref{eq:condconst1}) for the four field variables, which we simply denote by $U$, $V$, $C$ and $D$.

We start by considering the variables $U$, $C$, whose equations of motion (\ref{eq:condAk}, \ref{eq:condCk}) simplify to
\begin{subequations}
\label{eq:condsu2YM}
\begin{eqnarray}
U'' & = &
-\frac{2}{r}U' + \frac{1}{\mu _{S}r^2}\left(U\omega^2 - h\omega C + \xi\omega C\right),
\label{eq:condsu2A}
\\
C'' & = &
-\frac{\mu _{S}'}{\mu _{S}}C' + \frac{U h \omega}{\mu _{S}^2} + \frac{\omega^2 C}{\mu _{S}r^2} - \frac{\xi}{\mu _{S}^2}\left(U\omega + \xi C\right), \label{eq:condsu2C}
\end{eqnarray}
\end{subequations}
and for which the first order constraint (\ref{eq:condconst1a}) reduces to
\begin{equation}
0= h U' - U h' + \frac{\mu_{S}}{r^2}\left(\omega C' - C\omega'\right) + \xi U' .
\label{eq:condsu2}
\end{equation}
Following \cite{Gubser:2008wv} we take the expansions of $U$ and $C$ near the horizon to be
\begin{eqnarray}
U & = & \left( r-r_{h} \right)^{i\xi\rho + \lambda_U} \left[ x^{(0)} + x^{(1)}\left( r-r_{h} \right) + x^{(2)}\left( r-r_{h}  \right)^2 + ...\right],
\nonumber \\
C & = & \left( r-r_{h} \right)^{i\xi\rho + \lambda_C} \left[ y^{(0)} + y^{(1)}\left( r-r_{h} \right) + y^{(2)}\left( r-r_{h} \right)^2 + ...\right],
\label{eq:condsu2hor}
\end{eqnarray}
where $\rho $, $\lambda _{U}$, $\lambda _{C}$ and all $x^{(a)}$ and $y^{(a)}$ are constants.
Substituting (\ref{eq:condsu2hor}) into (\ref{eq:condsu2A}), for a nontrivial solution we require either $x^{(0)}=0$ with $\lambda _{U}=\lambda _{C}$, or else $\lambda _{U}=\lambda _{C}+1$.  These two cases give equivalent leading order behaviour for the perturbation $U$, but for notational convenience we shall set $\lambda _{U}=\lambda _{C}$ with $x^{(0)}=0$.
Turning now to (\ref{eq:condsu2C}), for a nontrivial solution the following equation must hold
\begin{equation}
\xi^2 + \frac{9}{\ell ^4}\left(\lambda_C^2 + 2i\xi\rho\lambda_C - \xi^2\rho^2\right) = 0.
\end{equation}
We must therefore take $\lambda _{C}=0$ for solutions with real, nonzero $\xi $.
We then have
\begin{equation}
\rho = \pm \frac{\ell^2}{3} = \pm \frac{1}{4\pi T},
\label{eq:condbcsu2rho}
\end{equation}
where $T$ is the Hawking temperature (\ref{eq:THawking}).
Following \cite{Gubser:2008wv}, we consider the ingoing solution and take the negative root in (\ref{eq:condbcsu2rho}).

The equations (\ref{eq:condsu2YM}, \ref{eq:condsu2}) are linear in $U$ and $C$, so we may rescale to give $y^{(0)}=1$.
The first order constraint (\ref{eq:condsu2}) fixes the coefficient $x^{(1)}$, giving the expansions
\begin{subequations}
\label{eq:condsu2ic}
\begin{eqnarray}
U & = & \left( r-r_{h} \right)^{1 - \frac{i\xi \ell^2}{3}} \left[  \frac{i\omega(r_{h})}{1 - \frac{i\xi \ell^2}{3}} + \mathcal{O}\left(r-r_{h}\right)\right], \nonumber \\
C & = & \left( r-r_{h} \right)^{-\frac{i\xi \ell^2}{3}} \left[ 1 + \mathcal{O}\left(r-r_{h}\right)\right].
\label{eq:condsu2bcAC}
\end{eqnarray}
The equations of motion (\ref{eq:condBk}, \ref{eq:condDk}) and constraint (\ref{eq:condconst1b}) for $V$ and $D$ are the same as those for $U$ and $C$ with the replacement $\xi \to -\xi $.
We therefore find the following expansions near the horizon
\begin{eqnarray}
 V& = & \left( r-r_{h} \right)^{1 - \frac{i\xi \ell^2}{3}} \left[  \frac{i\omega(r_{h})}{\frac{i\xi \ell^2}{3}-1} + \mathcal{O}\left(r-r_{h}\right)\right], \nonumber \\
D & = & \left( r-r_{h} \right)^{-\frac{i\xi \ell^2}{3}} \left[ 1 + \mathcal{O}\left(r-r_{h}\right)\right].
\label{eq:condsu2bcBD}
\end{eqnarray}
\end{subequations}

We are interested in the conductivity when $\zeta =1$, the $\zeta =0$ case having been studied in \cite{Gubser:2008wv}.
The conductivity for $\zeta =1$ has been studied previously \cite{Roberts:2008ns}, but using different coordinates.
To find the conductivity with respect to electric fields applied in the $x$-direction, we consider the behaviour of the perturbation $\delta h_{1}$ at large $r$ (similarly, the perturbation $\delta h_{2}$ at large $r$ is considered for electric fields applied in the $y$-direction).
Since the conductivity is an observable quantity, it must be gauge-invariant.

In the ${\mathfrak {su}}(2)$ case, there is a set of gauge transformations which leave the matrix structure of the gauge potential (\ref{eq:SccondA}) invariant.
Under an infinitesimal gauge transformation (\ref{eq:gaugetrans}) with $W$ given by
\begin{equation}
W = e^{-i\xi t} \left[ W_{1}F_{1} + W_{2}G_{1} + W_{3}H_{1} \right] ,
\end{equation}
where $W_{k}$, $k=1,2,3$ are scalar functions and the elements $F_{1}$, $G_{1}$ and $H_{1}$ of the Lie algebra ${\mathfrak {su}}(2)$ are given by (\ref{eq:Hdef}, \ref{eq:FGdef}), the components of the gauge potential (\ref{eq:SccondA}) transform as follows
\begin{eqnarray}
A_t & \to & -e^{-i\xi t} \left( \delta u + \epsilon h W_2 + i\xi \epsilon W_1 \right) F_{1}
-e^{-i\xi t} \left( \delta v + i\xi\epsilon W_2 - h W_1 \right) G_{1}
\nonumber \\ & &
 - \left( h + i\xi\epsilon e^{-i\xi t} W_3 \right) H_{1} ,
 \nonumber \\
A_x & \to & \left( \epsilon e^{-i\xi t}\partial_x W_1 - \omega \right) F_{1} + \epsilon e^{-i\xi t}\left( \partial_x W_2 - \omega W_3 \right) G_{1}
- e^{-i\xi t}\left( \delta h_{1} - \epsilon \partial_x W_3 - \epsilon \omega W_2 \right) H_{1} ,
\nonumber \\
A_y & \to & \epsilon e^{-i\xi t}\left( \partial_y W_1 + \omega W_3 \right) F_{1} + \left( \epsilon e^{-i\xi t}\partial_y W_2 - \omega \right) G_{1}
- e^{-i\xi t}\left( \delta h_2 - \epsilon \partial_y W_3 - \epsilon\omega W_1 \right) H_{1} ,
\nonumber \\
A_r & \to & \epsilon e^{-i\xi t}\left( \partial_r W_1 F_{1} + \partial_r W_1 G_{1} + \partial_r W_3 H_{1}\right) ,
\label{eq:su2condgauge}
\end{eqnarray}
where we have kept only terms first order in either $\epsilon $ or the perturbations.
To retain the matrix structure of the gauge potential (\ref{eq:SccondA}), it must be the case that the coefficients of $G_{1}$ in $A_{x}$, and of $F_{1}$ in $A_{y}$ vanish, giving
\begin{equation}
\partial _{x}W_{2} - \omega W_{3} = 0 = \partial _{y}W_{1} + \omega W_{3},
\end{equation}
which are satisfied if $W$ is constant and $W_{3}=0$.
In this case $A_{r}=0$ as required, and the transformation (\ref{eq:su2condgauge}) is equivalent to
\begin{eqnarray}
\delta u & \to & \delta u + \epsilon\left( h W_2 + i\xi W_1\right),
\nonumber \\
\delta v  & \to & \delta v + \epsilon \left( i\xi W_2 - h W_1 \right) ,
\nonumber \\
\delta h_1 & \to & \delta h_1 - \epsilon\omega W_2,
\nonumber \\
\delta h_2 & \to & \delta h_2 - \epsilon \omega W_1.
\label{eq:su2condgaugesimp}
\end{eqnarray}
We therefore consider the following quantities
\begin{eqnarray}
\delta \hat {h}_1  & =  & \delta h_1 + \frac{\omega\left( i\xi\delta v + h\delta u \right)}{h^2 - \xi^2},
\nonumber \\
\delta \hat{h}_2  & =  & \delta h_2 + \frac{\omega\left( i\xi\delta u - h\delta v \right)}{h^2 - \xi^2},
\label{eq:Sccondsu2hhat}
\end{eqnarray}
which are invariant under (\ref{eq:su2condgaugesimp}).

The conductivity in the $x$-direction can be computed following \cite{Gubser:2008wv}, by expanding $\delta {\hat {h}}_{1}$ near the boundary at large $r$.
In particular, if we have
\begin{equation}
\delta {\hat {h}}_{1} = {\mathcal {H}}_{1}^{(0)} + \frac {{\mathcal {H}}_{1}^{(1)}}{r} + \ldots
\end{equation}
for large $r$, then the conductivity in the $x$-direction is given by
\begin{equation}
\sigma _{xx} = -\frac {i}{\xi \ell ^{2}} \frac {{\mathcal {H}}_{1}^{(1)}}{{\mathcal {H}}_{1}^{(0)}}.
\label{eq:sigmaxxsu2}
\end{equation}
Similarly, if for large $r$ we have the expansion
\begin{equation}
\delta {\hat {h}}_{2} = {\mathcal {H}}_{2}^{(0)} + \frac {{\mathcal {H}}_{2}^{(1)}}{r} + \ldots ,
\end{equation}
then the conductivity in the $y$-direction is
\begin{equation}
\sigma _{yy} = -\frac {i}{\xi \ell ^{2}} \frac {{\mathcal {H}}_{2}^{(1)}}{{\mathcal {H}}_{2}^{(0)}}.
\label{eq:sigmayysu2}
\end{equation}

We now solve equations (\ref{eq:condsu2YM}) (and the corresponding equations for $V$ and $D$) numerically.
To do this, we first solve the equilibrium equations (\ref{eq:staticfieldequations}) as described in section \ref{sec:sols}.
We use the same method, namely a Bulirsh-Stoer algorithm \cite{Press:1992zz} implemented in C++, to then solve the equations for $U$, $V$, $C$ and $D$
subject to the initial conditions (\ref{eq:condsu2ic}), integrating outwards from $r-r_{h}\sim 10^{-7}$.
The conductivities (\ref{eq:sigmaxxsu2}, \ref{eq:sigmayysu2}) are computed from $U$, $V$, $C$, $D$ and their derivatives at large $r$ using the results
\begin{eqnarray}
\mathcal{H}_1^{(0)}
& = & \lim_{r \to \infty} \left\{ \frac{1}{2}\left( C + D \right) + \frac{\omega}{2\left( h^2 - \xi^2 \right) }\left[\xi \left( U - V \right)
+ h\left( U + V \right) \right] \right\}  ,
\nonumber \\
\mathcal{H}_1^{(1)}
& = & \lim_{r \to \infty} \left\{ -\frac{r^2}{2}\left( C' + D' \right)
- \frac{\omega r^{2}}{2\left( h^2 - \xi^2\right) }\left[ \xi \left( U' - V' \right) + \left[ h \left( U + V \right) \right] '
\right]	
\right. \nonumber \\ & & \left.
- \frac{r^{2}\left[ \left(h^2 - \xi^2 \right)\omega' - 2\omega hh'\right] }{2 \left( h^2 - \xi^2 \right)^2}\left[
\xi\left( U - V \right) + h \left( U + V \right)\right]\right\}, 	
\nonumber \\
\mathcal{H}_2^{(0)}
& = & \lim_{r \to \infty} \left\{ \frac{i}{2}\left( D - C \right) + \frac{\omega}{2\left( h^2 - \xi^2 \right) }\left[i\xi\left( U + V \right)
+ ih\left( U - V\right )\right] \right\},
\nonumber \\
\mathcal{H}_2^{(1)}
& = & \lim_{r \to \infty} \left\{ -\frac{ir^2}{2}\left( D' - C' \right)
 - \frac{\omega r^{2}}{2\left(h^2 - \xi^2\right) }\left[ i\xi\left( U' + V' \right)
 + \left[ ih\left( U - V \right) \right] ' \right]
 \right. \nonumber \\ & & \left.
- \frac{r^{2}\left[ \left( h^2 - \xi^2 \right) \omega' - 2\omega hh'\right] }{2\left( h^2 - \xi^2\right) ^2}\left[
i\xi\left( U + V\right) + ih\left( U - V\right)  \right] \right\} .	
\label{eq:calHsu2}
\end{eqnarray}

\begin{figure}
\begin{center}
\includegraphics[width=8cm,angle=270]{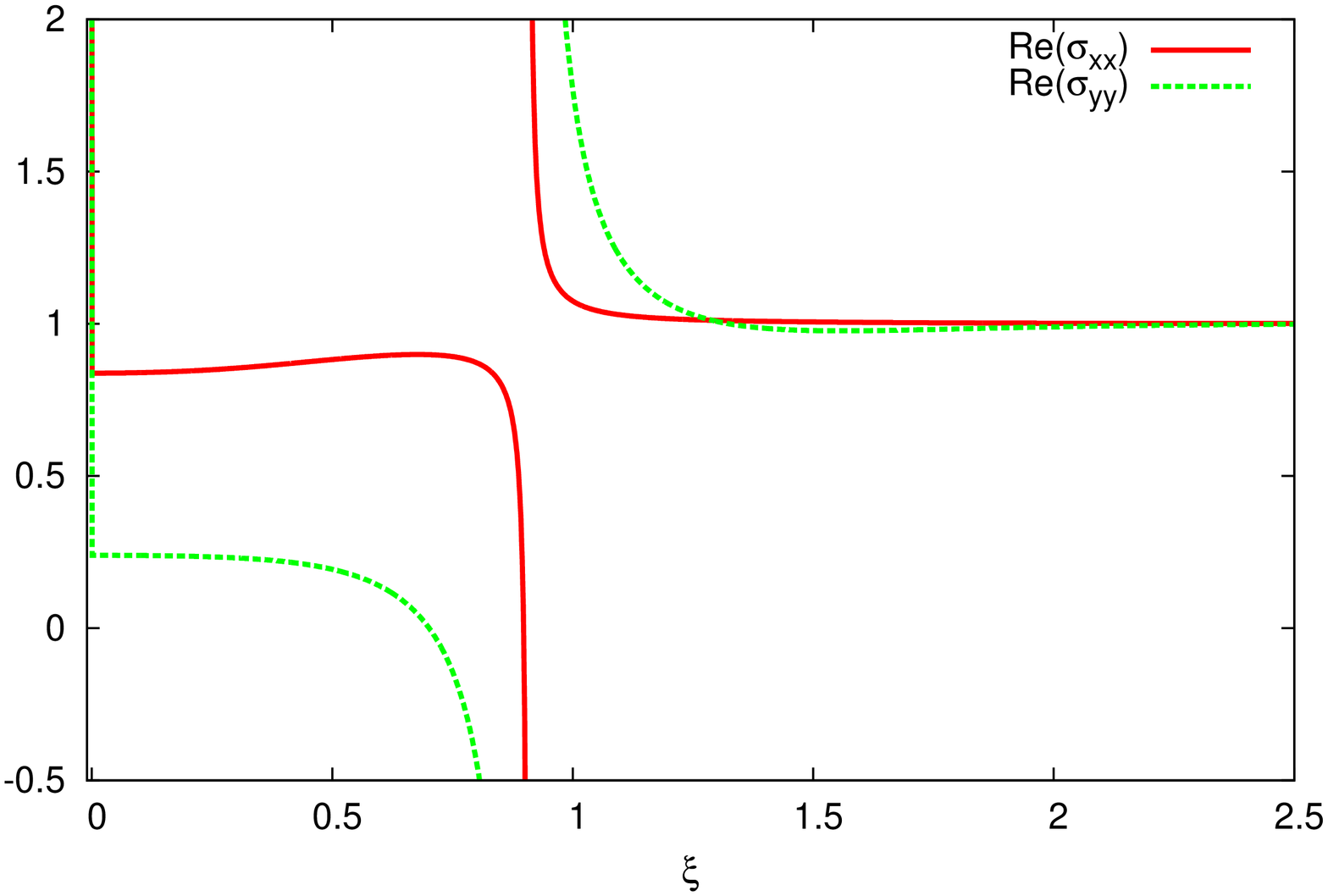}
\includegraphics[width=8cm,angle=270]{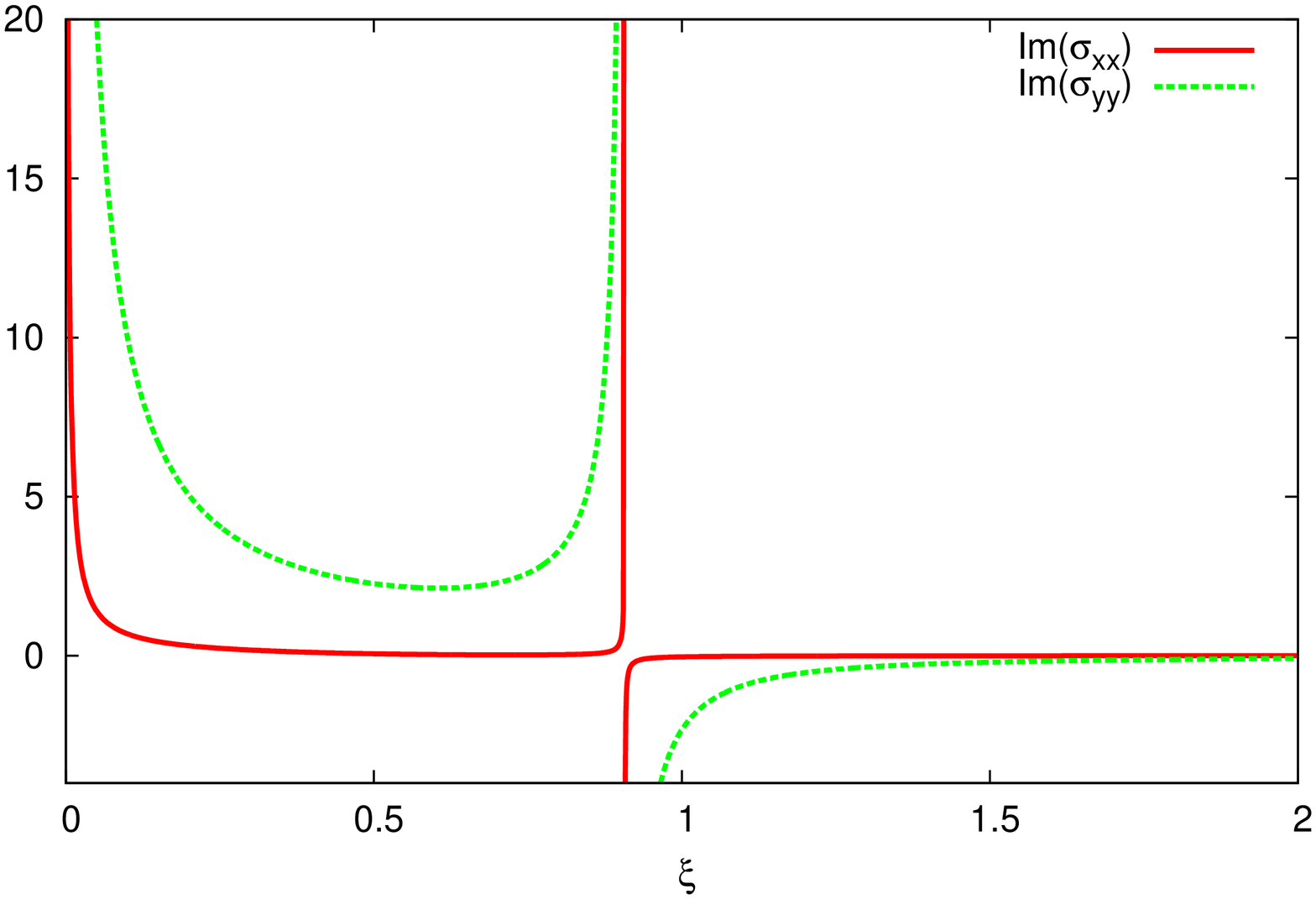}
\end{center}
\caption{Real part (top) and imaginary part (bottom) of the frequency-dependent conductivity for a planar ${\mathfrak {su}}(2)$ black hole with $\Lambda = -0.65$ and $\omega (r_{h})=0.1$.}
\label{fig:su2conductivity}
\end{figure}

In figure \ref{fig:su2conductivity} we show the real (top) and imaginary (bottom) parts of the conductivities $\sigma _{xx}$ (\ref{eq:sigmaxxsu2}) and $\sigma _{yy}$ (\ref{eq:sigmayysu2}) in the $x$- and $y$-directions respectively, for one particular planar ${\mathfrak {su}}(2)$ black hole. Qualitatively similar results are found for other ${\mathfrak {su}}(2)$ black holes.  As expected, there is a gap at low frequencies in both directions, that is, the real part of the low frequency conductivity is lower than the real part of the high frequency conductivity.
The gap is larger in $\sigma _{yy}$ than in $\sigma _{xx}$.
As in \cite{Gubser:2008wv}, we find a pole in the imaginary part of the conductivity as $\xi \rightarrow 0$, and hence a delta function at zero frequency in the real part of the conductivity, corresponding to infinite DC conductivity.
We also find a divergence in both the real and imaginary parts of the conductivity at nonzero frequency (similar behaviour is found in \cite{Roberts:2008ns}).  This effect arises due to the $\left( h^{2}-\xi ^{2}\right) ^{-1}$ term in ${\mathcal {H}}_{1}^{(0)}$ and the $\left( h^{2}-\xi ^{2}\right) ^{-2}$ term in ${\mathcal {H}}_{1}^{(1)}$ (\ref{eq:calHsu2}),  and occurs when $\xi = \lim _{r\rightarrow \infty }h(r)$, which is the chemical potential in the thermal CFT state.
The main purpose of this section is to compare the results in figure \ref{fig:su2conductivity} with the frequency-dependent conductivity in the ${\mathfrak {su}}(3)$ case, which is computed in the next subsection.

\subsection{Conductivity for ${\mathfrak {su}}(3)$ perturbations}
\label{sec:condsu3}

In the ${\mathfrak {su}}(3)$ case we have eight variables: $U_{k}$, $V_{k}$, $C_{k}$ and $D_{k}$ for $k=1,2$.
We begin by considering $U_{k}$ and $C_{k}$.
We have four equations of motion  (\ref{eq:condAk}, \ref{eq:condCk}), two first order constraints (\ref{eq:condconst1a}) and a single zeroth order constraint (\ref{eq:condconst0a}).
The near-horizon expansions for $U_{k}$ and $C_{k}$ are
\begin{eqnarray}
U_{k} & = & \left( r-r_{h} \right)^{i\xi\rho + \lambda_U} \left[ x_{k}^{(0)} + x_{k}^{(1)}\left( r-r_{h} \right) + x_{k}^{(2)}\left( r-r_{h}  \right)^2 + ...\right],
\nonumber \\
C_{k} & = & \left( r-r_{h} \right)^{i\xi\rho + \lambda_C} \left[ y_{k}^{(0)} + y_{k}^{(1)}\left( r-r_{h} \right) + y_{k}^{(2)}\left( r-r_{h} \right)^2 + ...\right] .
\label{eq:condsu3hor}
\end{eqnarray}
As in the ${\mathfrak {su}}(2)$ case, expanding the equations (\ref{eq:condAk}) for $k=1,2$ implies that either $\lambda _{U}=\lambda _{C}$ with $x_{1}^{(0)}=x_{2}^{(0)}=0$ or else $\lambda _{U}=\lambda _{C}+1$.
As before these are equivalent and we make the choice to set $\lambda _{U}=\lambda _{C}$.
Substituting (\ref{eq:condsu3hor}) into the second equation of motion (\ref{eq:condCk}) for $k=1,2$ gives, as in the ${\mathfrak {su}}(2)$ case,
$\lambda _{U}=\lambda _{C}=0$ and $\rho $ is given by (\ref{eq:condbcsu2rho}), where again we take the negative root so that we are considering ingoing solutions.
We use the fact that the equations (\ref{eq:condAk}, \ref{eq:condCk}, \ref{eq:condconst0a}, \ref{eq:condconst1a}) are linear in $U_{k}$ and $C_{k}$ to set
$y_{1}^{(0)}=1$ without loss of generality.
The two first order constraints (\ref{eq:condconst1a}) fix the constants $x_{1}^{(1)}$ and $x_{2}^{(1)}$, whilst the zeroth order constraint (\ref{eq:condconst0a}) gives $y_{2}^{(0)}$.
Altogether the expansions (\ref{eq:condsu3hor}) become
\begin{eqnarray}
U_1 & = & \left( r-r_{h} \right)^{1-\frac{i\xi \ell^2}{3}} \left[ \frac{i\omega_1(r_{h})}{1 - \frac{i\xi \ell^2}{3}}+ \mathcal{O}\left(r-r_{h}\right)\right],
\nonumber \\
U_2 & = & \left( r-r_{h} \right)^{1-\frac{i\xi \ell^2}{3}} \left[ \frac{i\omega_2(r_{h})}{1 - \frac{i\xi \ell^2}{3}}+ \mathcal{O}\left( r-r_{h} \right) \right],
\nonumber \\
C_1 & = & \left( r-r_{h} \right)^{-\frac{i\xi \ell^2}{3}} \left[ 1 + \mathcal{O}\left( r-r_{h} \right)\right],
\nonumber \\
C_2 & = & \left( r-r_{h} \right)^{-\frac{i\xi \ell^2}{3}} \left[ \sqrt{3} + \mathcal{O}\left(r-r_{h}\right)\right].
\label{eq:condsu3bcAC}
\end{eqnarray}
Following the same procedure for the $V_k$ and $D_k$ equations (\ref{eq:condBk}, \ref{eq:condDk}, \ref{eq:condconst0b}, \ref{eq:condconst1b}), we find
\begin{eqnarray}
V_1 & = & \left( r-r_{h} \right)^{1-\frac{i\xi \ell^2}{3}} \left[ \frac{i\omega_1(r_{h})}{\frac{i\xi \ell^2}{3}-1} + \mathcal{O}\left(r-r_{h}\right)\right],
\nonumber \\
V_2 & = & \left( r-r_{h} \right)^{1-\frac{i\xi \ell^2}{3}} \left[ \frac{i\omega_2(r_{h})}{\frac{i\xi \ell^2}{3}-1}+ \mathcal{O}\left( r-r_{h} \right) \right],
\nonumber \\
D_1 & = & \left( r-r_{h} \right)^{-\frac{i\xi \ell^2}{3}} \left[ 1 + \mathcal{O}\left( r-r_{h} \right)\right],
\nonumber \\
D_2 & = & \left( r-r_{h} \right)^{-\frac{i\xi \ell^2}{3}} \left[ \sqrt{3} + \mathcal{O}\left(r-r_{h}\right)\right].
\label{eq:condsu3bcBD}
\end{eqnarray}

In the ${\mathfrak {su}}(3)$ case, there are no residual gauge transformations which preserve the matrix structure of (\ref{eq:SccondA}), and hence to find the conductivities it is sufficient to consider the asymptotic values of the quantities $\delta h_{1,1}$, $\delta h_{1,2}$, $\delta h_{2,1}$ and $\delta h_{2,2}$.
However, the situation is more complicated than the ${\mathfrak {su}}(2)$ case because of the presence of two perturbations in both the $x$ and $y$ directions.

The conductivity is determined from \cite{Gubser:2008wv}
\begin{equation}
\left. {\mathcal {J}} \right| _{\text{bdy}} = i\xi \left(
\begin{array}{cccc}
\delta h_{1,1}^{*} & \delta h_{1,2}^{*} & \ldots & \delta v_{2}^{*}
\end{array}
\right)
{\mbox {\boldmath {$\sigma $}}}
\left( \begin{array}{c}
\delta h_{1,1} \\ \delta h_{1,2} \\ \vdots \\ \delta v_{2}
\end{array}
\right)
\label{eq:condsu3cond}
\end{equation}
where {\boldmath {$\sigma $}} is the conductivity matrix and $\left. {\mathcal {J}} \right| _{\text{bdy}}$ is the large $r$ limit of
\begin{eqnarray}
{\mathcal {J}} & = &
 r\left[ \delta u_{1}^{*}\partial _{r}\delta u_{1} + \delta u_{2}^{*} \partial _{r}\delta u_{2} + \delta v_{1}^{*}\partial _{r}\delta v_{1}
 + \delta v_{2}^{*}\partial _{r}\delta v_{2} \right]
 \nonumber \\  & &
- \mu _{S}(r) \left[ \delta h_{1,1}^{*}\partial _{r} \delta h_{1,1} + \delta h_{1,2}^{*} \partial _{r}\delta h_{1,2}
+ \delta h_{2,1}^{*} \partial _{r} \delta h_{2,1} + \delta h_{2,2}^{*} \partial _{r}\delta h_{2,2} \right] .
\end{eqnarray}
To find the conductivity $\sigma _{xx}$ in the $x$-direction, the relevant perturbations are $\delta h_{1,1}$ and $\delta h_{1,2}$.
We use the zeroth order constraints (\ref{eq:condconst0}) to write $\delta h_{1,2}$ in terms of $\delta h_{1,1}$ as follows:
\begin{equation}
\delta h_{1,2} = {\sqrt {3}} \delta h_{1,1} + \ldots ,
\end{equation}
where we have omitted terms involving $\delta u_{i}$ and $\delta v_{i}$ since they give off-diagonal terms in the conductivity matrix.
If the behaviour of $\delta h_{1,1}$ at large $r$ is given by
\begin{equation}
\delta h_{1,1} = {\mathcal {H}}_{1,1}^{(0)} + \frac {{\mathcal {H}}_{1,1}^{(1)}}{r} + \ldots ,
\end{equation}
then (\ref{eq:condsu3cond}) gives
\begin{equation}
\sigma _{xx} = -\frac {4i}{\xi \ell ^{2}} \frac {{\mathcal {H}}_{1,1}^{(1)}}{{\mathcal {H}}_{1,1}^{(0)}} .
\label{eq:sigmaxxsu3}
\end{equation}
Similarly we find
\begin{equation}
\sigma _{yy} = -\frac {4i}{\xi \ell ^{2}} \frac {{\mathcal {H}}_{2,1}^{(1)}}{{\mathcal {H}}_{2,1}^{(0)}},
\label{eq:sigmayysu3}
\end{equation}
where, for large $r$,
\begin{equation}
\delta h_{2,1} = {\mathcal {H}}_{2,1}^{(0)} + \frac {{\mathcal {H}}_{2,1}^{(1)}}{r} + \ldots .
\end{equation}

The first step in computing the conductivities (\ref{eq:sigmaxxsu3}, \ref{eq:sigmayysu3}) is to solve the field equations (\ref{eq:condYM}) subject to the constraints (\ref{eq:condconst0}, \ref{eq:condconst1}).  The four first order constraints (\ref{eq:condconst1}) are satisfied at the event horizon by our choice of initial conditions (\ref{eq:condsu3hor}) and are therefore satisfied everywhere since they propagate.
In fact, we can use (\ref{eq:condconst1}) as a check on the accuracy of our numerical integration.
We also need to implement the zeroth order constraints (\ref{eq:condconst0}), which we use to eliminate $U_{2}$ and $V_{2}$, writing them as follows:
\begin{eqnarray}
U_2  & = &
\frac{1}{\frac{3}{2}h_1 \omega_2 + \frac{\sqrt{3}}{2}h_2 \omega_1 + \xi \omega_1}\left\{U_1\omega_2\left(\sqrt{3}h_2 +\xi\omega_2\right) + \frac{\mu_{S}\omega_1\omega_2}{2r^2}\left(C_1 - \sqrt{3}C_2\right)\right\},
\nonumber \\
V_2 & = &
\frac{1}{\frac{3}{2}h_1 \omega_2 + \frac{\sqrt{3}}{2}h_2 \omega_1 - \xi \omega_1}\left\{V_1\omega_2\left(\sqrt{3}h_2 -\xi\omega_2\right) + \frac{\mu_{S}\omega_1\omega_2}{2r^2}\left(D_1 - \sqrt{3}D_2\right)\right\}.
\end{eqnarray}
The numerical method is the same as that implemented in the ${\mathfrak {su}}(2)$ case.  The conductivities (\ref{eq:sigmaxxsu3}, \ref{eq:sigmayysu3}) are
determined from $C_{1}$ and $D_{1}$ using (\ref{eq:ABCDdef}):
\begin{subequations}
\label{eq:sigmasu3}
\begin{eqnarray}
\sigma _{xx} & = & \lim _{r\rightarrow \infty } \frac {4ir^{2}}{\xi \ell ^{2}} \frac {C_{1}'+D_{1}'}{C_{1}+D_{1}} ,
\label{eq:sigmaxxsu3new}
\\
\sigma _{yy} & = & \lim _{r\rightarrow \infty } \frac {4r^{2}}{\xi \ell ^{2}} \frac {C_{1}'-D_{1}'}{C_{1}-D_{1}}.
\label{eq:sigmayysu3new}
\end{eqnarray}
\end{subequations}

\begin{figure}
\begin{center}
\includegraphics[width=8cm,angle=270]{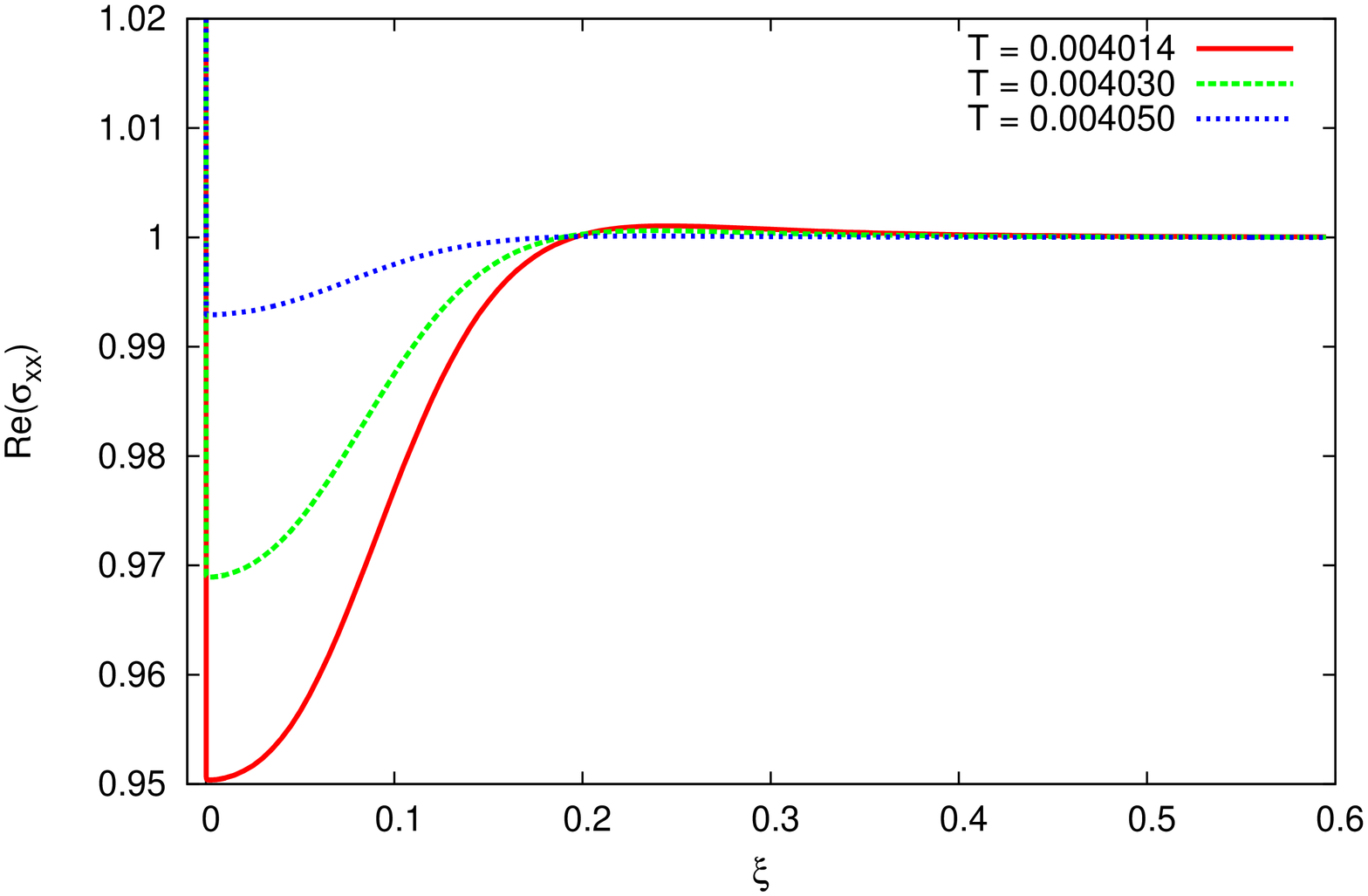}
\includegraphics[width=8cm,angle=270]{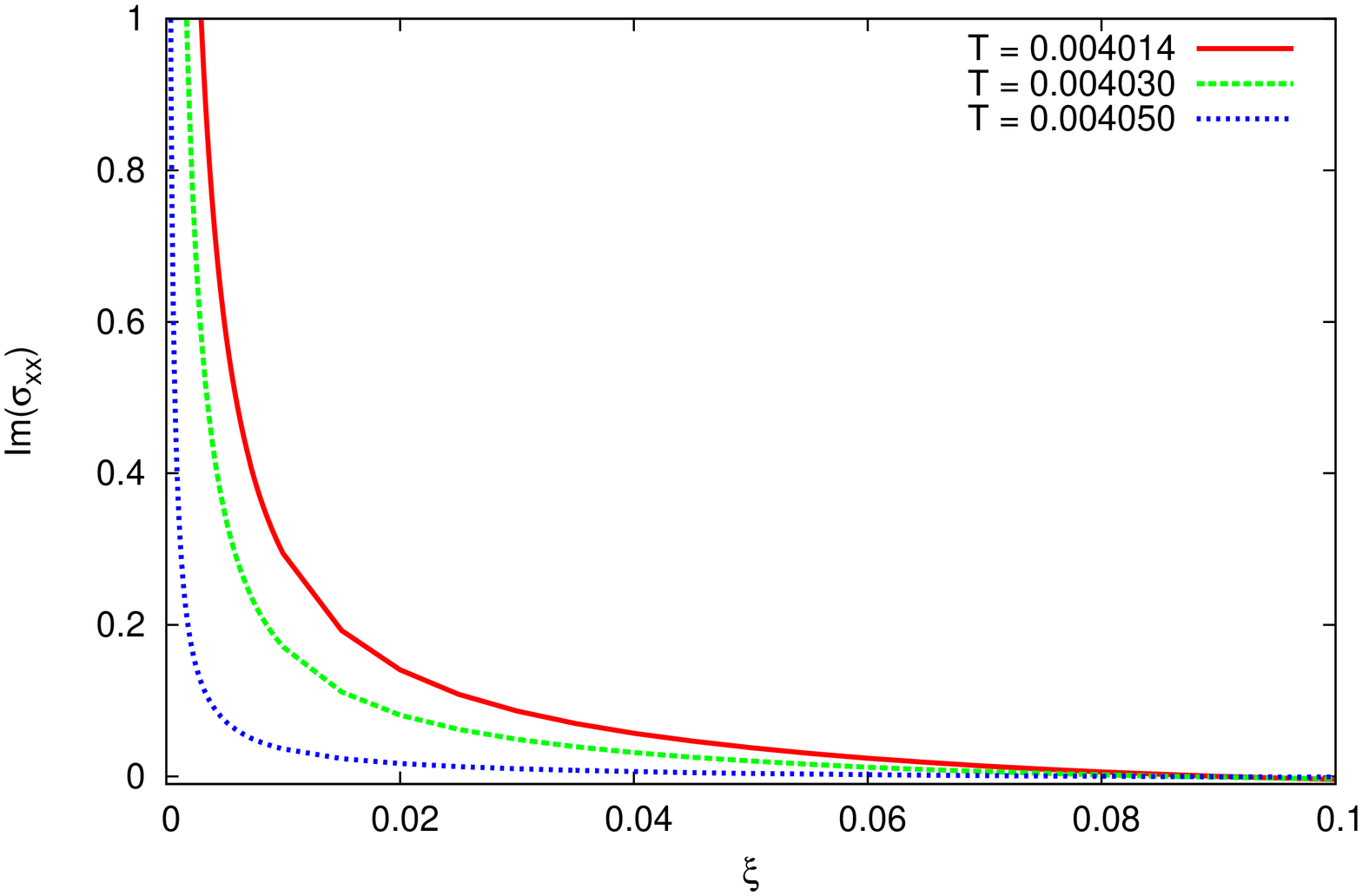}
\end{center}
\caption{Real part (top) and imaginary part (bottom) of the frequency-dependent conductivity for planar ${\mathfrak {su}}(3)$ black holes with $\ell ={\sqrt {-3/\Lambda }}=5$ and various values of the temperature.}
\label{fig:su3conductivity}
\end{figure}

In figure \ref{fig:su3conductivity} we plot the real (top) and imaginary (bottom) parts of the conductivity $\sigma _{xx}$ (\ref{eq:sigmaxxsu3new}) for a selection of ${\mathfrak {su}}(3)$ black holes.
Qualitatively similar results are found for other ${\mathfrak {su}}(3)$ black holes.
As in the ${\mathfrak {su}}(2)$ case, there is a gap in the real part of the conductivity at nonzero frequency, with higher conductivity at higher frequencies.  The gap increases as the temperature increases.
The real part of the conductivity is infinite in the zero frequency DC limit.
Unlike the ${\mathfrak {su}}(2)$ case shown in figure \ref{fig:su2conductivity}, there is no divergence in either the real or imaginary parts at nonzero frequency.
The imaginary part is large for small frequency $\xi $, and tends to zero at large $\xi $, as found in \cite{Gubser:2008wv} for the ${\mathfrak {su}}(2)$ case with $\zeta =0$.
We find qualitatively similar results for $\sigma _{yy}$ (\ref{eq:sigmayysu3new}), except that, as in the ${\mathfrak {su}}(2)$ case, the gap in the real part of the conductivity is larger for $\sigma _{yy}$ than it is for $\sigma _{xx}$.

\section{Conclusions}
\label{sec:conc}

In this paper we have studied dyonic planar hairy black hole solutions of four-dimensional ${\mathfrak {su}}(N)$ EYM theory in asymptotically AdS space-time.
We considered two possible ansatze for the ${\mathfrak {su}}(N)$ gauge field, generalizing the isotropic ${\mathfrak {su}}(2)$ $p+ip$-wave superconductor model of
\cite{Gubser:2008zu,Roberts:2008ns,Zeng:2010fs,Arias:2012py} and the anisotropic ${\mathfrak {su}}(2)$ $p$-wave superconductor model of
\cite{Gubser:2008wv,Zeng:2009dr,Basu:2009vv,Gubser:2010dm,Zeng:2010fs,Gangopadhyay:2012gx,Arias:2012py,Arias:2014msa,Herzog:2014tpa}.
When the gauge field has the generalized anisotropic form, we found that there are no genuinely ${\mathfrak {su}}(N)$ solutions, only embedded ${\mathfrak {su}}(2)$ solutions.
We therefore focussed our attention on the isotropic gauge field ansatz.

We examined the space of hairy black hole solutions for ${\mathfrak {su}}(2)$ and ${\mathfrak {su}}(3)$ gauge groups.
We then presented example solutions for which the magnetic part of the gauge potential vanishes on the AdS boundary and forms a condensate close to the planar event horizon.
Such black holes could be gravitational analogues of holographic superconductors, and we therefore explored some of the physical properties of our new solutions in this context.

First, we defined non-Abelian electric charges following \cite{Chrusciel:1987jr}, and hence a total electric charge.
Working in the canonical ensemble, we compared the free energy of a hairy black hole solution with a non-Abelian magnetic condensate with that of an embedded RN-AdS black hole having the same temperature and electric charge.
For all cases studied, the hairy black hole has lower free energy.
We also found that ${\mathfrak {su}}(3)$ hairy black holes have lower free energy than embedded ${\mathfrak {su}}(2)$ black holes with the same total electric charge.

It is anticipated that at a critical temperature $T_{C}$ there will be a phase transition between the embedded planar RN-AdS black holes (representing a normal phase with an unbroken Abelian gauge field symmetry) and the nontrivial hairy black hole (representing a superconducting phase in which the Abelian gauge field symmetry is broken).
At $T_{C}$, the RN-AdS black hole admits a static perturbation. In both the ${\mathfrak {su}}(2)$ and ${\mathfrak {su}}(3)$ cases, we found that above the critical temperature $T_{C}$ the only possible solution is the embedded planar RN-AdS black hole; below $T_{C}$ non-Abelian black holes exist.
At the critical temperature, it is therefore thermodynamically favourable for the RN-AdS black hole to decay into a non-Abelian hairy black hole with a nonzero condensate.

Working in the probe limit, with a fixed planar Schwarzschild-AdS space-time background, we also studied the frequency-dependent conductivity, by applying oscillating perturbations to the equilibrium probe gauge field.
The conductivities in the two directions in planes parallel to the horizon have very similar properties.
For both the ${\mathfrak {su}}(2)$ and ${\mathfrak {su}}(3)$ cases, the real part of the conductivity exhibits a gap, with smaller conductivity at low frequencies compared to high frequencies.
As the frequency of the perturbations tends to zero, there is a pole in the imaginary part of the conductivity and a delta function in the real part, corresponding to infinite DC conductivity, as expected in a superconductor.
For ${\mathfrak {su}}(2)$ black holes, there is an additional divergence in both the real and imaginary parts of the conductivity at a particular nonzero frequency (similar behaviour has been found previously \cite{Gubser:2008wv,Roberts:2008ns}).
This divergence disappears when we consider the larger ${\mathfrak {su}}(3)$ gauge group.
It would be interesting to investigate whether this behaviour persists when either an even larger gauge group is considered or the back-reaction of the gauge field on the space-time geometry is included.
Both generalizations of our work in this paper would yield highly complicated equations, so we leave this investigation for future research.

We have not considered the dynamical stability of our solutions.  Working in the probe limit, there are indications in \cite{Gubser:2008wv} for the ${\mathfrak {su}}(2)$ gauge group that the $p$-wave configurations with an
anisotropic gauge field ansatz are dynamically stable, but the $p+ip$-wave solutions (for which the gauge field has an isotropic ansatz) are unstable, and would likely decay to a $p$-wave configuration.
To explore this issue in more detail, it would be necessary to consider, in the fully coupled case, linearized perturbations of the metric and ${\mathfrak {su}}(N)$ gauge field.
The inclusion of a nontrivial electric part in the gauge field makes the analysis of the perturbation equations just in the ${\mathfrak {su}}(2)$ case more challenging than for purely magnetic gauge field configurations \cite{Nolan:2015vca}.
We therefore expect that the analysis of the perturbation equations for the dyonic ${\mathfrak {su}}(N)$ black holes discussed in this paper will be highly involved, and we leave this for future work.

\acknowledgments
We thank Tigran Tchrakian and Christopher Herzog for helpful discussions.
The work of B.L.S. is supported by UK EPSRC.
The work of E.W. is supported by the Lancaster-Manchester-Sheffield Consortium for Fundamental Physics under STFC grant ST/L000520/1.

\bibliographystyle{jhep}

\bibliography{suN}

\providecommand{\href}[2]{#2}\begingroup\raggedright\begin{thebibliography}{100}

\bibitem{Volkov:1998cc}
M.~S. Volkov and D.~V. Gal'tsov, {\it {Gravitating non-{A}belian solitons and
  black holes with {Y}ang-{M}ills fields}},  {\em Phys. Rept.} {\bf 319} (1999)
  1--83, [\href{http://arxiv.org/abs/hep-th/9810070}{{\tt hep-th/9810070}}].

\bibitem{Winstanley:2008ac}
E.~Winstanley, {\it {Classical {Y}ang-{M}ills black hole hair in anti-de
  {S}itter space}},  {\em Lect. Notes Phys.} {\bf 769} (2009) 49--87,
  [\href{http://arxiv.org/abs/0801.0527}{{\tt arXiv:0801.0527}}].

\bibitem{Winstanley:2015loa}
E.~Winstanley, {\it {A menagerie of hairy black holes}},
\newblock \href{http://arxiv.org/abs/1510.01669}{{\tt arXiv:1510.01669}}.

\bibitem{Volkov:2016ehx}
M.~S. Volkov, {\it {Hairy black holes in the XX-th and XXI-st centuries}},
  \href{http://arxiv.org/abs/1601.08230}{{\tt arXiv:1601.08230}}.

\bibitem{Galtsov:1989ip}
D.~V. Galtsov and A.~A. Ershov, {\it {Nonabelian baldness of colored black
  holes}},  {\em Phys. Lett.} {\bf A138} (1989) 160--164.

\bibitem{Bizon:1992pi}
P.~Bizon and O.~T. Popp, {\it {No hair theorem for spherical monopoles and
  dyons in SU(2) {E}instein {Y}ang-{M}ills theory}},  {\em Class. Quant. Grav.}
  {\bf 9} (1992) 193--205.

\bibitem{Volkov:1989fi}
M.~S. Volkov and D.~V. Galtsov, {\it {Non-{A}belian {E}instein {Y}ang-{M}ills
  black holes}},  {\em JETP Lett.} {\bf 50} (1989) 346--350.

\bibitem{Volkov:1990sva}
M.~S. Volkov and D.~V. Galtsov, {\it {Black holes in {E}instein {Y}ang-{M}ills
  theory}},  {\em Sov. J. Nucl. Phys.} {\bf 51} (1990) 747--753.

\bibitem{Bizon:1990sr}
P.~Bizon, {\it {Colored black holes}},  {\em Phys. Rev. Lett.} {\bf 64} (1990)
  2844--2847.

\bibitem{Kuenzle:1990is}
H.~P. Kuenzle and A.~K.~M. Masood-ul Alam, {\it {Spherically symmetric static
  SU(2) {E}instein {Y}ang-{M}ills fields}},  {\em J. Math. Phys.} {\bf 31}
  (1990) 928--935.

\bibitem{Galtsov:1991au}
D.~V. Galtsov and M.~S. Volkov, {\it {Charged non-{A}belian SU(3) {E}instein
  {Y}ang-{M}ills black holes}},  {\em Phys. Lett.} {\bf B274} (1992) 173--178.

\bibitem{Kleihaus:1995tk}
B.~Kleihaus, J.~Kunz, and A.~Sood, {\it {SU(3) {E}instein {Y}ang-{M}ills
  sphalerons and black holes}},  {\em Phys. Lett.} {\bf B354} (1995) 240--246,
  [\href{http://arxiv.org/abs/hep-th/9504053}{{\tt hep-th/9504053}}].

\bibitem{Kleihaus:1997rb}
B.~Kleihaus, J.~Kunz, and A.~Sood, {\it {Charged SU(N) {E}instein
  {Y}ang-{M}ills black holes}},  {\em Phys. Lett.} {\bf B418} (1998) 284--293,
  [\href{http://arxiv.org/abs/hep-th/9705179}{{\tt hep-th/9705179}}].

\bibitem{Kleihaus:1998qc}
B.~Kleihaus, J.~Kunz, A.~Sood, and M.~Wirschins, {\it {Sequences of globally
  regular and black hole solutions in SU(4) {E}instein {Y}ang-{M}ills theory}},
   {\em Phys. Rev.} {\bf D58} (1998) 084006,
  [\href{http://arxiv.org/abs/hep-th/9802143}{{\tt hep-th/9802143}}].

\bibitem{Mavromatos:1997zb}
N.~E. Mavromatos and E.~Winstanley, {\it {Existence theorems for hairy black
  holes in SU(N) {E}instein {Y}ang-{M}ills theories}},  {\em J. Math. Phys.}
  {\bf 39} (1998) 4849--4873, [\href{http://arxiv.org/abs/gr-qc/9712049}{{\tt
  gr-qc/9712049}}].

\bibitem{Ruan:2001dw}
W.~H. Ruan, {\it {Hairy black hole solutions to SU(3) {E}instein-{Y}ang-{M}ills
  equations}},  {\em Commun. Math. Phys.} {\bf 224} (2001) 373--397.

\bibitem{Straumann:1990as}
N.~Straumann and Z.~H. Zhou, {\it {Instability of a colored black hole
  solution}},  {\em Phys. Lett.} {\bf B243} (1990) 33--35.

\bibitem{Galtsov:1991nk}
D.~V. Galtsov and M.~S. Volkov, {\it {Instability of {E}instein {Y}ang-{M}ills
  black holes}},  {\em Phys. Lett.} {\bf A162} (1992) 144--148.

\bibitem{Volkov:1994dq}
M.~S. Volkov and D.~V. Galtsov, {\it {Odd parity negative modes of {E}instein
  {Y}ang-{M}ills black holes and sphalerons}},  {\em Phys. Lett.} {\bf B341}
  (1995) 279--285, [\href{http://arxiv.org/abs/hep-th/9409041}{{\tt
  hep-th/9409041}}].

\bibitem{Hod:2008ir}
S.~Hod, {\it {Lifetime of unstable hairy black holes}},  {\em Phys. Lett.} {\bf
  B661} (2008) 175--178, [\href{http://arxiv.org/abs/0803.0608}{{\tt
  arXiv:0803.0608}}].

\bibitem{Lavrelashvili:1994rp}
G.~V. Lavrelashvili and D.~Maison, {\it {A remark on the instability of the
  {B}artnik-{M}c{K}innon solutions}},  {\em Phys. Lett.} {\bf B343} (1995)
  214--217, [\href{http://arxiv.org/abs/hep-th/9409185}{{\tt hep-th/9409185}}].

\bibitem{Volkov:1995np}
M.~S. Volkov, O.~Brodbeck, G.~V. Lavrelashvili, and N.~Straumann, {\it {The
  number of sphaleron instabilities of the {B}artnik-{M}c{K}innon solitons and
  non-{A}belian black holes}},  {\em Phys. Lett.} {\bf B349} (1995) 438--442,
  [\href{http://arxiv.org/abs/hep-th/9502045}{{\tt hep-th/9502045}}].

\bibitem{Brodbeck:1994vu}
O.~Brodbeck and N.~Straumann, {\it {Instability proof for {E}instein
  {Y}ang-{M}ills solitons and black holes with arbitrary gauge groups}},  {\em
  J. Math. Phys.} {\bf 37} (1996) 1414--1433,
  [\href{http://arxiv.org/abs/gr-qc/9411058}{{\tt gr-qc/9411058}}].

\bibitem{Maldacena:1997re}
J.~M. Maldacena, {\it {The large N limit of superconformal field theories and
  supergravity}},  {\em Adv. Theor. Math. Phys.} {\bf 2} (1998) 231--252,
  [\href{http://arxiv.org/abs/hep-th/9711200}{{\tt hep-th/9711200}}].

\bibitem{Witten:1998qj}
E.~Witten, {\it {Anti-de {S}itter space and holography}},  {\em Adv. Theor.
  Math. Phys.} {\bf 2} (1998) 253--291,
  [\href{http://arxiv.org/abs/hep-th/9802150}{{\tt hep-th/9802150}}].

\bibitem{Gubser:1998bc}
S.~S. Gubser, I.~R. Klebanov, and A.~M. Polyakov, {\it {Gauge theory
  correlators from noncritical string theory}},  {\em Phys. Lett.} {\bf B428}
  (1998) 105--114, [\href{http://arxiv.org/abs/hep-th/9802109}{{\tt
  hep-th/9802109}}].

\bibitem{Aharony:1999ti}
O.~Aharony, S.~S. Gubser, J.~M. Maldacena, H.~Ooguri, and Y.~Oz, {\it {Large N
  field theories, string theory and gravity}},  {\em Phys. Rept.} {\bf 323}
  (2000) 183--386, [\href{http://arxiv.org/abs/hep-th/9905111}{{\tt
  hep-th/9905111}}].

\bibitem{Winstanley:1998sn}
E.~Winstanley, {\it {Existence of stable hairy black holes in SU(2) {E}instein
  {Y}ang-{M}ills theory with a negative cosmological constant}},  {\em Class.
  Quant. Grav.} {\bf 16} (1999) 1963--1978,
  [\href{http://arxiv.org/abs/gr-qc/9812064}{{\tt gr-qc/9812064}}].

\bibitem{Bjoraker:1999yd}
J.~Bjoraker and Y.~Hosotani, {\it {Stable monopole and dyon solutions in the
  {E}instein-{Y}ang-{M}ills theory in asymptotically Anti-de {S}itter space}},
  {\em Phys. Rev. Lett.} {\bf 84} (2000) 1853--1856,
  [\href{http://arxiv.org/abs/gr-qc/9906091}{{\tt gr-qc/9906091}}].

\bibitem{Bjoraker:2000qd}
J.~Bjoraker and Y.~Hosotani, {\it {Monopoles, dyons and black holes in the
  four-dimensional {E}instein-{Y}ang-{M}ills theory}},  {\em Phys. Rev.} {\bf
  D62} (2000) 043513, [\href{http://arxiv.org/abs/hep-th/0002098}{{\tt
  hep-th/0002098}}].

\bibitem{Baxter:2007at}
J.~E. Baxter, M.~Helbling, and E.~Winstanley, {\it {Abundant stable gauge field
  hair for black holes in anti-de {S}itter space}},  {\em Phys. Rev. Lett.}
  {\bf 100} (2008) 011301, [\href{http://arxiv.org/abs/0708.2356}{{\tt
  arXiv:0708.2356}}].

\bibitem{Baxter:2007au}
J.~E. Baxter, M.~Helbling, and E.~Winstanley, {\it {Soliton and black hole
  solutions of su(N) {E}instein-{Y}ang-{M}ills theory in anti-de {S}itter
  space}},  {\em Phys. Rev.} {\bf D76} (2007) 104017,
  [\href{http://arxiv.org/abs/0708.2357}{{\tt arXiv:0708.2357}}].

\bibitem{Baxter:2008pi}
J.~E. Baxter and E.~Winstanley, {\it {On the existence of soliton and hairy
  black hole solutions of su(N) {E}instein-{Y}ang-{M}ills theory with a
  negative cosmological constant}},  {\em Class. Quant. Grav.} {\bf 25} (2008)
  245014, [\href{http://arxiv.org/abs/0808.2977}{{\tt arXiv:0808.2977}}].

\bibitem{Baxter:2015gfa}
J.~E. Baxter and E.~Winstanley, {\it {On the stability of soliton and hairy
  black hole solutions of ${\mathfrak {su}}(N)$ {E}instein-{Y}ang-{M}ills
  theory with a negative cosmological constant}},  {\em J. Math. Phys.} {\bf
  57} (2016) 022506, [\href{http://arxiv.org/abs/1501.07541}{{\tt
  arXiv:1501.07541}}].

\bibitem{Sarbach:2001mc}
O.~Sarbach and E.~Winstanley, {\it {On the linear stability of solitons and
  hairy black holes with a negative cosmological constant: the odd parity
  sector}},  {\em Class. Quant. Grav.} {\bf 18} (2001) 2125--2146,
  [\href{http://arxiv.org/abs/gr-qc/0102033}{{\tt gr-qc/0102033}}].

\bibitem{Winstanley:2001bs}
E.~Winstanley and O.~Sarbach, {\it {On the linear stability of solitons and
  hairy black holes with a negative cosmological constant: the even parity
  sector}},  {\em Class. Quant. Grav.} {\bf 19} (2002) 689--724,
  [\href{http://arxiv.org/abs/gr-qc/0111039}{{\tt gr-qc/0111039}}].

\bibitem{Mann:2006jc}
R.~B. Mann, E.~Radu, and D.~H. Tchrakian, {\it {Non-Abelian solutions in
  {A}d{S}(4) and d=11 supergravity}},  {\em Phys. Rev.} {\bf D74} (2006)
  064015, [\href{http://arxiv.org/abs/hep-th/0606004}{{\tt hep-th/0606004}}].

\bibitem{Shepherd:2012sz}
B.~L. Shepherd and E.~Winstanley, {\it {Characterizing asymptotically anti-de
  {S}itter black holes with abundant stable gauge field hair}},  {\em Class.
  Quant. Grav.} {\bf 29} (2012) 155004,
  [\href{http://arxiv.org/abs/1202.1438}{{\tt arXiv:1202.1438}}].

\bibitem{Fan:2014ixa}
Z.-Y. Fan and H.~Lu, {\it {SU(2)-colored (A)d{S} black holes in conformal
  gravity}},  {\em JHEP} {\bf 02} (2015) 013,
  [\href{http://arxiv.org/abs/1411.5372}{{\tt arXiv:1411.5372}}].

\bibitem{Kichakova:2015lza}
O.~Kichakova, J.~Kunz, E.~Radu, and Y.~Shnir, {\it {Thermodynamic properties of
  asymptotically anti-de {S}itter black holes in $d = 4$
  {E}instein-{Y}ang-{M}ills theory}},  {\em Phys. Lett.} {\bf B747} (2015)
  205--211, [\href{http://arxiv.org/abs/1503.01268}{{\tt arXiv:1503.01268}}].

\bibitem{Nolan:2012ax}
B.~C. Nolan and E.~Winstanley, {\it {On the existence of dyons and dyonic black
  holes in {E}instein-{Y}ang-{M}ills theory}},  {\em Class. Quant. Grav.} {\bf
  29} (2012) 235024, [\href{http://arxiv.org/abs/1208.3589}{{\tt
  arXiv:1208.3589}}].

\bibitem{Shepherd:2015dse}
B.~L. Shepherd and E.~Winstanley, {\it {Dyons and dyonic black holes in
  ${\mathfrak {su}}(N)$ Einstein-Yang-Mills theory in anti-de Sitter
  spacetime}},  {\em Phys. Rev.} {\bf D93} (2016) 064064,
  [\href{http://arxiv.org/abs/1512.03010}{{\tt arXiv:1512.03010}}].

\bibitem{Nolan:2015vca}
B.~C. Nolan and E.~Winstanley, {\it {On the stability of dyons and dyonic black
  holes in {E}instein-{Y}ang-{M}ills theory}},  {\em Class. Quant. Grav.} {\bf
  33} (2016) 045003, [\href{http://arxiv.org/abs/1507.08915}{{\tt
  arXiv:1507.08915}}].

\bibitem{Baxter:2015tda}
J.~E. Baxter, {\it {Existence of topological hairy dyons and dyonic black holes
  in anti-de {S}itter ${\mathfrak {su}}(N)$ {E}instein-{Y}ang-{M}ills theory}},
   {\em J. Math. Phys.} {\bf 57} (2016) 022505,
  [\href{http://arxiv.org/abs/1507.05314}{{\tt arXiv:1507.05314}}].

\bibitem{Volkov:2006xt}
M.~S. Volkov, {\it {Gravitating non-{A}belian solitons and hairy black holes in
  higher dimensions}},  in {\em {Recent developments in theoretical and
  experimental general relativity, gravitation and relativistic field theories.
  Proceedings, 11th Marcel Grossmann Meeting, MG11, Berlin, Germany, July
  23-29, 2006.}}, pp.~1379--1396, 2006.
\newblock \href{http://arxiv.org/abs/hep-th/0612219}{{\tt hep-th/0612219}}.

\bibitem{Birmingham:1998nr}
D.~Birmingham, {\it {Topological black holes in {A}nti-de {S}itter space}},
  {\em Class. Quant. Grav.} {\bf 16} (1999) 1197--1205,
  [\href{http://arxiv.org/abs/hep-th/9808032}{{\tt hep-th/9808032}}].

\bibitem{Brill:1997mf}
D.~R. Brill, J.~Louko, and P.~Peldan, {\it {Thermodynamics of (3+1)-dimensional
  black holes with toroidal or higher genus horizons}},  {\em Phys. Rev.} {\bf
  D56} (1997) 3600--3610, [\href{http://arxiv.org/abs/gr-qc/9705012}{{\tt
  gr-qc/9705012}}].

\bibitem{Lemos:1994fn}
J.~P.~S. Lemos, {\it {Two-dimensional black holes and planar general
  relativity}},  {\em Class. Quant. Grav.} {\bf 12} (1995) 1081--1086,
  [\href{http://arxiv.org/abs/gr-qc/9407024}{{\tt gr-qc/9407024}}].

\bibitem{Lemos:1994xp}
J.~P.~S. Lemos, {\it {Cylindrical black hole in general relativity}},  {\em
  Phys. Lett.} {\bf B353} (1995) 46--51,
  [\href{http://arxiv.org/abs/gr-qc/9404041}{{\tt gr-qc/9404041}}].

\bibitem{Lemos:1995cm}
J.~P.~S. Lemos and V.~T. Zanchin, {\it {Rotating charged black string and
  three-dimensional black holes}},  {\em Phys. Rev.} {\bf D54} (1996)
  3840--3853, [\href{http://arxiv.org/abs/hep-th/9511188}{{\tt
  hep-th/9511188}}].

\bibitem{Vanzo:1997gw}
L.~Vanzo, {\it {Black holes with unusual topology}},  {\em Phys. Rev.} {\bf
  D56} (1997) 6475--6483, [\href{http://arxiv.org/abs/gr-qc/9705004}{{\tt
  gr-qc/9705004}}].

\bibitem{Cai:1996eg}
R.-G. Cai and Y.-Z. Zhang, {\it {Black plane solutions in four-dimensional
  space-times}},  {\em Phys. Rev.} {\bf D54} (1996) 4891--4898,
  [\href{http://arxiv.org/abs/gr-qc/9609065}{{\tt gr-qc/9609065}}].

\bibitem{Mann:1996gj}
R.~B. Mann, {\it {Pair production of topological anti-de Sitter black holes}},
  {\em Class. Quant. Grav.} {\bf 14} (1997) L109--L114,
  [\href{http://arxiv.org/abs/gr-qc/9607071}{{\tt gr-qc/9607071}}].

\bibitem{Smith:1997wx}
W.~L. Smith and R.~B. Mann, {\it {Formation of topological black holes from
  gravitational collapse}},  {\em Phys. Rev.} {\bf D56} (1997) 4942--4947,
  [\href{http://arxiv.org/abs/gr-qc/9703007}{{\tt gr-qc/9703007}}].

\bibitem{Mann:1997zn}
R.~B. Mann, {\it {Charged topological black hole pair creation}},  {\em Nucl.
  Phys.} {\bf B516} (1998) 357--381,
  [\href{http://arxiv.org/abs/hep-th/9705223}{{\tt hep-th/9705223}}].

\bibitem{VanderBij:2001ia}
J.~J. Van~der Bij and E.~Radu, {\it {New hairy black holes with negative
  cosmological constant}},  {\em Phys. Lett.} {\bf B536} (2002) 107--113,
  [\href{http://arxiv.org/abs/gr-qc/0107065}{{\tt gr-qc/0107065}}].

\bibitem{Baxter:2014nka}
J.~E. Baxter, {\it {On the existence of topological hairy black holes in
  $\mathfrak {su}(N)$ EYM theory with a negative cosmological constant}},  {\em
  Gen. Rel. Grav.} {\bf 47} (2015) 1829,
  [\href{http://arxiv.org/abs/1403.0171}{{\tt arXiv:1403.0171}}].

\bibitem{Baxter:2015ffm}
J.~E. Baxter and E.~Winstanley, {\it {Topological black holes in ${\mathfrak
  {su}}(N)$ {E}instein-{Y}ang-{M}ills theory with a negative cosmological
  constant}},  {\em Phys. Lett.} {\bf B753} (2016) 268--273,
  [\href{http://arxiv.org/abs/1511.04955}{{\tt arXiv:1511.04955}}].

\bibitem{Baxter:2015xfa}
J.~E. Baxter, {\it {Stable topological hairy black holes in $\mathfrak{su}(N)$
  EYM theory with $\Lambda<0$}},  \href{http://arxiv.org/abs/1507.03127}{{\tt
  arXiv:1507.03127}}.

\bibitem{Hartnoll:2009sz}
S.~A. Hartnoll, {\it {Lectures on holographic methods for condensed matter
  physics}},  {\em Class. Quant. Grav.} {\bf 26} (2009) 224002,
  [\href{http://arxiv.org/abs/0903.3246}{{\tt arXiv:0903.3246}}].

\bibitem{Herzog:2009xv}
C.~P. Herzog, {\it {Lectures on holographic superfluidity and
  superconductivity}},  {\em J. Phys.} {\bf A42} (2009) 343001,
  [\href{http://arxiv.org/abs/0904.1975}{{\tt arXiv:0904.1975}}].

\bibitem{Horowitz:2010gk}
G.~T. Horowitz, {\it {Introduction to holographic superconductors}},  {\em
  Lect. Notes Phys.} {\bf 828} (2011) 313--347,
  [\href{http://arxiv.org/abs/1002.1722}{{\tt arXiv:1002.1722}}].

\bibitem{Horowitz:2010nh}
G.~T. Horowitz, {\it {Surprising connections between general relativity and
  condensed matter}},  {\em Class. Quant. Grav.} {\bf 28} (2011) 114008,
  [\href{http://arxiv.org/abs/1010.2784}{{\tt arXiv:1010.2784}}].

\bibitem{Kaminski:2010zu}
M.~Kaminski, {\it {Flavor superconductivity and superfluidity}},  {\em Lect.
  Notes Phys.} {\bf 828} (2011) 349--393,
  [\href{http://arxiv.org/abs/1002.4886}{{\tt arXiv:1002.4886}}].

\bibitem{Sachdev:2011wg}
S.~Sachdev, {\it {What can gauge-gravity duality teach us about condensed
  matter physics?}},  {\em Ann. Rev. Condensed Matter Phys.} {\bf 3} (2012)
  9--33, [\href{http://arxiv.org/abs/1108.1197}{{\tt arXiv:1108.1197}}].

\bibitem{Benini:2012iq}
F.~Benini, {\it {Holography and condensed matter}},  {\em Fortsch. Phys.} {\bf
  60} (2012) 810--821, [\href{http://arxiv.org/abs/1202.6008}{{\tt
  arXiv:1202.6008}}].

\bibitem{Salvio:2013ja}
A.~Salvio, {\it {Superconductivity, superfluidity and holography}},  {\em J.
  Phys. Conf. Ser.} {\bf 442} (2013) 012040,
  [\href{http://arxiv.org/abs/1301.0201}{{\tt arXiv:1301.0201}}].

\bibitem{Musso:2014efa}
D.~Musso, {\it {Introductory notes on holographic superconductors}},  {\em PoS}
  {\bf Modave2013} (2013) 004, [\href{http://arxiv.org/abs/1401.1504}{{\tt
  arXiv:1401.1504}}].

\bibitem{Cai:2015cya}
R.-G. Cai, L.~Li, L.-F. Li, and R.-Q. Yang, {\it {Introduction to holographic
  superconductor models}},  {\em Sci. China Phys. Mech. Astron.} {\bf 58}
  (2015) 060401, [\href{http://arxiv.org/abs/1502.00437}{{\tt
  arXiv:1502.00437}}].

\bibitem{Gubser:2008zu}
S.~S. Gubser, {\it {Colorful horizons with charge in anti-de Sitter space}},
  {\em Phys. Rev. Lett.} {\bf 101} (2008) 191601,
  [\href{http://arxiv.org/abs/0803.3483}{{\tt arXiv:0803.3483}}].

\bibitem{Gubser:2008wv}
S.~S. Gubser and S.~S. Pufu, {\it {The gravity dual of a p-wave
  superconductor}},  {\em JHEP} {\bf 11} (2008) 033,
  [\href{http://arxiv.org/abs/0805.2960}{{\tt arXiv:0805.2960}}].

\bibitem{Arias:2012py}
R.~E. Arias and I.~S. Landea, {\it {Backreacting p-wave superconductors}},
  {\em JHEP} {\bf 01} (2013) 157, [\href{http://arxiv.org/abs/1210.6823}{{\tt
  arXiv:1210.6823}}].

\bibitem{Roberts:2008ns}
M.~M. Roberts and S.~A. Hartnoll, {\it {Pseudogap and time reversal breaking in
  a holographic superconductor}},  {\em JHEP} {\bf 08} (2008) 035,
  [\href{http://arxiv.org/abs/0805.3898}{{\tt arXiv:0805.3898}}].

\bibitem{Herzog:2014tpa}
C.~P. Herzog, K.-W. Huang, and R.~Vaz, {\it {Linear resistivity from
  non-{A}belian black holes}},  {\em JHEP} {\bf 11} (2014) 066,
  [\href{http://arxiv.org/abs/1405.3714}{{\tt arXiv:1405.3714}}].

\bibitem{Gangopadhyay:2012gx}
S.~Gangopadhyay and D.~Roychowdhury, {\it {Analytic study of properties of
  holographic p-wave superconductors}},  {\em JHEP} {\bf 08} (2012) 104,
  [\href{http://arxiv.org/abs/1207.5605}{{\tt arXiv:1207.5605}}].

\bibitem{Arias:2014msa}
R.~E. Arias and I.~S. Landea, {\it {Hydrodynamic modes of a holographic
  $p$-wave superfluid}},  {\em JHEP} {\bf 11} (2014) 047,
  [\href{http://arxiv.org/abs/1409.6357}{{\tt arXiv:1409.6357}}].

\bibitem{Zeng:2009dr}
H.-B. Zeng, Z.-Y. Fan, and H.-S. Zong, {\it {Superconducting coherence length
  and magnetic penetration depth of a p-wave holographic superconductor}},
  {\em Phys. Rev.} {\bf D81} (2010) 106001,
  [\href{http://arxiv.org/abs/0912.4928}{{\tt arXiv:0912.4928}}].

\bibitem{Zeng:2010fs}
H.-B. Zeng, W.-M. Sun, and H.-S. Zong, {\it {Supercurrent in p-wave holographic
  superconductor}},  {\em Phys. Rev.} {\bf D83} (2011) 046010,
  [\href{http://arxiv.org/abs/1010.5039}{{\tt arXiv:1010.5039}}].

\bibitem{Basu:2009vv}
P.~Basu, J.~He, A.~Mukherjee, and H.-H. Shieh, {\it {Hard-gapped holographic
  superconductors}},  {\em Phys. Lett.} {\bf B689} (2010) 45--50,
  [\href{http://arxiv.org/abs/0911.4999}{{\tt arXiv:0911.4999}}].

\bibitem{Gubser:2010dm}
S.~S. Gubser, F.~D. Rocha, and A.~Yarom, {\it {Fermion correlators in
  non-Abelian holographic superconductors}},  {\em JHEP} {\bf 11} (2010) 085,
  [\href{http://arxiv.org/abs/1002.4416}{{\tt arXiv:1002.4416}}].

\bibitem{Giordano:2016tws}
G.~L. Giordano, N.~E. Grandi, and A.~R. Lugo, {\it {Fermionic spectral
  functions in backreacting p-wave superconductors at finite temperature}},
  \href{http://arxiv.org/abs/1610.04268}{{\tt arXiv:1610.04268}}.

\bibitem{Ammon:2009xh}
M.~Ammon, J.~Erdmenger, V.~Grass, P.~Kerner, and A.~O'Bannon, {\it {On
  holographic p-wave superfluids with back-reaction}},  {\em Phys. Lett.} {\bf
  B686} (2010) 192--198, [\href{http://arxiv.org/abs/0912.3515}{{\tt
  arXiv:0912.3515}}].

\bibitem{Herzog:2009ci}
C.~P. Herzog and S.~S. Pufu, {\it {The second sound of SU(2)}},  {\em JHEP}
  {\bf 04} (2009) 126, [\href{http://arxiv.org/abs/0902.0409}{{\tt
  arXiv:0902.0409}}].

\bibitem{Akhavan:2010bf}
A.~Akhavan and M.~Alishahiha, {\it {P-wave holographic insulator/superconductor
  phase transition}},  {\em Phys. Rev.} {\bf D83} (2011) 086003,
  [\href{http://arxiv.org/abs/1011.6158}{{\tt arXiv:1011.6158}}].

\bibitem{Cai:2010zm}
R.-G. Cai, Z.-Y. Nie, and H.-Q. Zhang, {\it {Holographic phase transitions of
  p-wave superconductors in {G}auss-{B}onnet gravity with back-reaction}},
  {\em Phys. Rev.} {\bf D83} (2011) 066013,
  [\href{http://arxiv.org/abs/1012.5559}{{\tt arXiv:1012.5559}}].

\bibitem{Erdmenger:2012zu}
J.~Erdmenger, D.~Fernandez, and H.~Zeller, {\it {New transport properties of
  anisotropic holographic superfluids}},  {\em JHEP} {\bf 04} (2013) 049,
  [\href{http://arxiv.org/abs/1212.4838}{{\tt arXiv:1212.4838}}].

\bibitem{Manvelyan:2008sv}
R.~Manvelyan, E.~Radu, and D.~H. Tchrakian, {\it {New {A}d{S} non-{A}belian
  black holes with superconducting horizons}},  {\em Phys. Lett.} {\bf B677}
  (2009) 79--87, [\href{http://arxiv.org/abs/0812.3531}{{\tt
  arXiv:0812.3531}}].

\bibitem{Chrusciel:1987jr}
P.~T. Chrusciel and W.~Kondracki, {\it {Some global charges in classical
  {Yang-Mills} theory}},  {\em Phys. Rev.} {\bf D36} (1987) 1874--1881.

\bibitem{Brandt:1980em}
R.~A. Brandt and F.~Neri, {\it {Magnetic monopoles in $SU(n)$ gauge theories}},
   {\em Nucl. Phys.} {\bf B186} (1981) 84--108.

\bibitem{Forgacs:1979zs}
P.~Forgacs and N.~S. Manton, {\it {Space-time symmetries in gauge theories}},
  {\em Commun. Math. Phys.} {\bf 72} (1980) 15--35.

\bibitem{Bergmann:1978fi}
P.~G. Bergmann and E.~J. Flaherty, {\it {Symmetries in gauge theories}},  {\em
  J. Math. Phys.} {\bf 19} (1978) 212--214.

\bibitem{Harnad:1979in}
J.~P. Harnad, L.~Vinet, and S.~Shnider, {\it {Group actions on principal
  bundles and invariance conditions for gauge fields}},  {\em J. Math. Phys.}
  {\bf 21} (1980) 2719--2724.

\bibitem{Shepherd}
B.~L. Shepherd, {\em {E}instein-{Y}ang-{M}ills black holes in anti-de {S}itter
  space}.
\newblock PhD thesis, University of Sheffield, 2012.

\bibitem{Press:1992zz}
W.~H. Press, S.~A. Teukolsky, W.~T. Vetterling, and B.~P. Flannery, {\em
  {Numerical recipes in FORTRAN: The art of scientific computing}}.
\newblock {C}ambridge {U}niversity {P}ress, 1992.

\bibitem{Alken}
P.~Alken, M.~Booth, J.~Davies, M.~Galassi, B.~Gough, G.~Jungman, F.~Rossi, and
  J.~Theiler, {\em {GNU scientific library reference manual}}.
\newblock Network Theory Ltd., third~ed., 2009.

\bibitem{Chan:1980xv}
H.-M. Chan and S.~T. Tsou, {\it {On the characterization of monopoles in
  nonabelian gauge theories}},  {\em Phys. Lett.} {\bf B95} (1980) 395--400.

\bibitem{Creighton:1995au}
J.~D.~E. Creighton and R.~B. Mann, {\it {Quasilocal thermodynamics of dilaton
  gravity coupled to gauge fields}},  {\em Phys. Rev.} {\bf D52} (1995)
  4569--4587, [\href{http://arxiv.org/abs/gr-qc/9505007}{{\tt gr-qc/9505007}}].

\bibitem{Goddard:1976qe}
P.~Goddard, J.~Nuyts, and D.~I. Olive, {\it {Gauge theories and magnetic
  charge}},  {\em Nucl. Phys.} {\bf B125} (1977) 1--28.

\bibitem{Kleihaus:2001ti}
B.~Kleihaus, J.~Kunz, A.~Sood, and M.~Wirschins, {\it {Horizon properties of
  {E}instein-{Y}ang-{M}ills black holes}},  {\em Phys. Rev.} {\bf D65} (2002)
  061502, [\href{http://arxiv.org/abs/gr-qc/0110084}{{\tt gr-qc/0110084}}].

\bibitem{Sudarsky:1992ty}
D.~Sudarsky and R.~M. Wald, {\it {Extrema of mass, stationarity, and staticity,
  and solutions to the {E}instein {Y}ang-{M}ills equations}},  {\em Phys. Rev.}
  {\bf D46} (1992) 1453--1474.

\bibitem{Tafel:1982zs}
J.~Tafel and A.~Trautman, {\it {Can poles change color?}},  {\em J. Math.
  Phys.} {\bf 24} (1983) 1087--1092.

\bibitem{Oh:1987kb}
C.~H. Oh, C.~P. Soo, and C.~H. Lai, {\it {Global gauge transformations and
  conserved, gauge invariant electric and magnetic charges in {Y}ang-{M}ills
  gauge theories}},  {\em Phys. Rev.} {\bf D36} (1987) 2532--2538.

\bibitem{Corichi:1999nw}
A.~Corichi and D.~Sudarsky, {\it {Mass of colored black holes}},  {\em Phys.
  Rev.} {\bf D61} (2000) 101501,
  [\href{http://arxiv.org/abs/gr-qc/9912032}{{\tt gr-qc/9912032}}].

\bibitem{Kleihaus:2002ee}
B.~Kleihaus, J.~Kunz, and F.~Navarro-Lerida, {\it {Rotating
  {E}instein-{Y}ang-{M}ills black holes}},  {\em Phys. Rev.} {\bf D66} (2002)
  104001, [\href{http://arxiv.org/abs/gr-qc/0207042}{{\tt gr-qc/0207042}}].

\bibitem{Balasubramanian:1999re}
V.~Balasubramanian and P.~Kraus, {\it {A stress tensor for anti-de {S}itter
  gravity}},  {\em Commun. Math. Phys.} {\bf 208} (1999) 413--428,
  [\href{http://arxiv.org/abs/hep-th/9902121}{{\tt hep-th/9902121}}].

\end{thebibliography}\endgroup

\end{document}